\documentclass[useAMS,usenatbib]{mn2e}
\DeclareTextSymbol{\degre}{OT1}{23}
\pdfoutput=1
\usepackage{multirow}
\usepackage{graphicx}
\usepackage{epstopdf}
\usepackage{epsfig}
\usepackage{gensymb}
\usepackage{enumerate}
\usepackage{subfigure}
\usepackage{morefloats}
\usepackage{color,soul}

\usepackage{caption}
\usepackage[percent]{overpic}
\usepackage{multirow}
\usepackage{mathtools}
\usepackage{amsmath}
\usepackage{float}
\usepackage{pdflscape}

\usepackage{boxhandler}

\title[Imaging Spectroscopy of 4285 HII region candidates.]
{NGC628 with SITELLE : I. Imaging Spectroscopy of 4285 HII region candidates.}
\author[Rousseau-Nepton et al.]
{L. Rousseau-Nepton$^{1,2,3}$\thanks{E-mail: r-nepton@cfht.hawaii.edu}, C. Robert$^{1,2,3}$, R. P. Martin$^3$, L. Drissen$^{1,2,3}$, \newauthor and T. Martin$^2$ \\
1 Canada-France-Hawaii Telescope, Kamuela, HI, 96743, USA \\
2 D\'epartement de physique, de g\'enie physique et d'optique,
Universit\'e Laval, Qu\'ebec, QC, G1V 0A6, CA \\ 
\,\,\,\,\,Centre de Recherche en Astrophysique du Qu\'ebec \\
3 Department of Physics and Astronomy, University of Hawaii at Hilo, Hilo, HI, 96720, USA}
\begin{document}

\date{Accepted 2018. Received 2017; in original form 2018}

\pagerange{\pageref{firstpage}--\pageref{lastpage}} \pubyear{2018}

\maketitle

\label{firstpage}

\begin{abstract}

This is the first paper of a series dedicated to nebular physics and the chemical evolution of nearby galaxies by investigating large samples of HII regions with the CFHT imaging spectrograph SITELLE. We present a technique adapted to imaging spectroscopy to identify and extract parameters from 4285 HII region candidates found in the disc of NGC\,628. Using both the spatial and spectral capabilities of SITELLE, our technique enables the extraction of the position, dust extinction, velocity, H$\alpha$ profile, diffuse ionized gas (DIG) background, luminosity, size, morphological type, and the emission line fluxes for individual spaxels and the integrated spectrum for each region. We have produced a well-sampled HII region luminosity function and studied its variation with galactocentric radius and level of the DIG background. We found a slope $\alpha$ of $-$1.12$\pm$0.03 with no evidence of a break at high luminosity. Based on the width of the region profile, bright regions are rather compact, while faint regions are seen over a wide range of sizes. The radius function reveals a slope of $-$1.81$\pm$0.02. BPT diagrams of the individual spaxels and integrated line ratios confirm that most detections are HII regions. Also, maps of the line ratios show complex variations of the ionisation conditions within HII regions. All this information is compiled in a new catalogue for HII regions. The objective of this database is to provide a complete sample which will be used to study the whole parameter space covered by the physical conditions in active star-forming regions.

\end{abstract}


\section{INTRODUCTION}

The study of the ionized gas components in galaxies, for instance the population of HII regions as well as the diffuse ionized gas (DIG, also referred as the diffuse warm ionized medium WIM), helps gather important clues on the mechanisms that drives star formation and galaxy evolution. The gas emission lines are sensitive to the physical conditions in the interstellar medium (ISM) and their relative intensities probe underlying mechanisms involved in ionisation processes. One predominant example related to the ISM ionisation is the recent star formation sites where massive stars act as the main source of ionizing photons. Gas ionisation can also be triggered by other sources such as the energetic emission released by an active galactic nucleus (e.g. \citealt{h08}; \citealt{da14}), the global evolved stars contribution to the heating of the ISM (e.g. \citealt{f11}), different sources of small- and large-scale shocks in the ISM (e.g. \citealt{a08}), etc.

The gas strong emission lines can be observed at cosmological distances, allowing the study of various star-forming environments through time. Establishing reliable HII region and DIG physical characteristics are therefore key elements in probing galaxy's chemo-dynamical evolution and improving star formation models. The extraction of these physical characteristics often involves photoionisation codes, like MAPPINGS (\citealt{s93}; \citealt{n13}) and CLOUDY (\citealt{f98}). These codes are used to model the complex interaction between the ionizing spectrum of different sources and the ISM. To constrain photoionisation models, one needs spatially resolved star-forming region spectra including a set of different emission lines. 

From a historical perspective, imagers and Fabry-Perot instruments were often used to observe simultaneously a large number of HII regions in nearby galaxies. Imagers combined with different narrow-band filters allow us to target a few important emission lines and continuum windows (used to isolate the emission line flux from the presence of the underlying stellar and nebular continuum; \citealt{m09}, \citealt{s11}). Nevertheless, imaging techniques are limited by the number of emission lines accessible and are also affected by problems related to the spectral shift of the lines due to nebular and galactic dynamics, which prevent the accurate measurement of the lines (as mentioned by\citealt{y96}, \citealt{h01}, and \citealt{k08}). High-resolution spectra produced from Fabry-Perot interferometers can provide additional information on stellar and gas dynamics. However, as their spectral range is limited, Fabry-Perot interferometers were mostly used for the observation of one emission or absorption line (e.g. \citealt{f07} and \citealt{e08}). For this reason, most of the complex spectral analysis of extragalactic HII regions have been performed using slit spectrograph data, as they enable accurate measurements of multiple emission lines over a wide spectral window (e.g. \citealt{b15}). More recently, integral field spectrographs (IFS) using dispersive techniques offered a breakthrough by giving simultaneously spectral and spatial information on extragalactic HII regions (e.g. \citealt{k16}). Still, the incomplete spatial coverage of these instruments is not convenient for a detailed study of extragalactic HII regions and to extract some of their intrinsic properties (e.g. total luminosity and morphology). 

Imaging Fourier Transform Spectrograph (IFTS) have advantages over these other intruments. The multiplex advantage has to be mentioned first as multiple scans, or spectra, can be performed at a faster rate than most dispersive spectrographs. Secondly, the throughput of an IFTS is significantly higher than slit spectrograph while nothing limits the amount of light getting in the cameras and fewer mirrors are required. Finally, the sampling of the scans can be adjusted to reach a desired spectral resolution. Nevertheless, for an IFST working in the visible, constraining the bandpass using narrow- to wide-band filters is necessary to reach an appropriate spectral resolution in a reasonable amount of time (a few hours) and limit the photon noise. All things considered, the large field-of-view resulting from the multiplex advantage, the high throughput, and the complete spatial coverage make the IFST an ideal instrument to study extended object, and more particularly here, large nearby galaxies. 

Some fundamental problems have been outlined from the analysis of extragalactic HII and DIG regions in nearby galaxies with integral field spectroscopy using disperse units:\,\begin{itemize}
\vspace{-0.25cm}
\item[{1)}]{The lack of spatial resolution is preventing us from resolving individual HII regions and is imposing a great limitation on the scale of the physical processes studied. Even when analyzing the large scale properties of galaxies, it has been shown that the gas abundance [O/H] gradient (derived from emission line diagnostics) is affected by a low spatial resolution, resulting in an artificial flattening of the gradients with respect to the redshift (\citealt{y13}; \citealt{m14}). In addition, the lack of spatial resolution induces an attenuation on the extinction correction factor obtained from the Balmer decrement in integrated spectra as the extinction is biased toward the brightest and less dusty regions (\citealt{v14}). As a result, the derived star formation rate (SFR) and stellar mass content of HII regions are only upper limits.} 
\item[{2)}]{A variable contribution from the superimposed DIG emission is often pointed out as a source of uncertainties when trying to extract the HII region properties (\citealt{b09}; \citealt{k16}; \citealt{s17}).}
\item[{3)}]{A degeneracy effect between the age of the HII regions and the under-sampling of the massive end of the initial mass function (IMF) precludes proper modeling of the ionized gas (\citealt{v10}). Without resolving the stellar content, additional information on the ionisation structure surrounding the stars is required using photoionisation models to determine the precise stellar content of the regions.}
\item[{4)}]{Most of the diagnostic tools proposed to gather the abundances and other ionized gas properties are designed for integrated spectra (i.e. at low spatial resolution) and are calibrated using a small, sometimes biased, sample of HII regions with well-defined properties. They are therefore limited tools, not representative of a large variety of environments within galaxies, and are not suited for a desired detailed analysis of resolved HII regions.}
\end{itemize}\,\vspace{-0.25cm} 

To overcome these numerous issues, we have conducted new observations of the nearby spiral galaxy NGC\,628 with SITELLE, the Canada-France-Hawaii IFTS. SITELLE allows a high-spatial resolution (35 pc) over the whole galaxy disc (FOV: 11$^\prime$$\times$11$^\prime$) with full coverage (100\% filling factor). NGC\,628 is a well-known spiral galaxy seen almost face-on. It has been widely observed by others with various instruments. For instance, \cite{r11} and \cite{s11} performed a spectral analysis and studied the global properties (abundance gradients) of NGC\,628 using a mosaic of the PPAK IFS ($\sim$6$^\prime$ circular aperture) with individual fiber spectra and 94 selected apertures. \cite{b15} used the Multi Object Double Spectrograph (MODS), as part of the CHAOS project, to target 62 bright HII regions over the galaxy disc and investigated the abundance dispersion using direct temperature measurements from the auroral line. \cite{g15} used multiple Hubble Space Telescope broad band UV and optical images over two 2.7$^\prime$$\times$2.7$^\prime$ fields in the disc of NGC\,628 to identify 1392 young stellar cluster candidates ($<$\,100\,Myr) and studied their size, mass, and age, and the concentration of the stars embedded in the clusters. More recently, with the Multi-Unit Spectroscopic Explorer (MUSE), \cite{k16} sampled 391 HII regions at the spatial resolution of 35\,pc, analyzed the DIG contribution, and studied the differences between the HII region properties in the arms and inter-arm regions. These previous studies have targeted different science goals and their contribution to the understanding of star formation in galaxies is notable. Complementary to these studies, we believe that SITELLE has the potential to overcome several of the problems related to sample selection biases, the lack of spatial resolution, the contribution of the DIG, as well as to explore line ratio variations in the ISM with greater details. \begin{figure*}
\begin{center}
\includegraphics[width=7in]{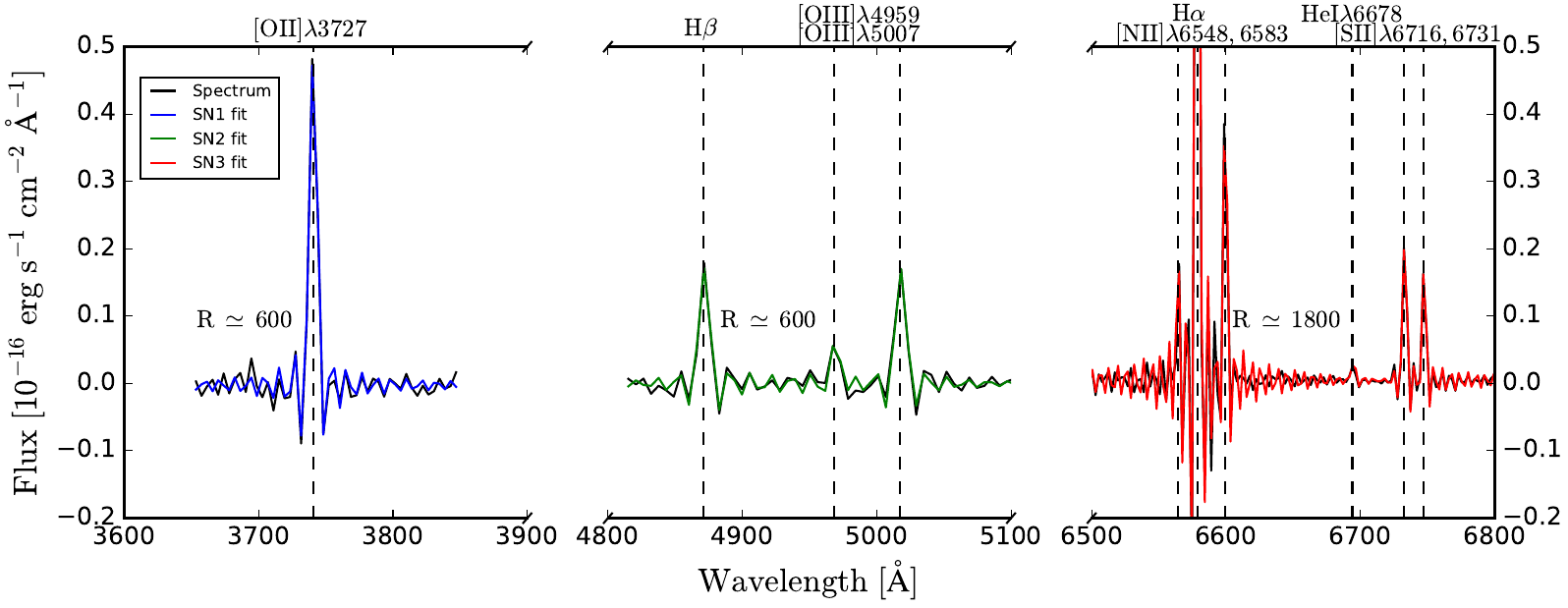} 
\caption{Continuum subtracted spectrum of an HII region in NGC\,628 extracted from the datacubes using a circular aperture with a radius of 1.6$^{\prime\prime}$, centered at RA 01h36m53.1s and DEC +15$\degree$48$^\prime$04.8$^{\prime\prime}$. The fits obtained with ORCS for the emission lines ($\S$\,\,3) are shown. Note that SITELLE's line profiles are fitted using a sine cardinal function.}
\label{spectrum}
\end{center}
\end{figure*}

The spectral datacubes obtained for this study with SITELLE triggered the development of new analysis tools tailored to its unique capabilities. In this paper, we present our approach for the systematic analysis of the emission line spectra of a large sample of regions observed with a high-spatial resolution. Our results are meticulously compiled in a general catalogue for the HII regions which is made available to the entire astronomical community. Our immediate goal is to demonstrate the impact of the spatial resolution, mainly through the evaluation of the DIG background and the effect of sampling on characterizing the HII regions.

This paper is the first of a series where the systematic study of many nearby galaxies will be done with SITELLE to enlarge the domain of HII region physical parameters analysed. All the data gathered for the HII regions will be added to the catalogue. It aims to be the largest and most detailed spectroscopic study of star-forming region properties in different galactic environments, and represents an important statistical tool to study: small-scale processes in chemical enrichment and mixing; the balance between ionizing photons absorbed by the gas, the UV flux observed from a young population, and the overall extinction; the DIG in relation with HII regions; the star formation efficiency and rate directly from the gas and stellar population characteristics; and the massive stars IMF in different environments. This catalogue is tailored for the development of new diagnostics with photoionisation codes and for the study of HII regions and their content at different spatial resolution. 

Sections 2 and 3 describe the observations and the technique used to properly measure the gas emission lines. Section\,4 focuses on the emission region definition, while Section\,5 presents the luminosity function, morphology, and size of the regions. Section\,6 presents the emission line ratios measured in individual spaxel within the emission regions, as well as global ratios for each region for a comparison with the literature. The results, along with the database content, are summarized in Section\,7.

\section{OBSERVATIONS AND DATA REDUCTION}

NGC\,628 was observed with SITELLE (Spectro-Imageur \`a Transform\'ee de Fourier pour l'\'Etude en Long et en Large des raies d'\'Emission) at the Canada-France-Hawaii Telescope during the instrument commissioning and science verification runs in August 2015 and January 2016. Table \ref{op} gives the main observing parameters. SITELLE is a Michelson interferometer inserted, with an inclination, into the collimated beam of an astronomical camera system (\citealt{d14}). Its has a large field-of-view (FOV: 11$^\prime$$\times$11$^\prime$) with complete spatial coverage, a high-resolution power up to R\,$\simeq$10\,000, and a broad wavelength range from 3500 to 9000\,\AA\,\,with an excellent efficiency all the way to the blue part of the spectrum. SITELLE is equipped with two E2V detectors (2048$\times$2064 pixels, resulting in a mean pixel scale of 0.321$^{\prime\prime}$$\times$0.321$^{\prime\prime}$). Filters are necessary with SITELLE in order to reduce the noise in a selected bandpass. A raw datacube from SITELLE is composed of a sequence of images (one for each position of the movable mirror) that, when taken together, results in more than 4 million interferograms (one interferogram for each pixel of the detectors). After the data reduction, SITELLE produces 4 million spectra in a single datacube for a selected filter. \begin{table}
\centering
\caption{Observing Parameters}
\scriptsize
\label{op}
\begin{tabular}{lccc}
\hline
\bf{Filter} & \bf{SN1} & \bf{SN2} & \bf{SN3} \\
\hline
Spectral range [\AA] & 3640-3850 & 4840-5120 & 6480-6860 \\
Observing date & 2016/01/13 & 2016/01/11 & 2015/08/09 \\
Exposure time/step [s] & 73.0 & 103.3 & 25.0 \\
Number of steps & 105 & 138 & 323 \\ 
Total exposure time [h] & 2.13 & 3.96 & 2.24 \\
Mean resolution R & $\sim$ 600 & $\sim$ 600 & $\sim$ 1800 \\
Image Quality [arcsec]& 1.28 & 0.96 & 0.90 \\
\hline
\end{tabular}
\end{table}

\begin{figure*}
\begin{center}
\includegraphics[width=6.6in]{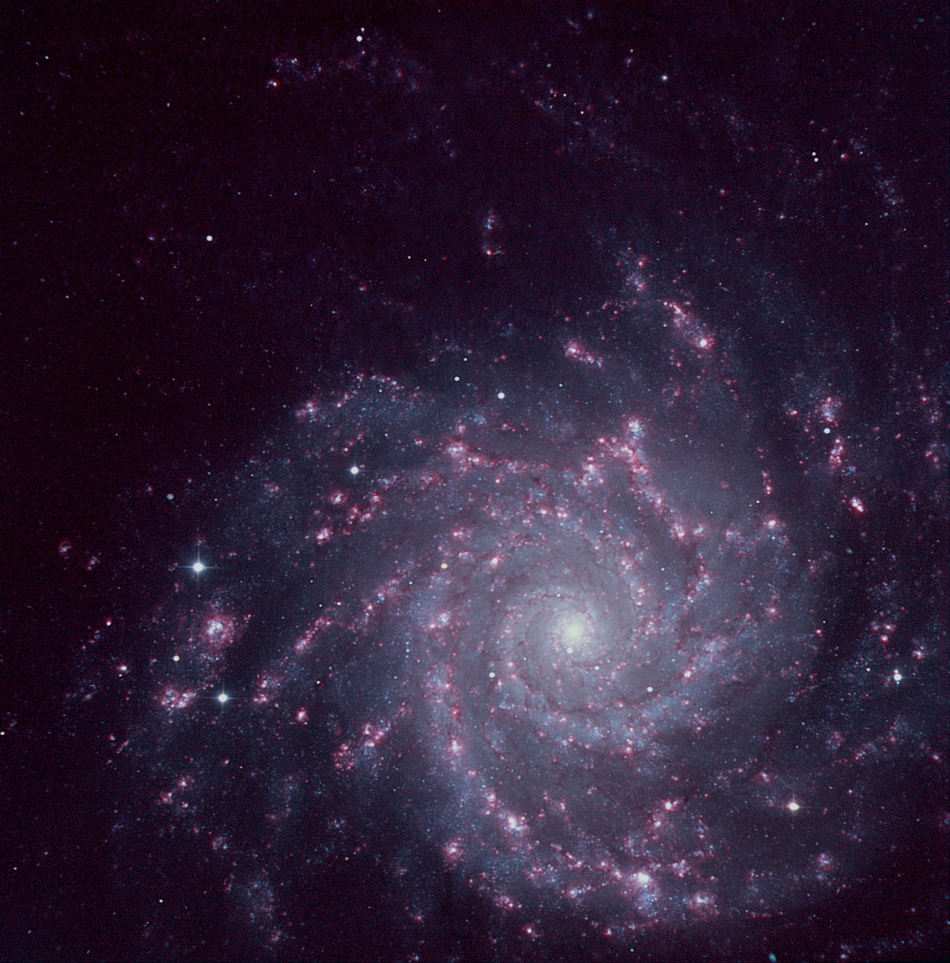} 
\caption{SITELLE's deep image of NGC\,628. For each pixel, the information from the three filters was summed together along with the H$\alpha$ intensity map. Adding the H$\alpha$ map highlights the ionized gas emission regions (in red on the image). North is up and East is left. The FOV is 11$^\prime$$\times$11$^\prime$ centred on RA 01h36m46.2s and DEC +15$\degree$48$^\prime$42.8$^{\prime\prime}$.}
\label{ALL}
\end{center}
\end{figure*}

For this project, three filters have been used : SN1, SN2, and SN3; Table \ref{op} gives the spectral range and the resolution adopted for each filter. These configurations enabled the measurement of multiple strong emission lines: [OII]$\lambda$3727, H$\beta$\,(4861\,\AA), [OIII]$\lambda$4959, [OIII]$\lambda$5007, [NII]$\lambda$6548, H$\alpha$\,(6563\,\AA), [NII]$\lambda$6583, HeI$\lambda$6678, [SII]$\lambda$6716, and [SII]$\lambda$6731. Figure\,\,\ref{spectrum} shows the integrated spectrum in all three filters for one HII region. Figure\,\,\ref{ALL} shows the deep image produced by adding, for each pixel, the whole signal from the three filters together with an enhanced contribution of the H$\alpha$ intensity map extracted from the line fitting procedure (as described in $\S$\,\,3). SITELLE's FOV was slightly offset from the galactic centre of NGC\,628 in order to cover a vast portion of the disc as well as the extended northern spiral arm; the FOV was centered at RA 01h36m46.2s and DEC +15$\degree$48$^\prime$42.8$^{\prime\prime}$.  

The data reduction was performed using the fully-parallelized reduction software ORBS (data-release 1) specifically developed for SITELLE (\citealt{mmm15}). The data reduction process requires additional calibrations from the instrument. Typical flat (using a white light source) were acquired during each night and a pre-scan section of the CCD was preserved to measure the bias contribution in each exposure. The flux calibration was performed using a datacube of the standard star GD\,71 secured with each filter to define the global transmission function (taking into account the minimum airmass value for the standard star and the galaxy, which is 1.002 for both objects). The flux calibrated spectrum of NGC\,628 is in good agreement with those from other studies (as shown in \citealt{md17}). 
The spectral calibration was done using a Helium-Neon laser datacube acquired during daytime. For the SN3 filter (at higher R), greater precision on the spectral calibration was obtained using the centroid position of sky lines ($\S$\,\,3.1). Additionally, three flat datacubes were acquired using a white light source (one for each filter) in order to apply high-order phase correction in the data reduction process (\citealt{mdm18}). To match the image quality obtained for the different datacubes (see Tab. \ref{op}) and attenuate the impact of a small misalignment of the datacubes, we applied a Gaussian convolution (with a FWHM\,$=$\,3\,pixels) on all the images within the datacubes. We did not correct for the Milky Way extinction knowing that it is very small (E(B$-$V)\,=\,0.062 according to NED\footnote{The NASA/IPAC Extragalactic Database (NED) is operated by the Jet Propulsion Laboratory, California Institute of Technology, under contract with the National Aeronautics and Space Administration.}).

\section{MEASUREMENTS}
\label{msrt}

A few manipulations are required before proceeding with the accurate measurements of the emission lines. These steps include the refinement of the spectral calibration for the SN3 datacube, sky subtraction, corrections for the astrometry and image distortion, and subtraction of the galaxy stellar contributions along the line of sight of each spaxel. Fitting of the emission lines is then performed. These operations are explained in the following subsections. 

\subsection{Spectral Calibration Refinement} 
\label{scr}
The spectral resolution of the SN3 filter was pushed further on, compared to the other filters (see Tab.\,\ref{op}), to allow us to fully separate the [SII] doublet, but also to enable the use of the strong H$\alpha$ emission line to extract the ionized gas kinematics. At this resolution, higher precision for the spectral calibration of the SN3 datacube can be achieved by fitting, over the FOV, an isolated and strong sky line available in the filter (a similar technique than the one used by \citealt{mmm16} and \citealt{s16}; this technique is now a standard procedure and is well documented in \citealt{mdm18}). Since the Helium-Neon laser datacube used for the spectral calibration was obtained with the telescope at zenith, some telescope and instrument flexures affecting the calibration may occur while pointing at the target. These flexures are then corrected by using sky-line centroid positions observed simultaneously. For this task, we selected the OH sky line at 6498.729\,\AA\,\,because of its proximity to the H$\alpha$ line while it remains uncontaminated by emission or absorption lines from other sources. The sky line signal is enhanced by summing its flux over boxes of 10$\times$10 spaxels. To get the spectral calibration correction, the sky line was fitted (see the description of line measurement technique in $\S$\,\,\ref{lfproc}) in order to obtain its centroid shift with respect to the expected rest frame wavelength. A constant shift of 83.2\,km\,s$^{-1}$ was determined over the FOV and considered for the correction of the velocity map (as expected according to Martin et al. 2016, 2018).

\subsection{Sky Subtraction} 

\begin{figure}
\begin{center}
\includegraphics[width=3.3in]{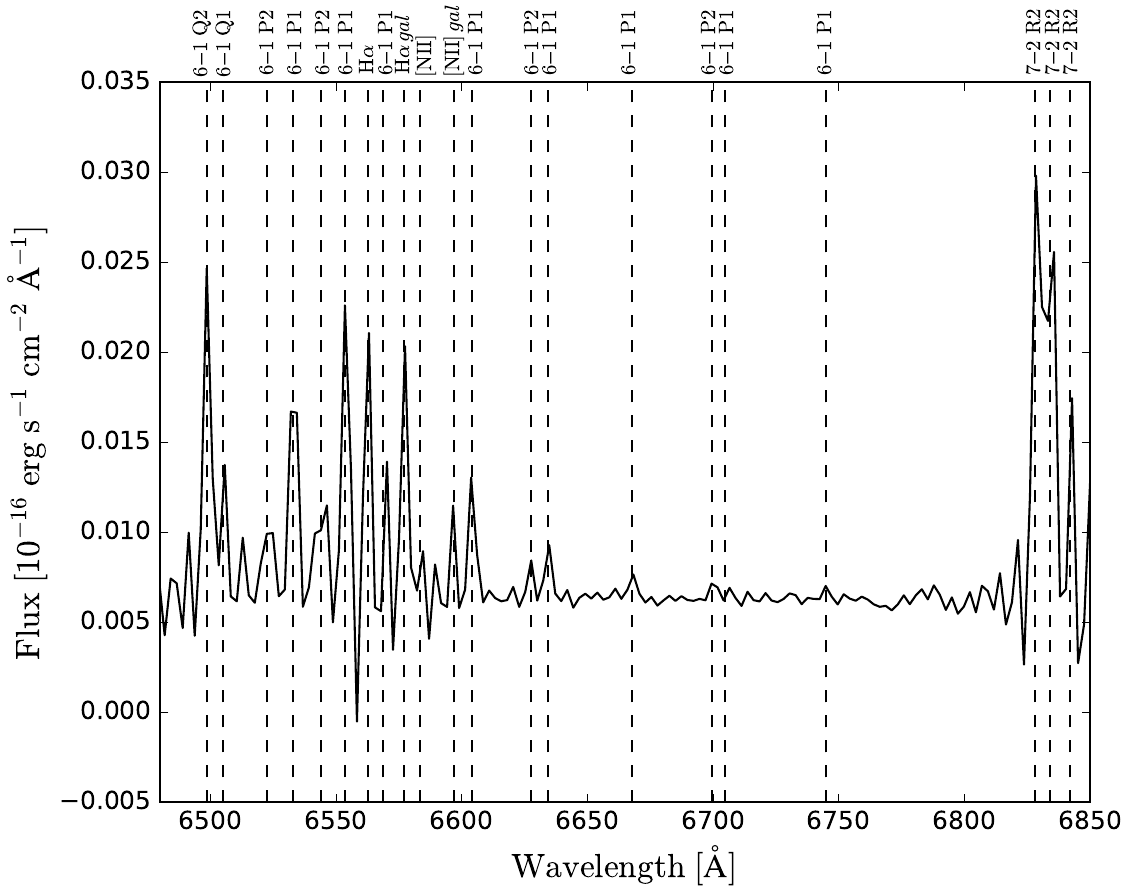} 
\caption{The median sky spectrum in the SN3 datacube. A total of 40\,000 spaxels have been combined in a region centred at the position RA 01h37m04.2s and DEC +15\degree49$^\prime$23.3$^{\prime\prime}$. It shows multiple known OH sky lines (marked by dashed vertical lines).}
\label{sp_sky}
\end{center}
\end{figure}

For each datacube, sky subtraction was performed using the median sky spectrum extracted from a region located as far as possible from the galaxy disc. A total of 40\,000 spaxels (200$\times$200 spaxels) were used centred at the position RA 01h37m04.2s and DEC +15\degree49$^\prime$23.3$^{\prime\prime}$. The sky spectrum in the SN3 datacube is shown in Figure\,\,\ref{sp_sky}. 

\subsection{Astrometry and Distortion Corrections} 

The astrometric solution for each datacube was calculated using bright stars in the FOV of the deep image and applied using ccmap and ccsetwcs (STSDAS/IRAF\footnote{IRAF is distributed by the National Optical Astronomy Observatories, which are operated by the Association of Universities for Research in Astronomy, Inc., under cooperative agreement with the National Science Foundation.}). We selected a TNX astrometric solution with a polynomial degree of 4. The accuracy of the solutions is below 0.324$^{\prime\prime}$ ($\sim$\,1 pixel) for most regions (all except a small group of $\sim$ 20 regions located in the SE corner of the FOV) detected in Section\,\,\ref{Proc} for all filters. Ultimately distortions were corrected using remap (WCStools) and all line flux maps (after the line fitting procedure described in $\S$\,\,3.5) were adjusted to correspond to the astrometric solution of the SN3 deep image. This step is crucial to measure reliable line ratios extracted from lines lying in different filters. \begin{figure}
\begin{center}
\includegraphics[width=3.3in]{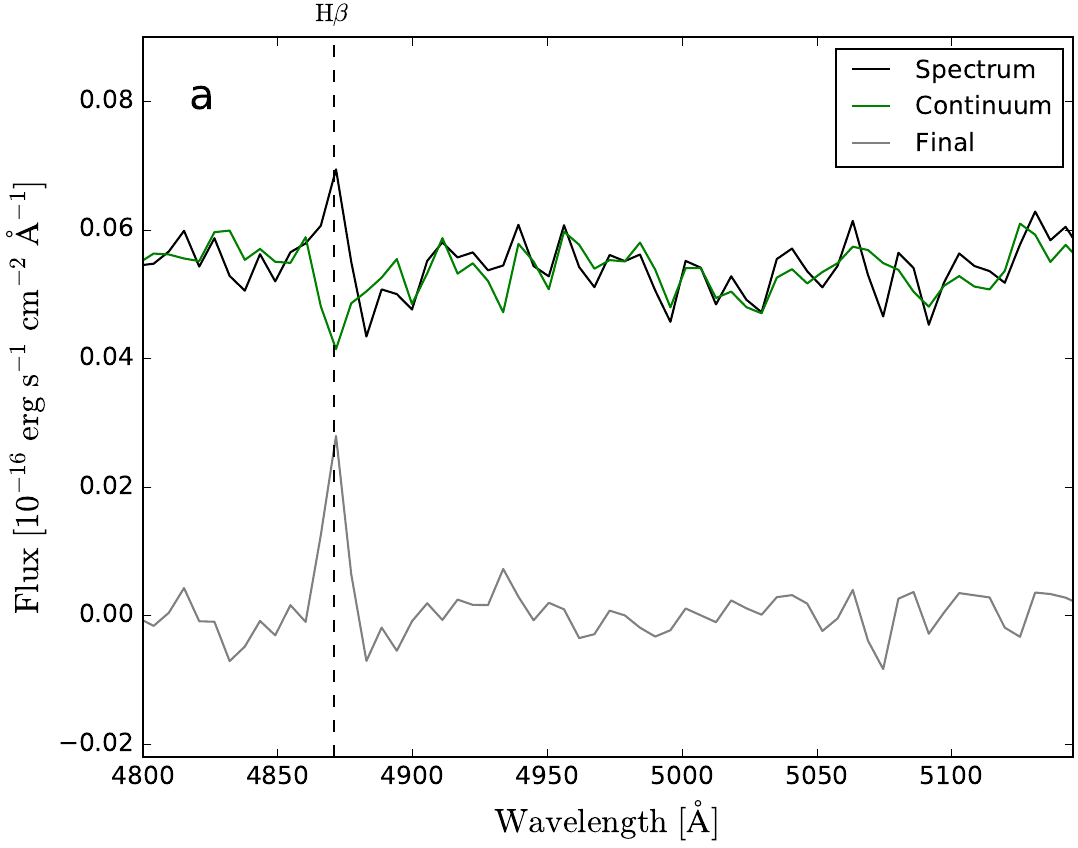} 
\includegraphics[width=3.3in]{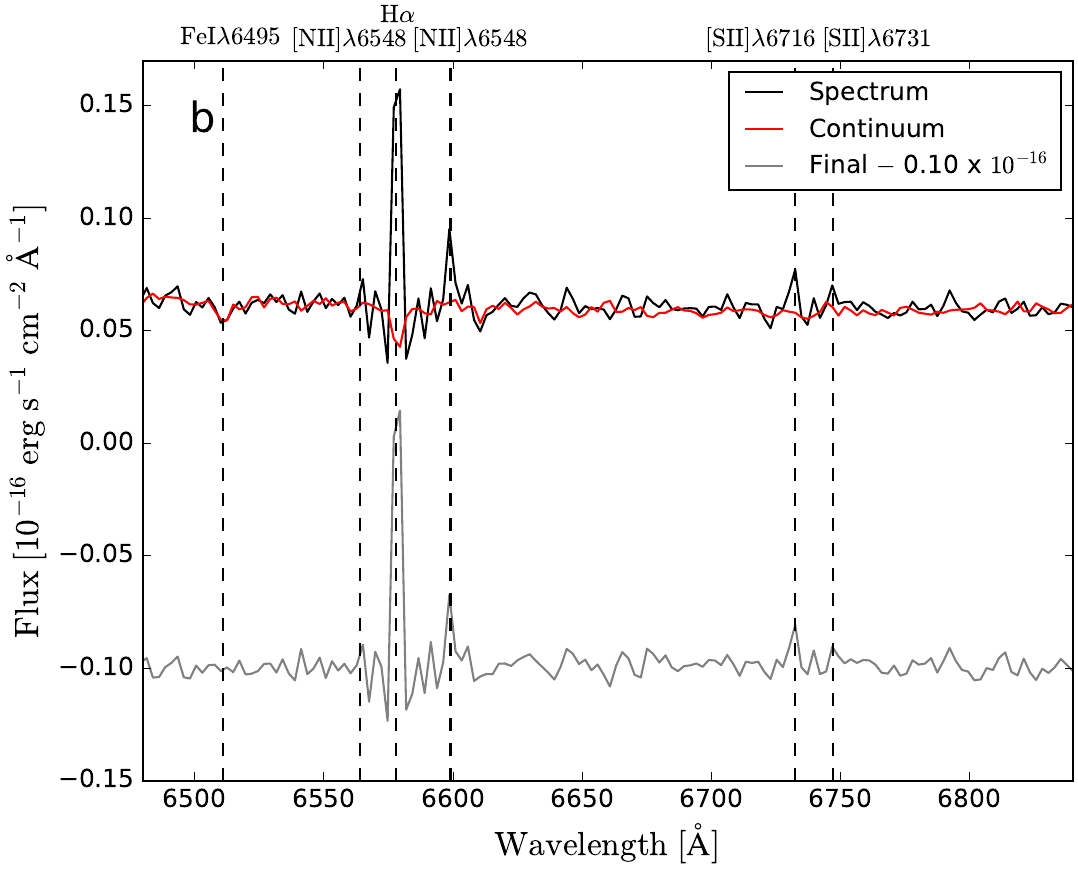} 
\caption{The spectrum for one pixel in a faint HII region close to the galaxy centre extracted at RA 01h36m41.5s and DEC +15$\degree$46$^{\prime}$57.0$^{\prime\prime}$ from the SN2 (a) and SN3 (b) datacubes. The reference spectrum used to represent the galaxy stellar populations on the line of sight of the HII region is shown shifted to the pixel velocity and scaled to the level of the continuum. The final spectrum obtained after subtraction is shown at the bottom. The SN3 final spectrum has been shifted downward by 0.10$\times10^{-16}$ for clarity.}
\label{sp_cont_HII_pixel2}
\end{center}
\end{figure} 
\begin{figure}
\begin{center}
\includegraphics[width=3.3in]{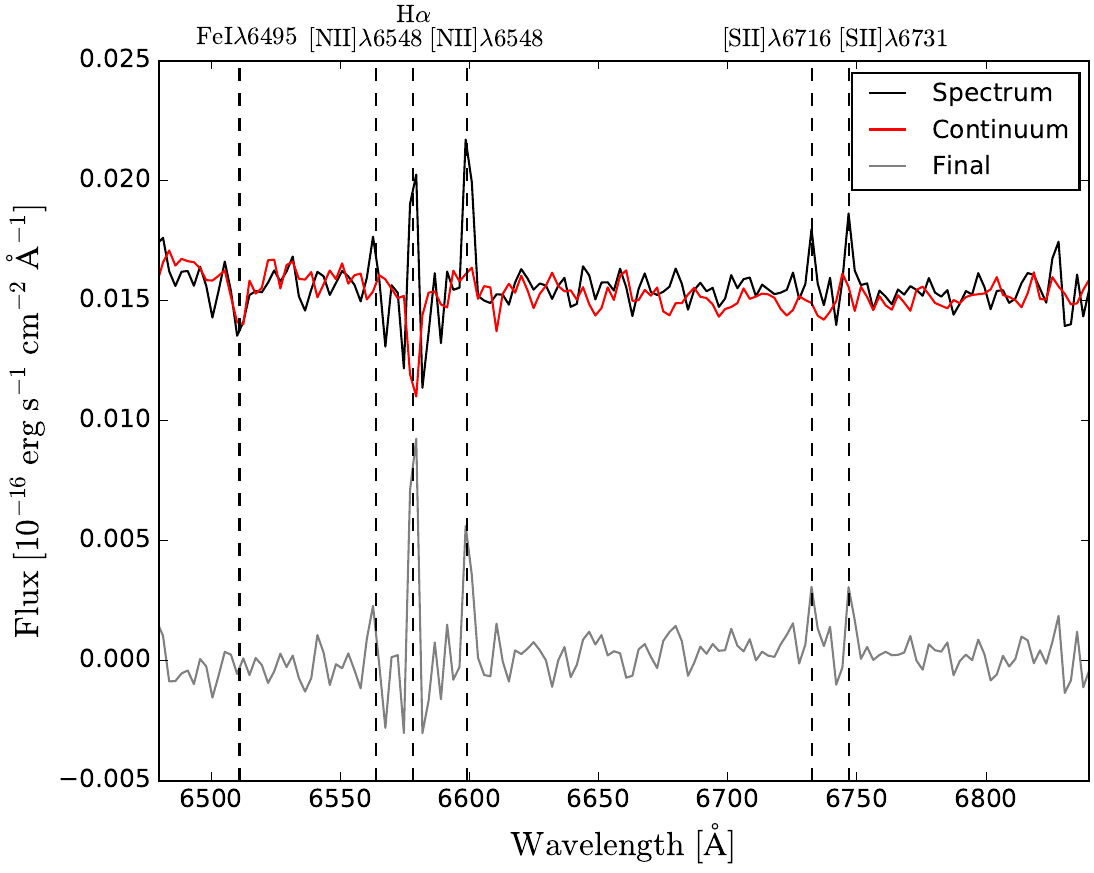} 
\caption{The spectrum of the DIG extracted from the SN3 datacube using a circular aperture with a radius of 8.0$^{\prime\prime}$, centred at RA 01h36m43.3s, DEC +15$\degree$46$^{\prime}$57.9$^{\prime\prime}$. The reference spectrum used to represent the galaxy stellar populations along the line of sight of the region is shown shifted to the pixel velocity and scaled to the level of the continuum. The final spectrum obtained after subtraction is shown at the bottom.}
\label{sp_cont_DIG_25}
\end{center}
\end{figure}

\subsection{Subtraction of the Stellar Population Contribution } 
\label{substel}

Stellar populations in galaxy discs are responsible for continuum emission and absorption features that add up to the spectrum of the ionized gas regions. In the visible domain, emission lines (and most importantly Balmer lines) are then a combination of a stellar absorption component (from the older galaxy stellar populations and, if the case, from the young stellar population embedded within an HII region) and the emission line from the ionized gas itself. Depending on the galaxy mass and type, and also on where an HII region is located (e.g. far out in the disc or near the galaxy centre), the measurement of its emission lines can be significantly affected by the disc and bulge older stellar populations (\citealt{r17}). Although \cite{gd05} indicate, for example, that the equivalent width of the H$\beta$ absorption line can reach 5\,\AA\ for a young population below 10\,Myr, the young cluster absorption features may also be important after its superposition with the old population, depending on the spatial resolution and morphology of the young cluster relative to the emission region. 

We first created a reference spectrum for the old disc/bulge stellar population for each datacube considering a small circular aperture of 5.5$^{\prime\prime}$ centered on the galaxy continuum emission peak (emission line regions are not detected in this aperture). The choice of this central aperture (compared to specific regions in the galaxy disc) allows us to minimise the noise in the reference spectrum. This reference spectrum also offers the advantage of containing the exact information about the instrumental lines profile (i.e. a sinc profile) as compared to a stellar population model. A comparison of spectra created at different galactocentric radii does not show a significant variation in the absorption lines strength and width when the noise is taken into account. This is consistent with the study of the stellar population characteristics and the velocity dispersion done by Rousseau-Nepton (2017) using long-slit spectroscopy data (between 3500 and 7500\,\AA) across the disc and nucleus of NGC 628. It supports the idea that our reference spectrum is a good representation of the old galaxy stellar contribution all over the disc, considering an adjustment of its velocity, using a preliminary interpolated velocity map, and an adjustment of its continuum level. For this last step we decided to simply match the continuum level of the old population with the continuum level of each spaxel studied, assuming that it will be a sufficient correction for the young population contribution, if present. For NGC\,628, SITELLE gives us a spatial resolution of 35\,pc, which allows us to see, with limited details, the spatial distribution of the ionized gas relative to the young stellar clusters: a young cluster (considering variation in the continuum level) may sometimes be seen sitting in the centre of an emission region, or it is not seen at all, or it is sitting at the edge of it. Not to mention as well that a young cluster and its related emission region may not share the same extinction; this is a complex problem that we will address in another paper. Nevertheless, here we ran some tests to see what would be the error on the emission line flux if we had specifically considered young population spectra. To do so, we combined the average old population spectrum with young (6 and 10\,Myr) population models from Starburst99 (\citealt{l99}) in various proportions. For example, for HII regions at a large galactocentric radius (we considered regions at 6 and 14\,kpc), if we suppose that the young population contributes to 100\% of the continuum, relative to the case where 100\% of the continuum is associated with the old population, we see a maximum variation of 5\% and 1\% in the 
H$\beta$ and H$\alpha$ flux, respectively, if the region has weak emission lines and presents a high continuum level. We find a similar variation of the line flux for HII regions at a small galactic radius if we consider that the young population accounts for 50\% of the continuum, relative to the case where 100\% of the continuum is associated with the old population. These are extreme examples for the integrated flux from HII regions, considering that while the old population is present in all the spaxels where the emission lines are measured, it is probably not the case for the young population clusters. But it is more realistic when studying individual spaxels. Nevertheless, the flux variations measured in our tests are small and we are therefore confident that our flux measurements are good when considering the old galaxy stellar population with its continuum level scaled with the one of the individual spaxels or the integrated regions.

Figures \ref{sp_cont_HII_pixel2} and \ref{sp_cont_DIG_25} show the effect of the subtraction of the stellar population reference spectrum for the H$\beta$ (SN2) and H$\alpha$ (SN3) emission line of one spaxel selected in the centre of a faint HII region, and for the H$\alpha$ (SN3) emission line in a DIG region. The stellar population reference spectrum is superimposed to the spectrum of these regions in the figures. The final spectrum obtained after the subtraction is also shown. It is important to realize that the oscillations around the emission lines are mostly the signature of the sinc instrumental profile. These oscillations naturally display large negative side-lobes on each side of the line centre, which must not be confused with the shape of an absorption component. These figures clearly show that it is important to subtract the stellar population prior to measure the emission lines. For weak emission regions, an emission component in the lines, especially for H$\beta$, may suddenly show up after the subtraction of the stellar population spectrum.

\begin{figure}
\begin{center}
\includegraphics[width=3.3in]{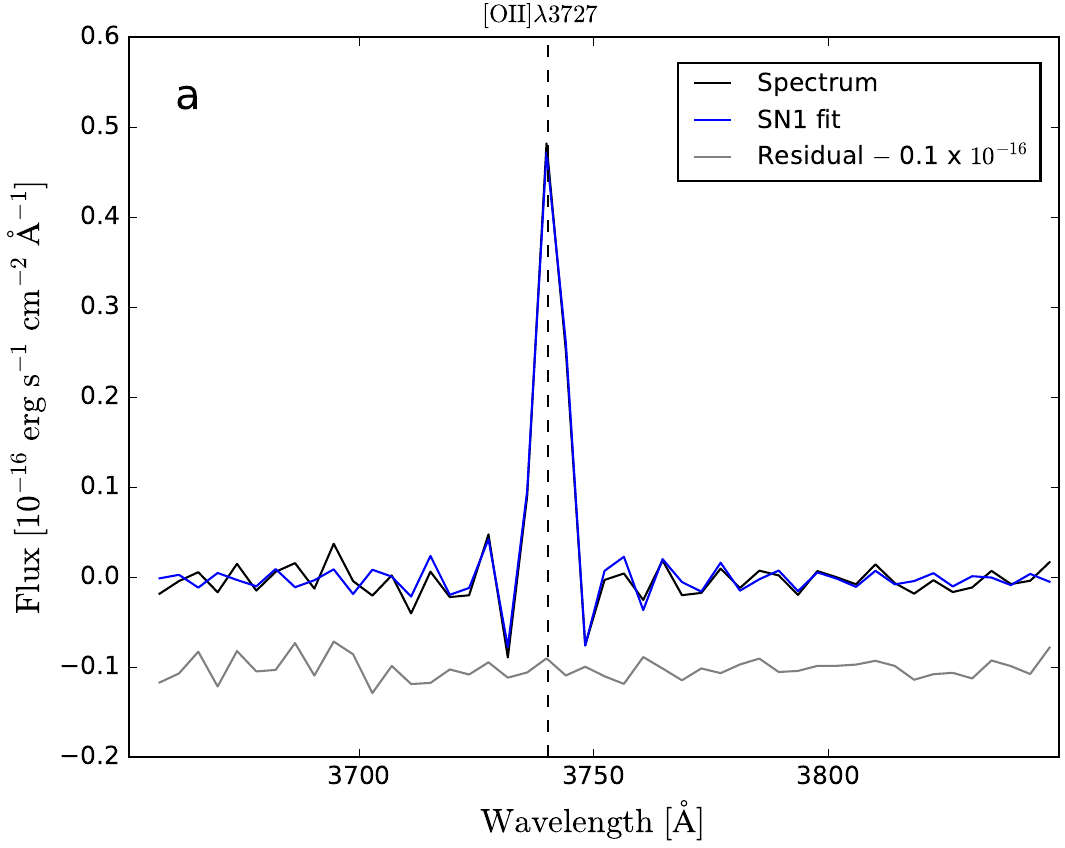} 
\includegraphics[width=3.3in]{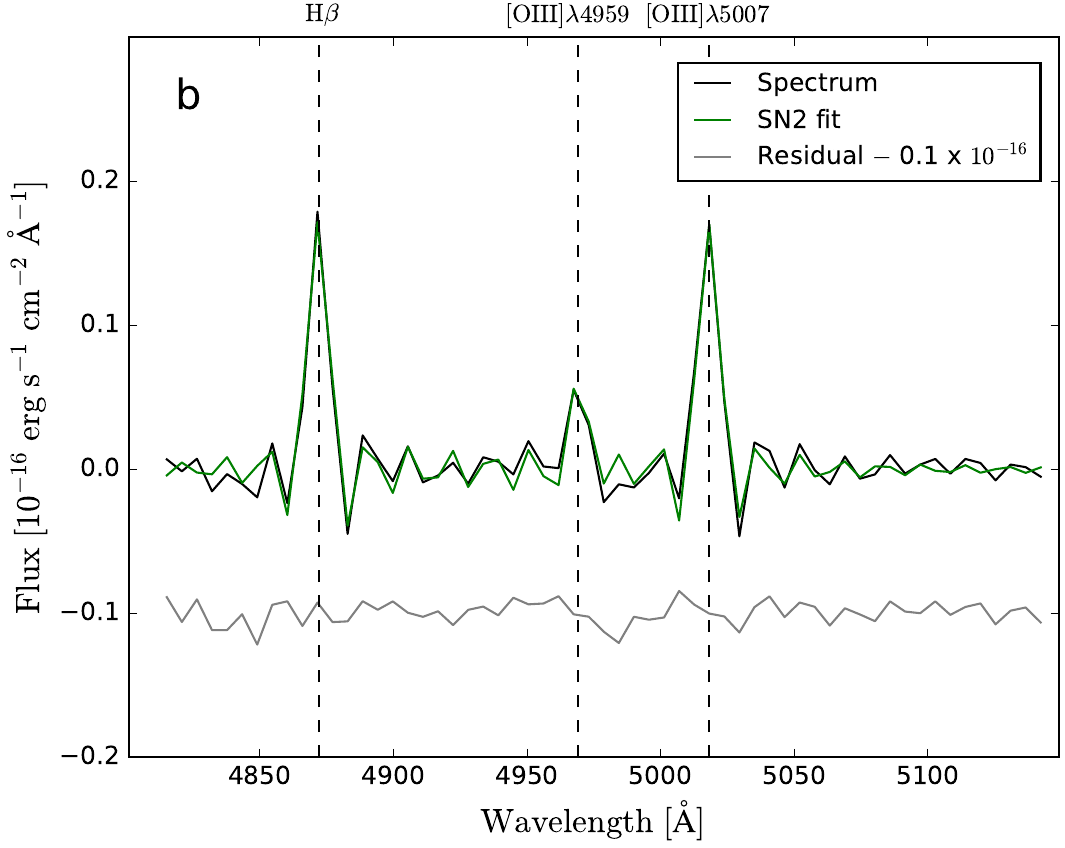} 
\includegraphics[width=3.3in]{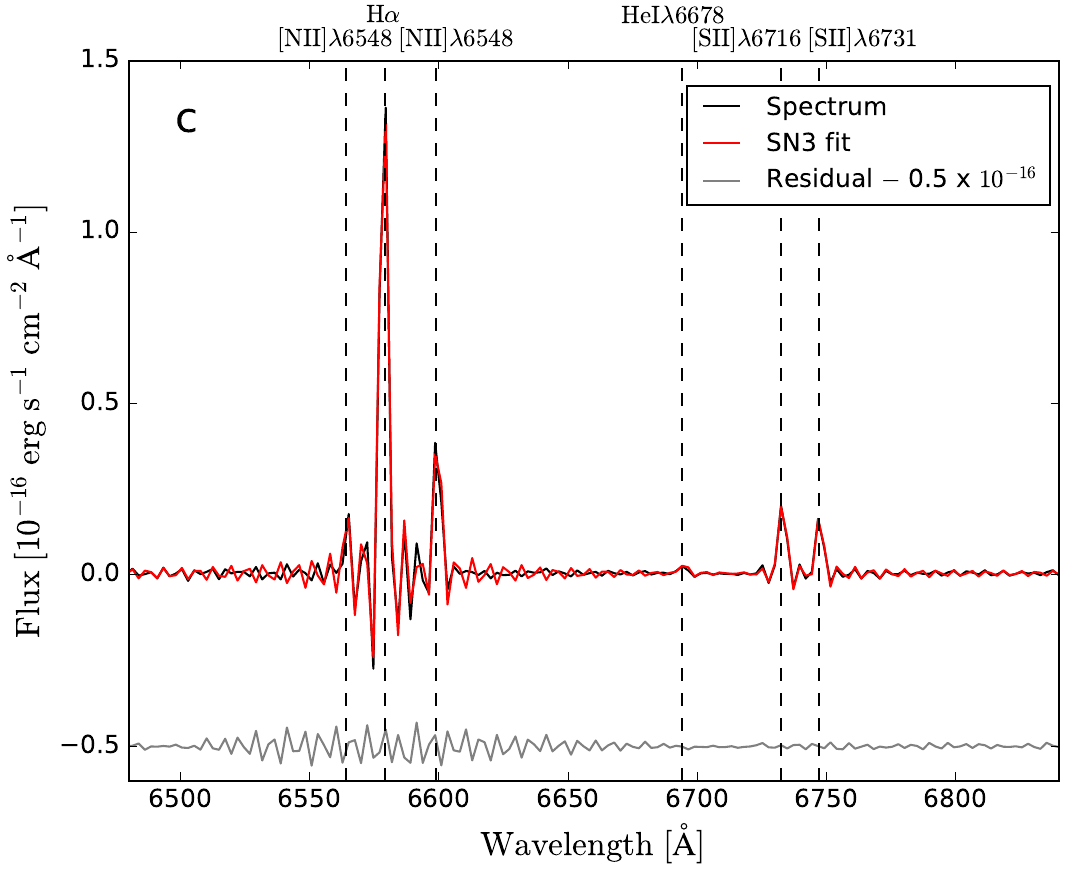} 
\caption{The spectrum of an HII region combining the spaxels within a circular 
aperture with a radius of 1.6$^{\prime\prime}$ centered at RA 01h36m53.1s and DEC +15$\degree$48$^\prime$04.8$^{\prime\prime}$. The SN1 filter is presented in~a, SN2 in b, and SN3 in c. The fit obtained with ORCS and the residual after the fit subtraction are shown. Residual spectra have been shifted by a constant value for clarity (see the legends).}
\label{spectrum1}
\end{center}
\end{figure}

\subsection{Line Fitting Procedure and the Final Velocity Map} 
\label{lfproc}
For each spaxel in all three datacubes, the emission lines are fitted simultaneously using
the extraction software ORCS (\citealt{mm15}\footnote{https://sourceforge.net/projects/orb-orcs/}). This software was specifically created for SITELLE and therefore uses sinc-shaped line profiles. ORCS output map for each line includes the maximum intensity $I_0$, FWHM, velocity, continuum level, all the corresponding uncertainties, the SNR, as well as the standard deviation $\chi^2$ of the fit. These parameters have been used to get the flux and its uncertainty map for all lines (the latest version of ORCS automatically gives all these quantities). Line flux uncertainties were here calculated using the relation for a non-resolved line (i.e. below the Nyquist criteria) in the case of an FTS instrument according to \cite{la92}; they consist of the product of the line flux with its ratio $\chi^2$/$I_0$. The uncertainty for line flux ratios were then put equal to half of the difference between the maximum and minimum ratio values obtained considering the line flux uncertainties.


Examples of the line fits are shown in Figure\,\,\ref{spectrum1} for all three filters. The fluctuation seen in the residual underneath H$\alpha$ is caused by the subtraction of the galaxy stellar contribution which, as we discussed in the previous section, may not be perfectly well adapted (in content and velocity dispersion). But, as seen here, the impact on the line fitting is negligible ($<$\,1$\,\%$). Appendix~\ref{flma} contains the resulting flux maps for all the lines (Fig.~\ref{SN2SN1_flux_reg} and \ref{SN3_flux_reg}).

Appendix \ref{flma} also presents the final velocity map (Fig.~\ref{v}) obtained by fitting simultaneously all the SN3 emission lines with ORCS. As explained in the appendix, the velocity precision achieved here is often better than 30~km~s$^{-1}$, i.e. the limit when SNR$_{{\rm H}\alpha} = 3$.

\section{REGION DEFINITION}
\label{Proc}

Although one can study individual spaxel within an ionized gas region in a nearby galaxy like NGC\,628 with SITELLE, some properties must be investigated using their integrated flux (e.g. the region luminosity function) and a description of their morphology (e.g. their size). It is therefore necessary to distinguish which pixels belong to a given region. A new code is presented here for identifying the position of the ionizing source associated to a region and its boundaries. Compared to other existing codes, like those based on the H$\alpha$-continuum-subtracted images and a standard-fixed-threshold-photometry method, such as the percent-of-peak-photometry method (\citealt{m96}), or HIIphot (\citealt{t00,t02}), this new code gives us flexibility in defining regions and their DIG background. 

Our method proceeds with multiple steps : 1)\,\,the identification of the emission peaks, 2)\,\,the determination of the zone of influence around each emission peak, and 3)\,\,the definition of the outer limit of a region and its DIG background. These operations are done prior to the extinction correction, to avoid introducing unnecessary noise. The different steps of our method are described in the following 
subsections. 

\begin{figure*}
\begin{center}
\includegraphics[width=7in]{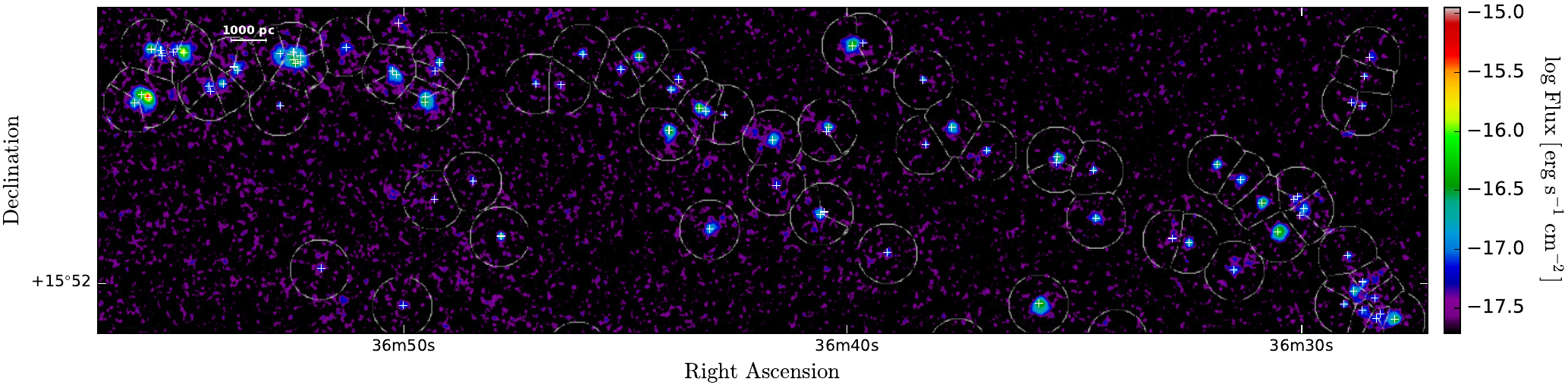} 
\includegraphics[width=6.87in]{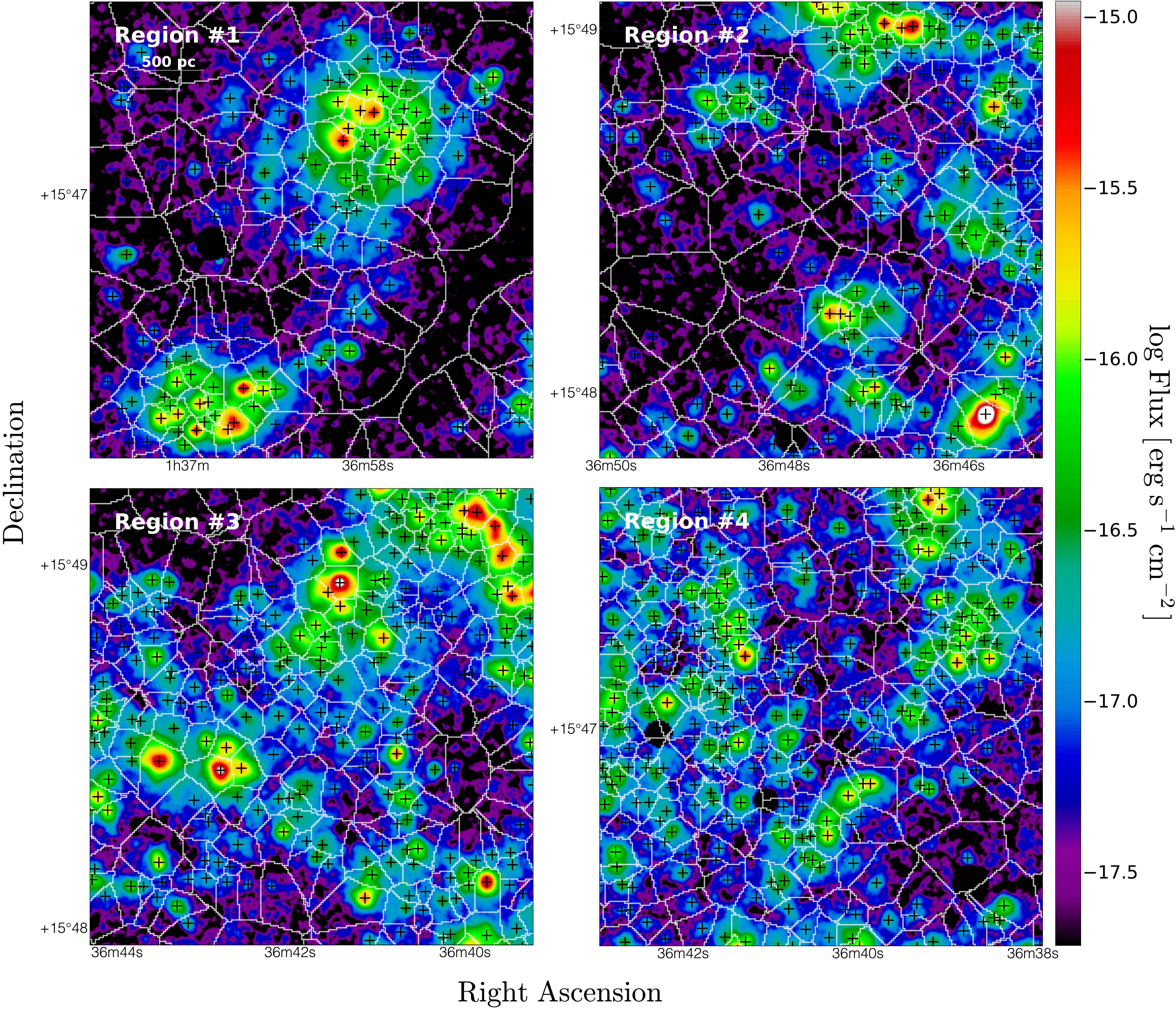} 
\caption{Examples of ionizing sources and their zone of influence in the galaxy northern arm (top) and in four reference regions of the galaxy (bottom) drawn over the H$\alpha$+H$\beta$+[OIII]$\lambda$$\lambda$4959,5007 continuum subtracted image. The location of the four reference regions may be seen on the H$\alpha$ image in Figure\,\ref{SN3_flux_reg}. The centroid position of each emission peak detected is identified with a cross. The white contours define the zones of influence surrounding the emission peaks.}
\label{flux_reg}
\end{center}
\end{figure*}

\subsection{Definition of the Peaks}

The first step involves adopting a definition for the emission peaks and identifying their positions. We use a combination of the H$\alpha$, H$\beta$, and [OIII]$\lambda$4959,5007 continuum subtracted images. This choice is motivated by the increased SNR of the combined image. Moreover, as found from our high-spatial resolution image, the maximal intensity of these three lines coincides with the centre of the emission regions. An emission peak is identified for a pixel when: 1)\,\,the pixel intensity is greater than the intensity of at least 5 immediate surrounding pixels (i.e. all these pixels must be in contact with the maximum intensity pixel), and 2)\,\,the total intensity in a 3$\times$3 pixels box centred on the emission peak is above the adopted detection threshold. This threshold is fixed by the 3$\sigma$ noise level of the H$\alpha$+H$\beta$+[OIII] flux map (a slightly different value is adopted for each of the CCD quadrant according to their individual noise level). Furthermore, if two emission peaks are separated by a distance smaller than the image quality of the observation, only the brightest peak is preserved. For each so-defined emission peak, two values of its mean flux are evaluated considering a circular aperture with a radius of 1 and 2 pixels centered on the emission peak. The peak is rejected when the mean flux does not significantly decrease between the small and large aperture (i.e when a drop less than $\sim$\,2$\%$ between adjacent pixels is seen for the bright regions, with a flux F$_{H\alpha}$\,$\simeq$1$\times$10$^{-15}$\,erg\,s$^{-1}$\,cm$^{-2}$, and when a drop less than $\sim$\,5$\%$ is seen for the faint regions). This criterion allows us to remove false identifications inside a bright ionized gas region, that are likely to be associated with small surface brightness variations, while eliminating false detections due to random noise fluctuations. To find the accurate position of the centroid, a 2D Gaussian model is fitted on the remaining emission peaks using the IDL procedure MPFIT2DPEAK. A total of 4285 emission peaks is finally detected for NGC\,628 with our extraction technique. After a visual verification of a representative group of emission peaks, we estimate that only a small fraction, $\sim$\,0.5$\%$, of these detections are false detections (not an ionized gas region). It is also important to note that most of the remaining regions are HII regions while a few percentage may be pure DIG regions, supernova remnants, and planetary nebulae, as it will be demonstrated from their location within the BPT diagrams in Section~7. 

\subsection{Zone of Influence}

Figure\,\,\ref{flux_reg} shows the emission peaks and their zone of influence for a section of the northern arm and for four reference regions selected in the galaxy disc. We name the area that is influenced by the ionizing photons from an emission peak, the zone of influence. It is defined by studying the distance between each pixel and all its surrounding emission peaks. In order to consider all the possible sizes for the zone of influence, we looked for emission peaks within a maximum distance of 425\,pc around each pixels (corresponding to 30\,pixels). This distance is greater than the maximal outer limit distance found for the regions of NGC\,628, as seen in the next paragraph. Most of the time a pixel is then associated with the closest emission peak. In the case of multiple emission peaks at the same distance (considering a distance uncertainty given by the image quality), the pixel is then associated with the brightest emission peak. The most distant pixels associated with an emission peak define the boundary of the zone of influence for this ionizing source.

Figure\,\,\ref{outer_DIG} shows an example, for a typical region, of the H$\alpha$ flux of all the individual pixels included inside the zone of influence as a function of their distance from the emission peak centroid. Although the average H$\alpha$ flux profile for most regions presents a smooth drop, the dispersion around the mean profile can, in some cases, be large
(see \S~\ref{morpho}). This dispersion is explained by the asymmetry of the emission region, inhomogeneous DIG emission, and also crowding and density effects (in these last cases, pixels are not associated with the right emission peak). 

\subsection{Outer Limit of a Region and the DIG Background}

\begin{figure}
\begin{center}
\includegraphics[width=3.3in]{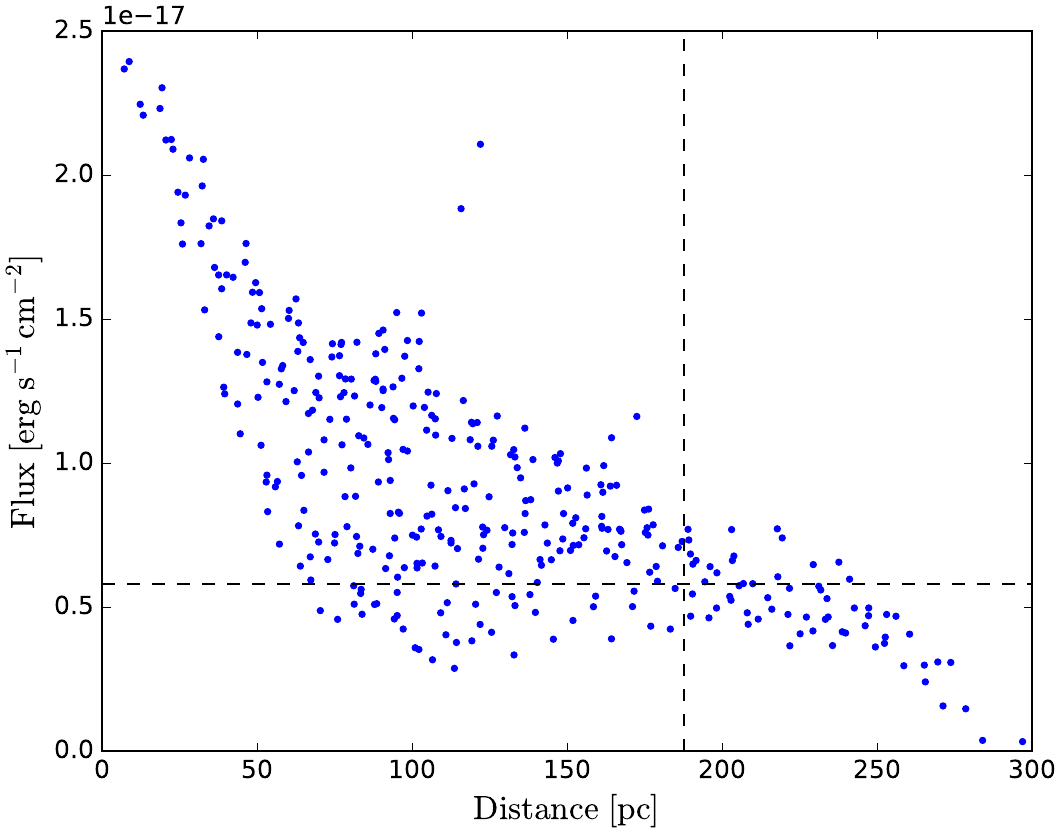}
\caption{The H$\alpha$ flux for the individual pixels of the HII region candidate ID 2688. Each pixel included within the zone of influence is represented by a dot. The vertical line indicates the outer limit of the region found from the total flux profile (Fig.\,\,\ref{outer_DIG2}) and the horizontal line gives the intensity level of the DIG background.}
\label{outer_DIG}
\end{center}
\end{figure}

\begin{figure}
\begin{center}
\includegraphics[width=3.3in]{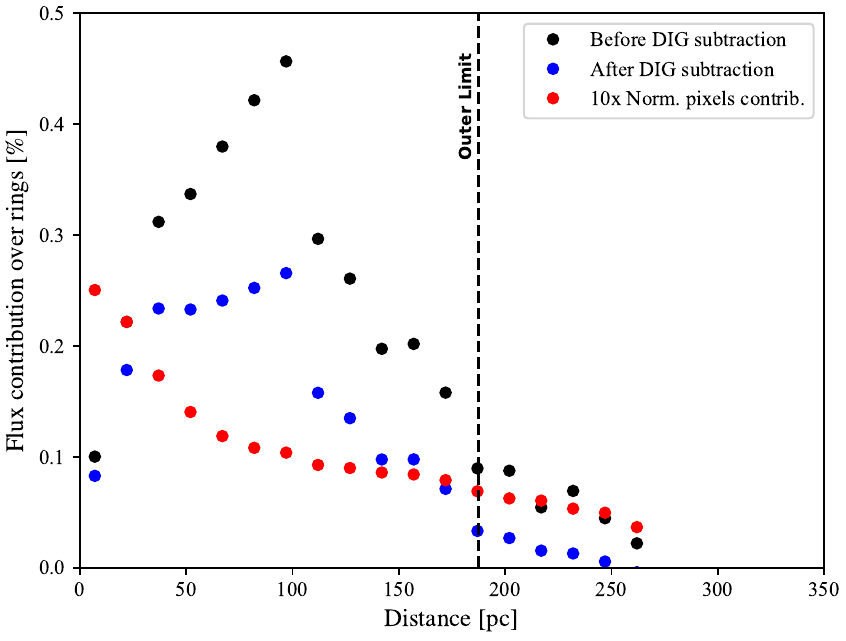} 
\caption{The H$\alpha$ total flux profile of the HII region candidate ID 2688. The total flux observed (black dots) is calculated within annuli centred on the ionizing source within the zone of influence. The vertical line indicates the outer limit, where the slope of the total flux profile decreases by less than 2\,$\%$. Blue dots show the total flux profile after the subtraction of the DIG background. The mean pixel contribution to the flux profile is shown in red (enhanced by a factor of 10).}
\label{outer_DIG2}
\end{center}
\end{figure}

Within the zone of influence of an ionizing source, the intensity profile changes rapidly (as shown in Fig.\,\,\ref{outer_DIG}). Figure\,\,\ref{outer_DIG2} shows the total flux profile for the region presented in Figure~\ref{outer_DIG}. The total flux profile is calculated by summing the flux of the pixels within a circular annulus 25\,pc thick centered on the emission peak. This fixed 25\,pc annulus, corresponding to $\sim$2 pixels, allows us to properly sample the profile (e.g. its slope) with a sufficient amount of pixels. The outer limit of the region is then defined at a distance where the slope of the total flux profile decreases by less than 2\,\% within the zone of influence. The DIG background surface brightness (DIG$_{\rm SB}$) is given by the median intensity value for all the pixels included in an annulus 50\,pc thick centered on the position of the outer limit. The zone of influence and the DIG background are indicated for the example in Figure~\ref{outer_DIG}. The 2\,\% threshold was selected after a few iterations: a larger value overestimates the DIG background for numerous HII regions, whereas a smaller value is not sensitive to variations of the slope. With these constrains, an outer limits is not found for only three regions; in these cases, the outer limit is simply set at the radius of the zone of influence. For our sample of regions, the average outer limit is 71\,pc (with a standard deviation of 33\,pc) and the smallest value is 21\,pc. In Figure\,\,\ref{outer_DIG2}, the effect of the DIG background subtraction on the total intensity profile is also shown. 

\section{MORPHOLOGY, SIZE, AND LUMINOSITY FUNCTION}

\subsection{Extinction Correction}

To evaluate the colour excess ${E(B-V)}$, we used the following equation: \begin{equation}
\label{ebveq}
E(B-V) = \frac{2.5}{1.07} \log \left(\frac {F_{H\alpha}/F_{H\beta\,obs}}{2.87}\right),
\end{equation}
\noindent with ${A_\lambda = 2.5log{I_\lambda}/{I_{\lambda0}}}$, and $E(\beta-\alpha)$/$E(B-V)$ = 1.07 (extracted from the extinction curve of \citealt{c89}), and where $F_{H_\alpha}/F_{H_\beta\,obs}$ is the observed flux ratio of the two main Balmer lines. We adopted a theoretical $F_{H_\alpha}/F_{H_\beta}$ ratio of 2.87 (as expected for the Case B of \citealt{o89}, at a temperature of 10\,000 K). 
As discussed at the end of this subsection, the value selected here are for an average HII region.

\begin{figure}
\begin{center}
\includegraphics[width=3.3in]{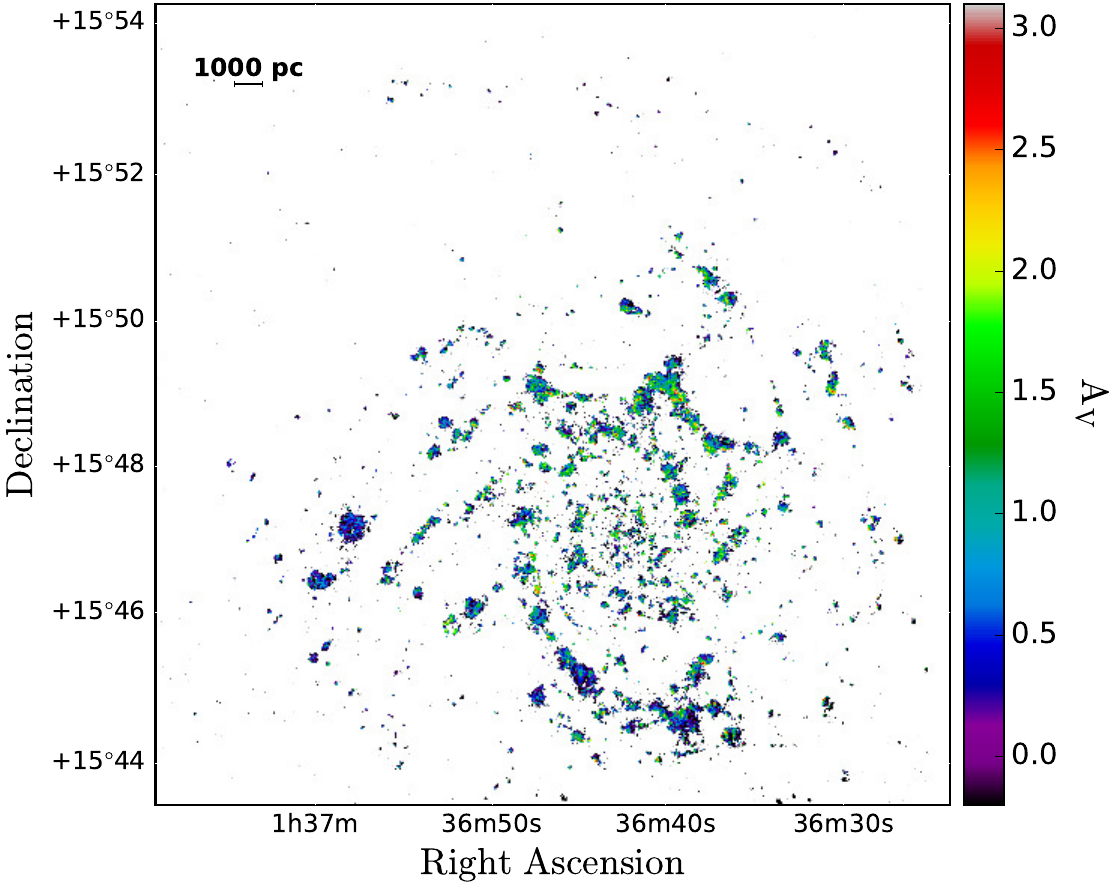} 
\caption{The extinction map obtained from SITELLE's H$\alpha$ and H$\beta$ line ratio for each spaxel. Only spaxels with a 3$\sigma$ detection limit for both lines are shown. The three giant HII regions, out in the disc to the South and East, display on average a smaller extinction value. }
\label{av}
\end{center}
\end{figure}

\begin{figure}
\begin{center}
\includegraphics[width=3.3in]{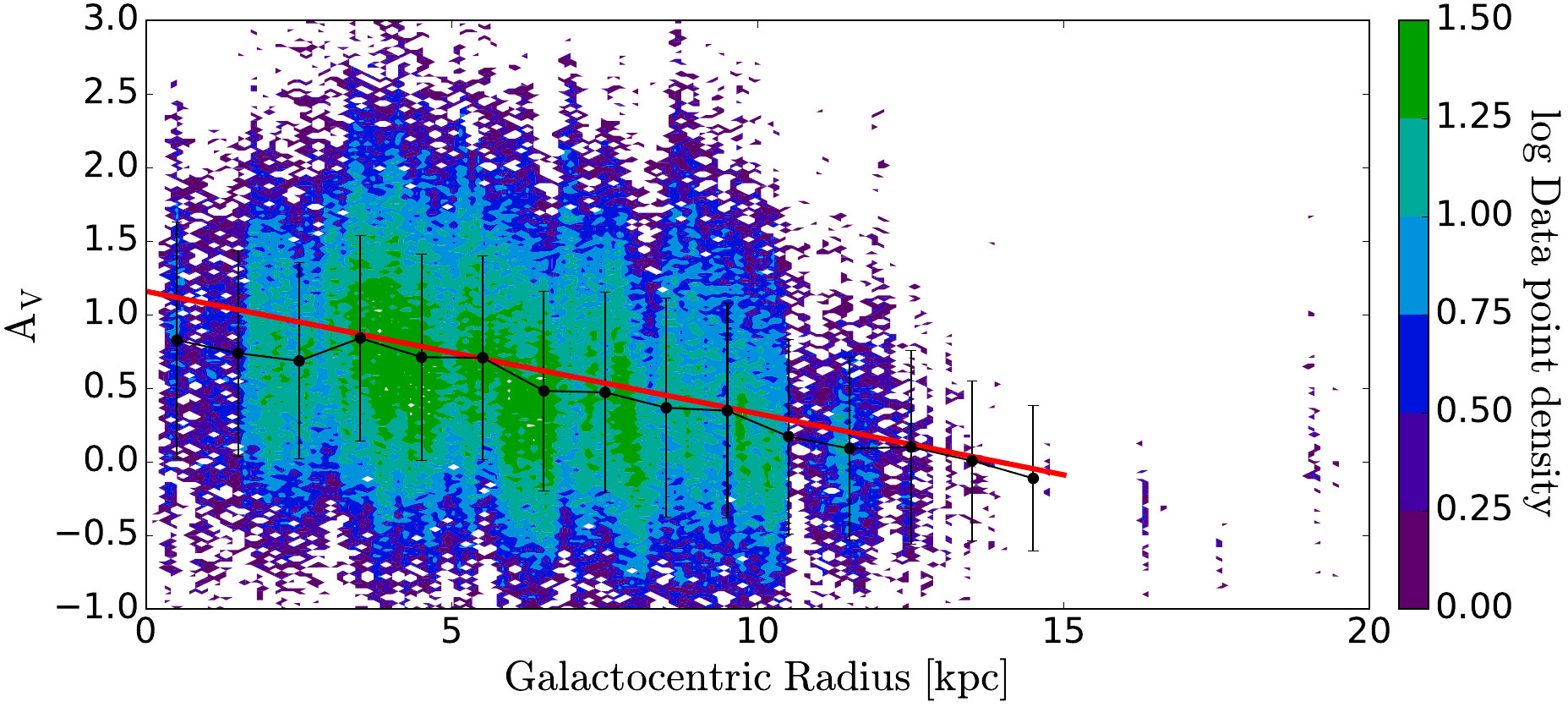} 
\caption{NGC\,628 extinction gradient. The extinction for all the spaxels (with SNR$_{H\alpha}$ and SNR$_{H\beta}$ $>$ 3) is plotted as a function of their galactocentric radius. The plot is colour coded according to the spaxel density. The mean values, using a bin size of 1\,kpc up to a radius of 15\,kpc, and their uncertainty (given by the standard deviation) are shown in black. The linear fit through the mean values is shown in red.}
\label{av2}
\end{center}
\end{figure}

Using the previous equation and the relation ${A_{V} = 3.1\,E(B-V)}$, we obtained a pixel-by-pixel SITELLE extinction map for NGC 628. This map is shown in Figure\,\,\ref{av}, where only the spaxels with a 3$\sigma$ detection for both lines have been considered. Figure\,\,\ref{av2} presents the extinction as a function of the galactocentric radius ($R_G$) for all the spaxels (the observed radii were deprojected using a galaxy inclination of 21$\degree$ from NED). The extinction is clearly decreasing as we look further out in the disc. Fitting a linear function through the data between the center and a galactocentric radius of 15\,kpc, we obtained the following relation: \begin{equation}
\label{avpoly}
 A_{V} = - 0.083R_G + 1.16. \end{equation} 
 

The highest density points in Figure\,\,\ref{av2} (in green) show an extinction value in good agreement with \cite{b15}, where $A_{V} \simeq 0$ to 1.28\,mag for a sample of 45 HII regions. Our mean value, considering only the inner part of the galaxy ($R_G \leq 9$\,kpc), is 0.86$\pm$0.81\,mag. This value is very close to the mean value of 0.79 obtained by \cite{c12} derived from a sample of 209 HII regions observed with narrow-band imaging.
Within the uncertainties, it is also in agreement with the result of \cite{s11}, 1.24$\pm$ 0.76\,mag, gathered using an IFS over the same area of the galaxy disc.

A close examination of the extinction map as displayed in Figure\,\,\ref{av} suggests significant fluctuations in the extinction within individual regions. Furthermore, $\sim$\,10$\%$ of the spaxels have an observed ratio $F_{H\alpha}$/$F_{H\beta}$ below the theoretical value (this is seen in Fig.\,\,\ref{av2} for A$_V$\,$<$\,0). Some tests were done to investigate the cause of these variations and the low flux ratios. 
First, the subtraction of the stellar population reference spectrum was considered. From our tests, using a combination of young burst (6 to 10 Myr) with the old stellar population in various proportions of the continuum level (\S~3.4), we get a maximum effect of $-$0.2 mag for A$_V$, which is of interest to explain some of the variations seen in the extinction map of individual regions (Fig.~\ref{av}), but below the dispersion seen in Figure~\ref{av2}.
Next, by combining many spaxels within a region, we used various aperture sizes to see the impact of atmospheric refraction on the measured flux ratio. The accuracy of the flux calibration and the alignment of the datacubes were also tested. Although these effects can increase the flux uncertainty, we find that they cannot be held responsible for all the spaxels with a Balmer line flux ratio below the theoretical value. A simple explanation often proposed for low Balmer line flux ratios invokes the low SNR of the H$\beta$ line. But this is not sufficient here as we estimate that only $\sim$\,4\,$\%$ of the spaxels with a ratio $F_{H\alpha}$/$F_{H\beta}$ below the theoretical value have SNR$_{H\beta}$\,$<$\,10 (including the effect associated to atmospheric refraction, flux calibration, and alignment of the datacubes). Therefore, some of the low Balmer line flux ratios, along with the extinction variation seen inside a region itself, could be real and possibly related to the change of the physical parameters (e.g. the electron temperature and density) within that region. For an HII region, the theoretical Balmer line flux ratio can reach 2.75 at 20\,000~K according to \cite{o89}. Depending on the morphology of the HII regions, the dust backscattering (\citealt{o17}) can also contribute to the observed variation of the flux ratio. In regions highly dominated by the DIG, the electron temperature and density would be very different than in an HII region (Hafner et al. 1999; Elwert \& Dettmar 2005). SITELLE's data is therefore revealing the detailed physical conditions within emission regions and their close surroundings.

Using the extinction law of \cite{c89} and considering that our emission regions are mainly HII regions, we corrected all our flux maps with the extinction calculated fror each spaxel or with the relation given by Equation\,2 for spaxels where no precise extinction was measured (for A$_V$\,$<$\,0 and with SNR$_{H\alpha}$ or SNR$_{H\beta}$\,$<$\,3). As the extinction correction is subject to uncertainties, line ratios (as presented in $\S$\,6) built using lines that are far away from each other (i.e. lines from different filters) also have greater uncertainties.

\subsection{Morphology of the Emission Regions }
\label{morpho}

Constraining the morphology and the size of the emission regions requires modeling their intensity profiles in more details. The morphological classification of HII regions is often performed by eye, but with a large sample of candidates as obtained here, an automated approach was considered. At the spatial resolution of our survey ($\sim$\,0.8$^{\prime\prime}$ or 35\,pc), we were mostly interested in obtaining the general shape of the regions and in separating compact from extended regions. 

\begin{figure}
\begin{center}
\includegraphics[width=3.3in]{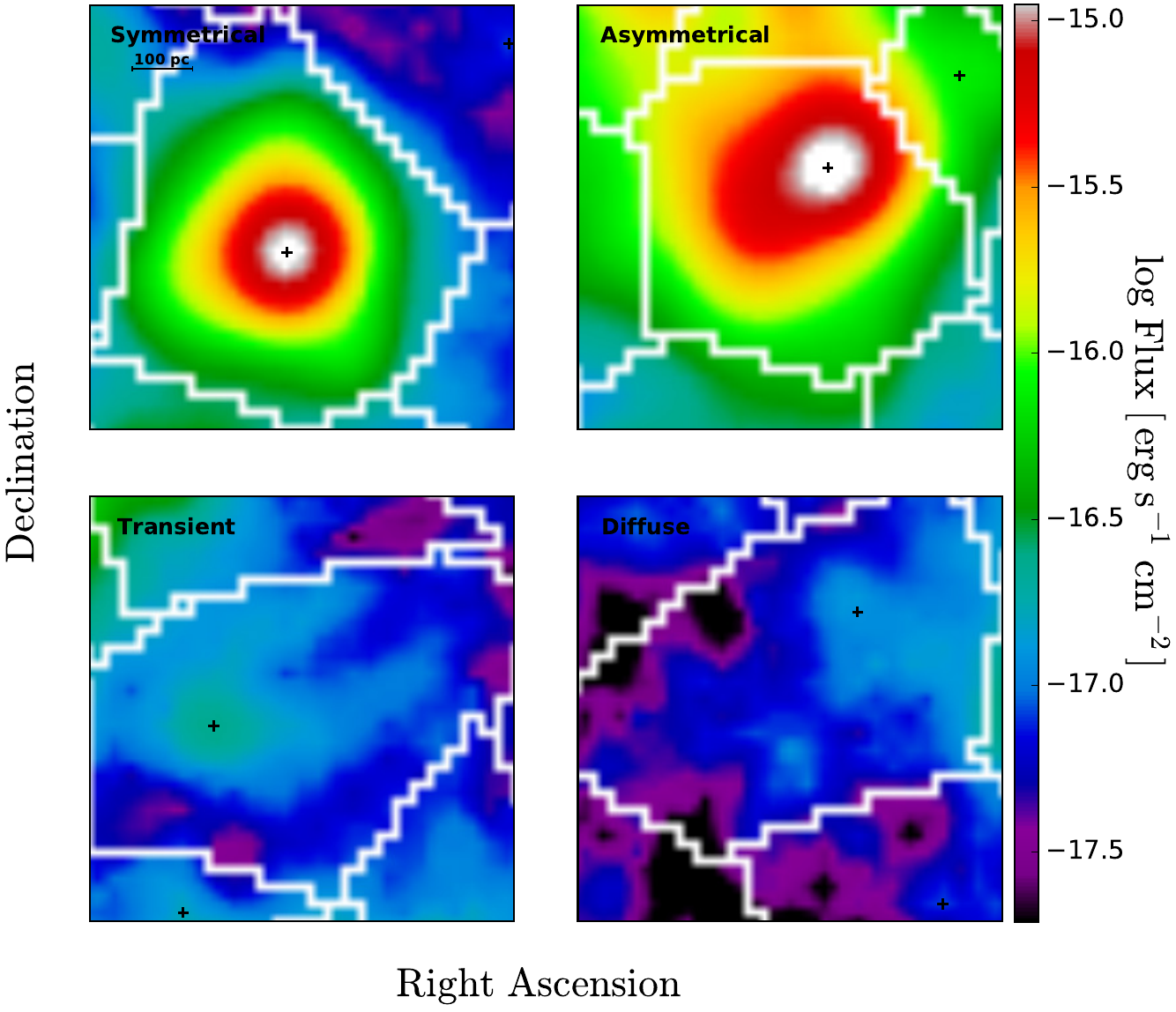} 
\caption{Examples of emission regions with different morphological classification. The four categories of regions are represented using the H$\alpha$ image; the categories are identified in the left corner of each image. The ID of the regions selected, from top-left to bottom-right, are : 2437, 2987, 2688, and 834.}
\label{class}
\end{center}
\end{figure}

\begin{figure}
\begin{center}
\subfigure{\includegraphics[width=1.6in]{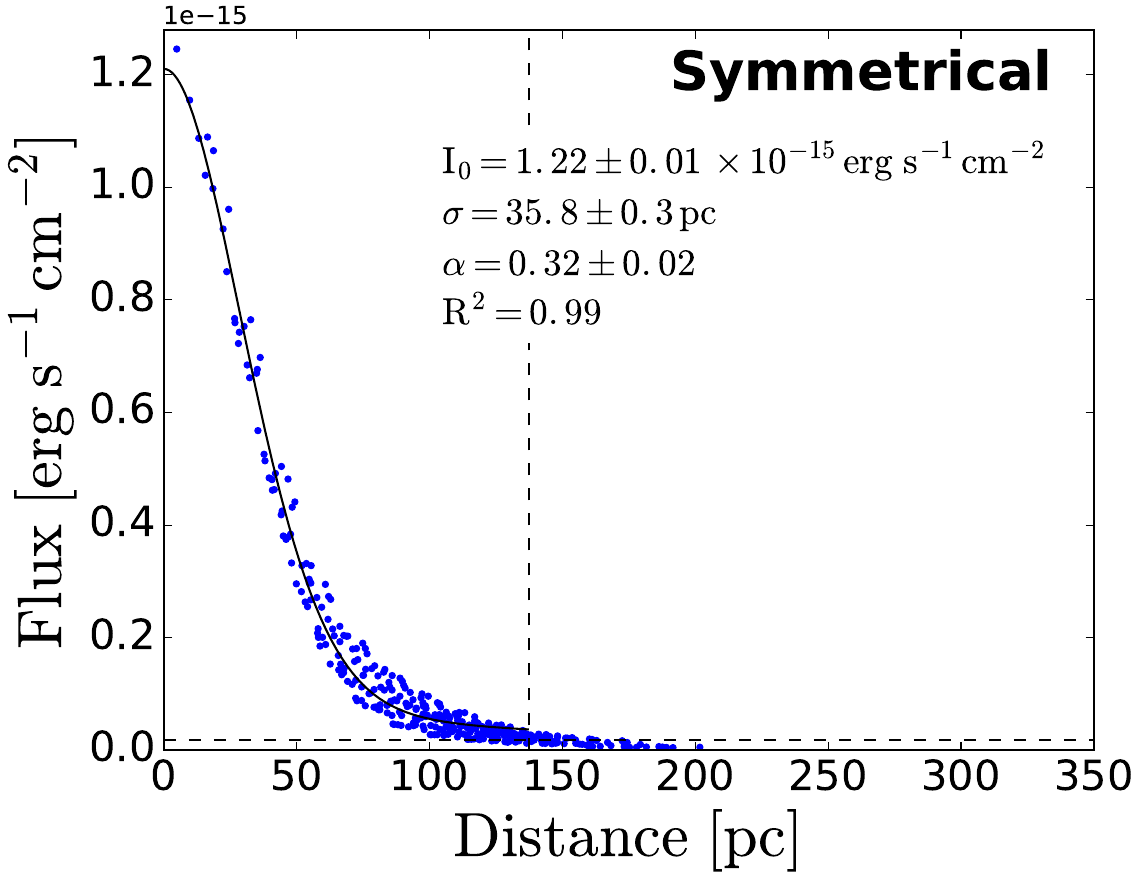}}
\subfigure{\includegraphics[width=1.6in]{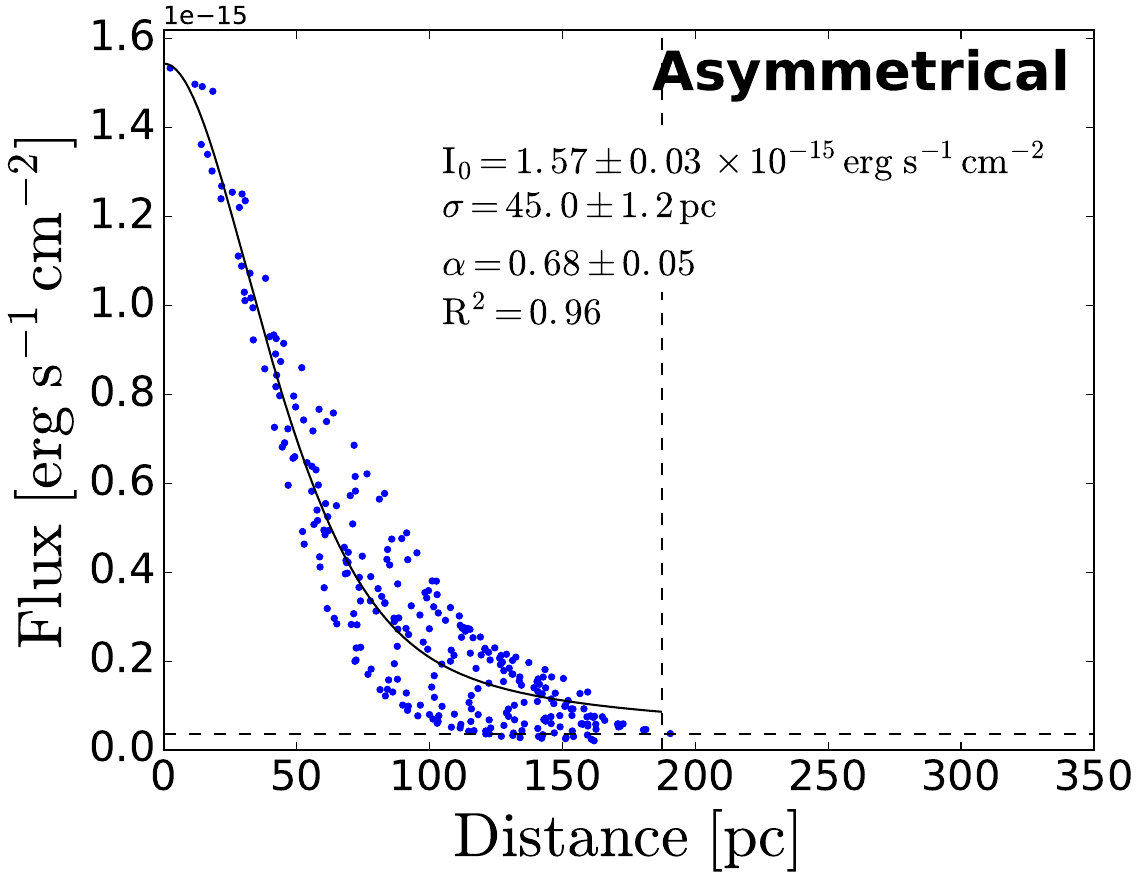}} 
\subfigure{\includegraphics[width=1.6in]{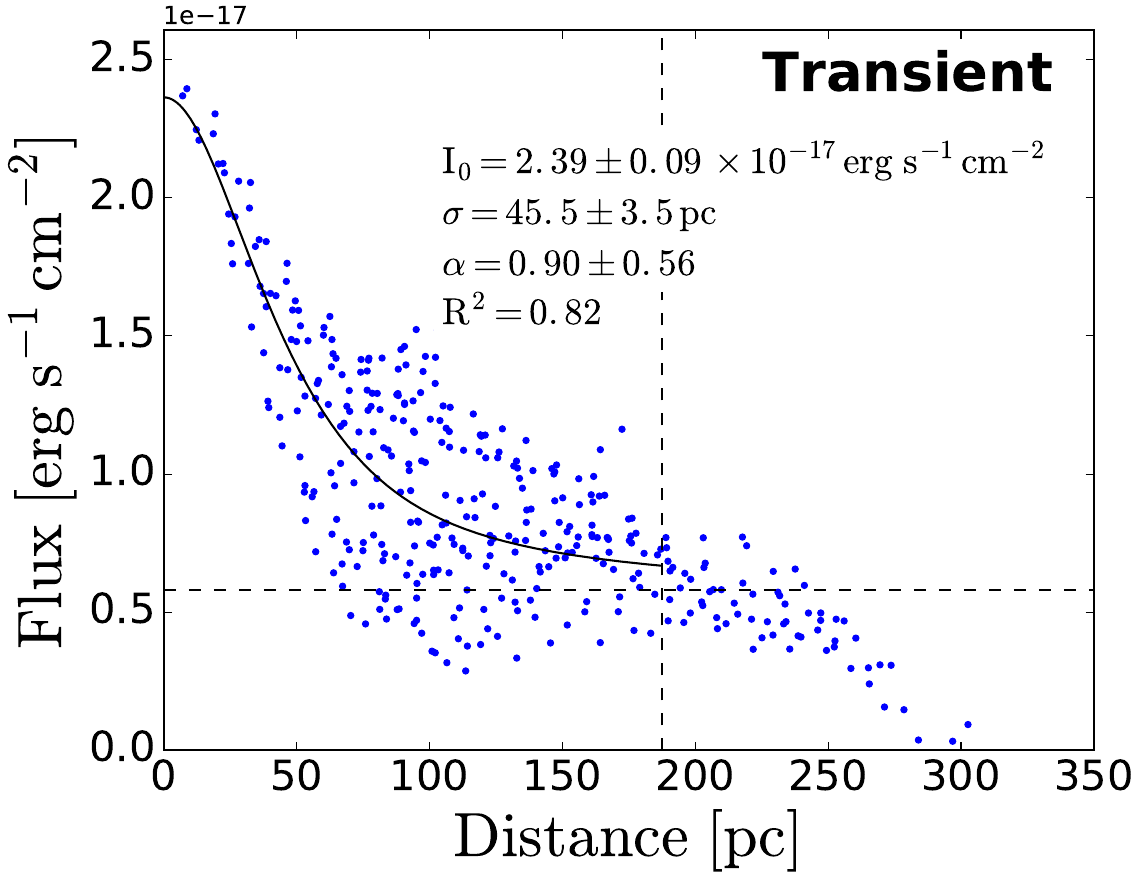}} 
\subfigure{\includegraphics[width=1.6in]{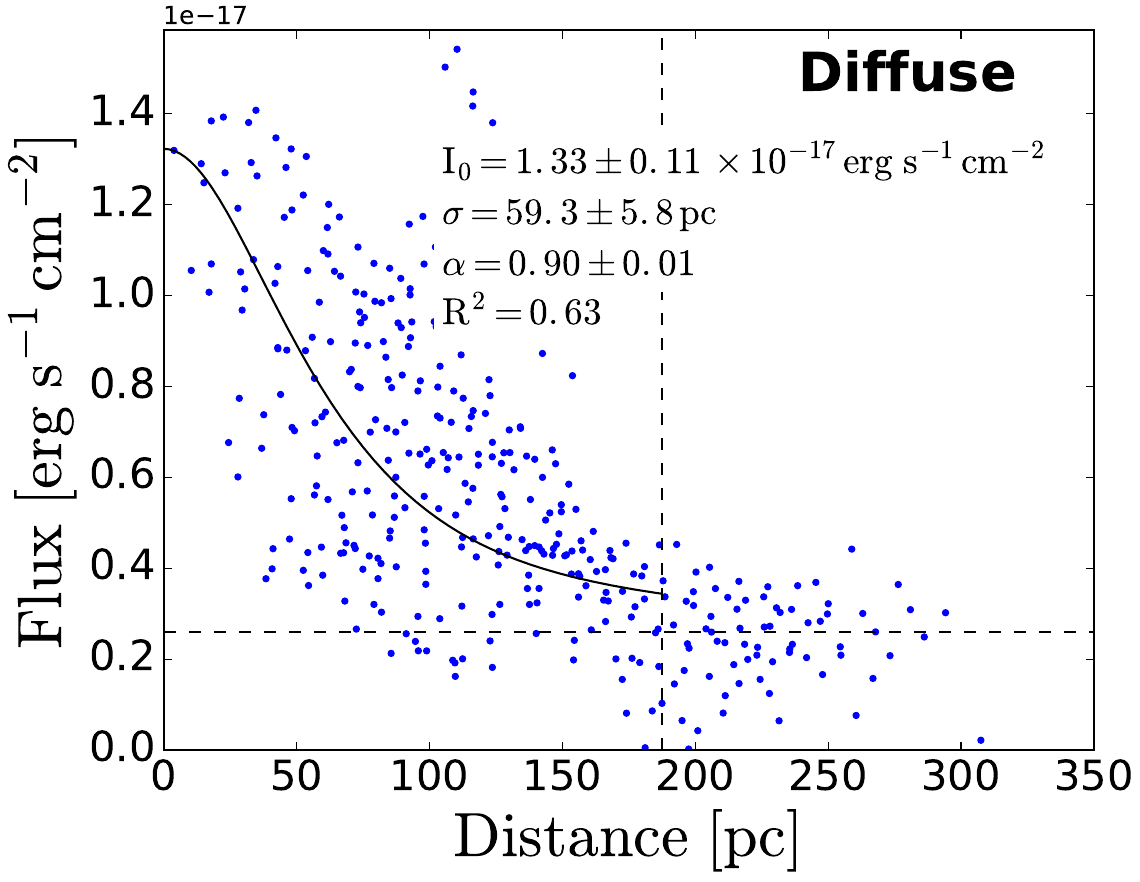}} 
\caption{The H$\alpha$ flux profile for the emission regions presented in Figure\,\,\ref{class}. The intensity for all the pixels included in the zone of influence are plotted as a function of their distance to the emission peak. The vertical line represents the outer limit and the horizontal line represents the DIG background. The black curve is the best $\chi^2$ fit obtained using a pseudo-Voigt profile. The categories are identified in the right corner of each plot along with the fit parameters of the Equation 3 and the correlation coefficient R$^2$. The ID of the regions selected, from top-left to bottom-right, are : 2437, 2987, 2688, and 834.}
\label{class_prof}
\end{center}
\end{figure}

To establish the general shape of the emission regions, we divided them into four categories: symmetrical, asymmetrical, transient, and diffuse. Figure\,\,\ref{class} presents an example for each category. Before explaining our criteria for the classification, we can see that the symmetrical region looks almost spherical on the H$\alpha$ image and presents very low dispersion in its luminosity profile, as shown in Figure\,\,\ref{class_prof}. The ionizing source (i.e. possibly a young stellar population here) of this region is likely to be concentrated in the centre with the surrounding gas equally distributed. The asymmetrical region shows an important emission peak but also a large variety of shapes diverging from the spherical case. In this category, the ionizing source is expected to be either more dispersed or to encompass multiple components, and/or the surrounding gas is not equally distributed in all directions, and/or emitting pixels have been associated with the wrong peak due to a crowding effect. The diffuse region is less peaked and its intensity profile presents even more dispersion; a fraction of the regions in this category may be more easily related to DIG regions. The dispersion seen for transient regions impedes our ability to distinguish between an asymmetrical region and a region with diffuse structures such as large filaments of ionized gas. 

The HII region morphologies are fully quantified using the intensity profile. We selected a pseudo-Voigt function to fit the profiles (a linear combination of a Gaussian and a Lorentzian profile) using the Nelder-Mead method (\citealt{n65}). The pseudo-Voigt profile is defined by the equation: 

\begin{equation}
\label{pseudov}
f(x, A, x_0, \sigma, \alpha) = \frac{1-\alpha}{\sqrt{2\pi}\sigma_g}Ae^{\frac{-(x-x_0)^2}{2\,\sigma_g^2}}+\frac{\alpha A}{\pi}\frac{\sigma}{(x-x_0)^2+\sigma^2},
\end{equation} 


\noindent where the constant $A$ is the amplitude, $\sigma_g = \sigma/\sqrt{2\,\mathrm{ln}2}$, and 2$\sigma$\,=\,FWHM. The constant $\alpha$ is a fraction revealing the relative importance of the Gaussian and the Lorentzian terms to the profile. The centroid position ($x_0$) was fixed to 0 as it is already constrained by our peak identification procedure. The use of a pseudo-Voigt function enables us to reproduce the observed H$\alpha$ flux profile very well in the case of a peaked emission with a spherical morphology (the pseudo-Voigt profile is shown in Fig.\,\,\ref{class_prof} for the example selected). The fraction $\alpha$ varies significantly from one region to another, and allows us to recover the outer part of the profile more accurately than by using a simple Gaussian profile. A correlation coefficient (R$^2$) for the fit was also evaluated for each region by considering the whole profile from the centre to the outer boundary. Only 13$\%$ of the regions have a correlation coefficient R$^2$\,$<$\,0.5 and 7$\%$ have a correlation coefficient R$^2$\,$<$\,0.3. 

For each emission region, we used the pseudo-Voigt H$\alpha$ intensity profile to evaluate the average pixel dispersion as a function of the distance over three 15\,pc annuli: in the very centre, at the position of 1$\sigma$, and at the outer limit. We also evaluated the average pixel dispersion within these same annuli by using the mean value instead of the fitted profile. This second method was required for specific regions where the pseudo-Voigt profile could not properly fit the peak or the background intensity. The smallest dispersion values, evaluated using either the mean or the pseudo-Voigt fitted profile, were considered to assign a morphological category to an emission region. The dispersion thresholds for all four categories were first selected arbitrarily (by eye) to finally follow the rules listed in Table \ref{rules}. \begin{table}
\centering
\caption{Dispersion thresholds for the morphological classification. The dispersion thresholds are given in \% of the region peak value.}
\scriptsize
\label{rules}
\begin{tabular}{lcccc}
\hline
\bf{Category} & \bf{Centre$^{a}$} & \bf{Sigma$^{a}$} & \bf{Outer Limit$^{a}$} & \bf{Regions} \\
\hline
Symmetrical & $\le$ 6\,\% & $\le$ 6\,\% & none & 448 \\
Asymmetrical & $\le$ 6\,\% & $>$ 6\,\% & $\le$ 3\,\% & 772 \\
Transient & $\le$ 6\,\% & $>$ 6\,\% & $>$ 3\,\% & 1785 \\
Diffuse & $\ge$ 6\,\% & $>$ 6\,\% & none & 1280 \\ 
\hline
\end{tabular}
\end{table} 

\begin{figure}
\begin{center}
\includegraphics[width=3.3in]{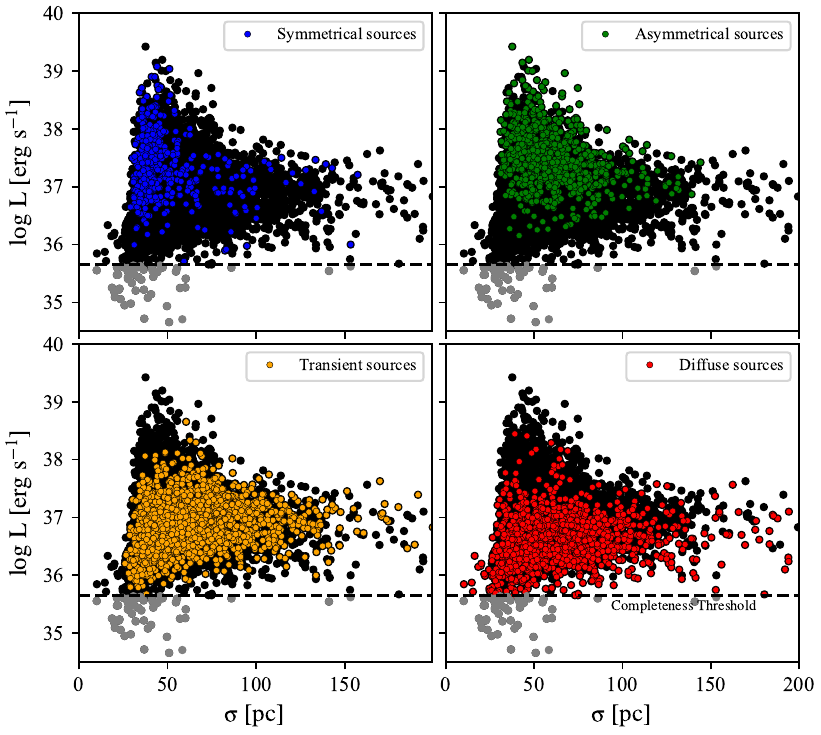}
\caption{The emission region H$\alpha$ luminosity distribution for the four morphological categories. The integrated H$\alpha$ luminosity is plotted as a function of the region width $\sigma$ given by the pseudo-Voigt profile for each region. The different plots outline the different morphology categories (in colour), as identified in the right corner, superimposed over the distribution for all the categories (in black). The horizontal dashed line is the luminosity detection limit.}
\label{reg_size}
\end{center}
\end{figure}

Figure\,\,\ref{reg_size} shows, for each morphological category individually, the integrated H$\alpha$ luminosity of the emission regions as a function of the $\sigma$ value of their pseudo-Voigt fitted profile. The symmetrical morphology category contains on average regions that are bright and more compact (with an outer limit between 21 and 173\,pc), while the diffuse category contains regions that are among the faintest ones and among the most extended ones (covering the whole domain of radius up to the limit of 300\,pc). Asymmetric regions, which may contain multiple sources, are among the brightest ones. 

\begin{figure}
\begin{center}
\includegraphics[width=3.3in]{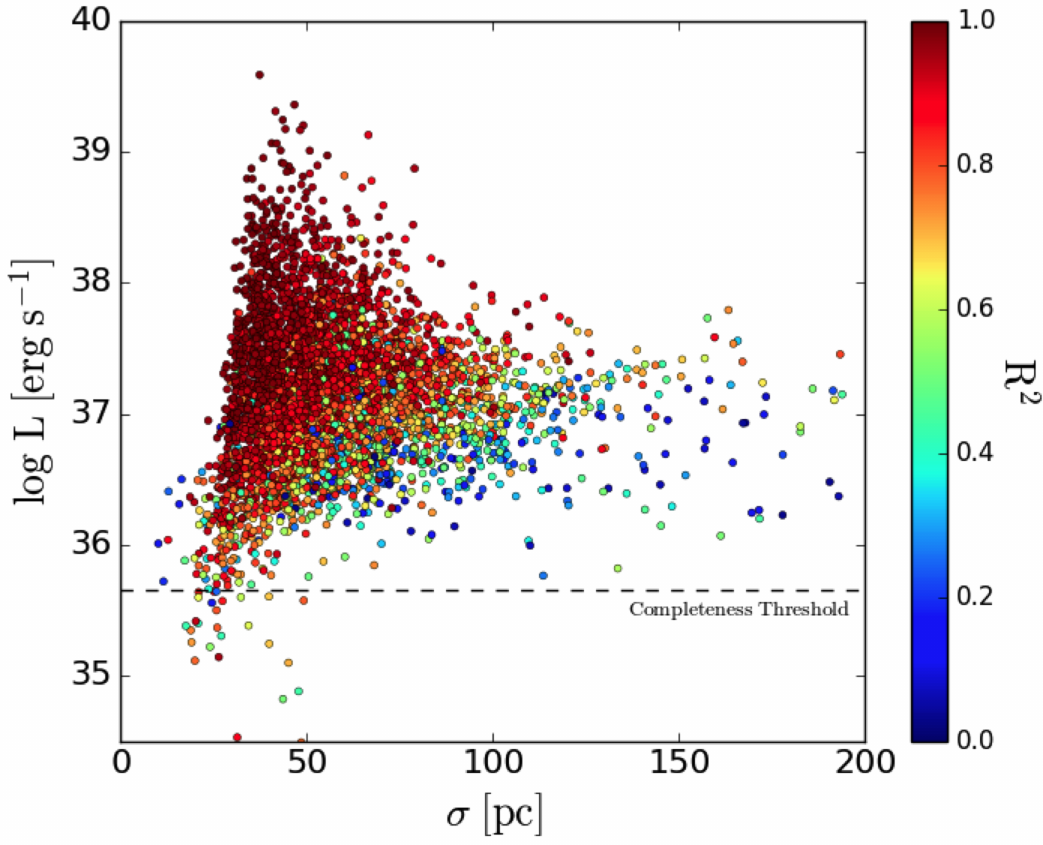}
\caption{The emission region H$\alpha$ luminosity distribution taking into account the 
correlation coefficient R$^2$ for the pseudo-Voigt fitted profile. The integrated H$\alpha$ luminosity is plotted as a function of the region half-width $\sigma$ given by the pseudo-Voigt fitted profile for each region. Each region is colour coded according to R$^2$, as indicated to the right. The horizontal dashed line is the luminosity detection limit. }
\label{reg_size2}
\end{center}
\end{figure}

Figure\,\,\ref{reg_size2} shows the integrated H$\alpha$ luminosity of the emission regions as a function of the $\sigma$ value of their pseudo-Voigt fitted profile and their corresponding correlation coefficient. As suspected for the diffuse category, low-luminosity and extended regions have a weaker correlation coefficient. Furthermore, Figures \ref{reg_size} and \ref{reg_size2} clearly show a lack of extended-high luminosity regions and reveal a possible maximum envelope for a correlation between the compactness and integrated H$\alpha$ luminosity of the regions. Multiple mechanisms could explain this behavior. Firstly, considering that our sample of emitting regions is dominated by HII regions (this is well supported by the BPT diagrams presented in Section 6), the most massive interstellar clouds are the progenitors of the most luminous regions and these clouds are known to be denser than the smaller ones (almost a linear function between the mass of the clouds and the ISM density; \citealt{mmmm15}). Due to the global equilibrium between the ISM density and the ionizing flux of the source, the high density of these clouds could explain why the massive regions are smaller in size than the low mass, less dense, regions. Secondly, the number of ionizing photons (Q$_H$) emitted by massive stars in the clusters is well constraint by the IMF while more dispersion is observed in low mass clusters where the ionizing stars content reflects a more stochastic sampling of the IMF (\citealt{mmmm15}). Additionally, the very bright HII regions observed here (log\,L$_{H\alpha}$\,$>$\,38.5) could be density bounded rather than ionisation bounded. This would also explain that
bright HII regions have on average smaller $\sigma$ value compared to fainter regions. In the same way, the fainter regions observed (between log\,L$_{H\alpha}$\,=\,37 and 38.5) can clearly be more extended; ionisation bounded regions in a lower density medium could produce this result. 

\subsection{Size of the Emission Regions}
\label{sz}
 
The size of HII regions can be compared to spherical models of uniform density (i.e. Str\"{o}mgren spheres; \citealt{o06}). This spherical model was chosen here for its simplicity but is certainly not representative of all the observed regions (\citealt{w05}). The radius of a Str\"{o}mgren sphere is given by : \begin{equation}
\label{strom}
R_{S} ^3= \frac{3Q_H}{4 \pi \epsilon {n_e}^{2} \alpha_{B}(H,T_e)},
\end{equation}
 \noindent where $Q_H$ is the number of ionizing photons (with E\,$>$\,13.6\,eV), $\epsilon$ is the volume filling factor of the ionized gas, $n_e$ is the electron density, and $\alpha_{B}$($H$, $T_e$) is the recombination coefficient of the hydrogen atom for a given electron temperature ($T_e$). Typical values of $\epsilon$ for large HII regions are between 10$^{-3}$ and 10$^{-4}$ (according to \citealt{g04} and \citealt{c13}) and $\alpha_{B}$($H$, 10\,000\,K)$\,\,=\,\,2.59$$\times$10$^{-13}$cm$^{3}$\,s$^{-1}$ (\citealt{o89}). The number of ionizing photons was estimated using the relation
given by \cite{mmmm15}: \begin{equation}
\label{mass_qh}
 \begin{split}
 \mathrm{log}(Q_H) = &\,\,5.0754 + 42.034x \\
 & - 15.0797x^{2} + 2.4439x^{3} - 0.1474x^{4}, \\
 \end{split}
\end{equation}

\noindent where $x$ is a function of the cluster stellar mass : $x\,=\,$log$(M_{cl}/M_\odot)$. 
The luminosity of a region was calculated using the equation of \cite{o06}, $L_{{\rm H}\alpha}=1.36\times10^{-12} Q_H$. It is important to realize that with the Str\"{o}mgren sphere model selected here, the regions are ionisation bounded and their luminosities are therefore upper limits.

Leaving aside emission regions with a poor quality H$\alpha$ flux profile (i.e. with a correlation coefficient R$^2 < 0.6$), we estimate the radius $R_{HII}$ of the 3466 remaining emission regions, 
considering that they are mainly HII regions. To convert the pseudo-Voigt profile half-width $\sigma$ into a radius, the uniform spherical models intensity profiles were convoluted with a Gaussian to account for differences in image quality and manipulations during data reduction. Then, the H$\alpha$ intensity profile half-width $\sigma_g$ of the models were obtained by fitting a Gaussian. A relation between $\sigma_g$ and $R_S$ was derived and transposed to our measurements (see the insurt in Fig.~\ref{reg_size3}) to get the corresponding radius $R_{HII}$. 

\def\Put(#1,#2)#3{\leavevmode\makebox(0,0){\put(#1,#2){#3}}}
\begin{figure}
\begin{center}
\includegraphics[width=3.3in]{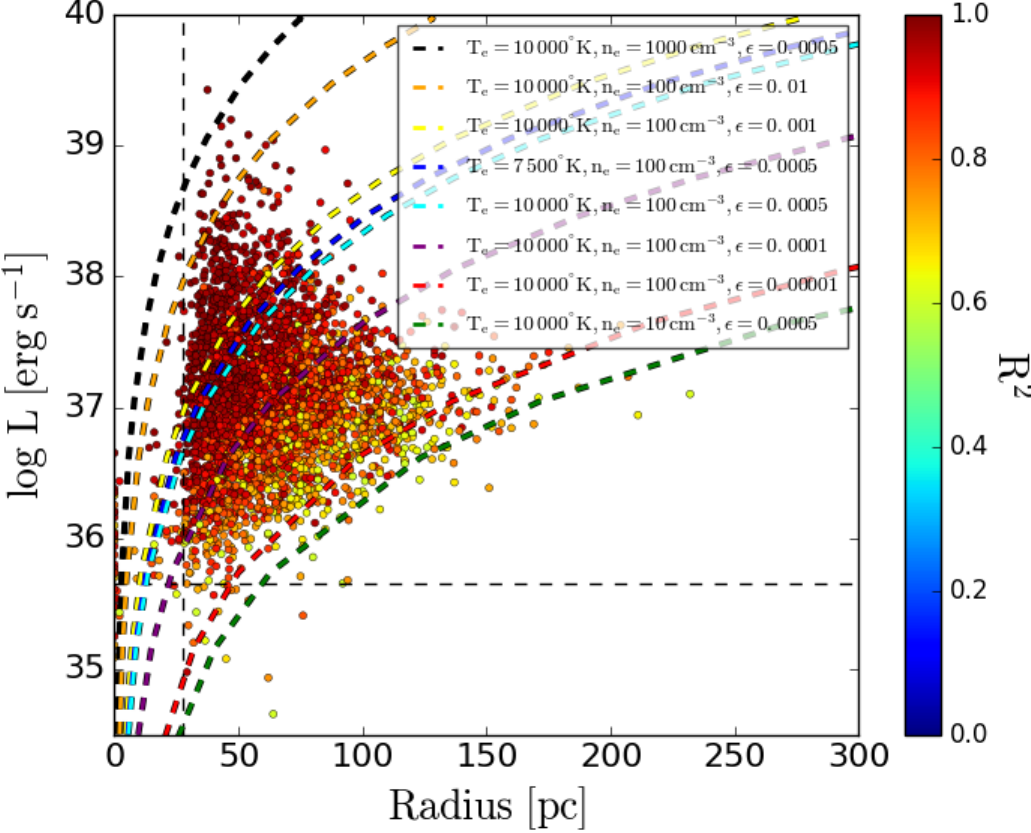}
\Put(15,132){\includegraphics[width=0.8in]{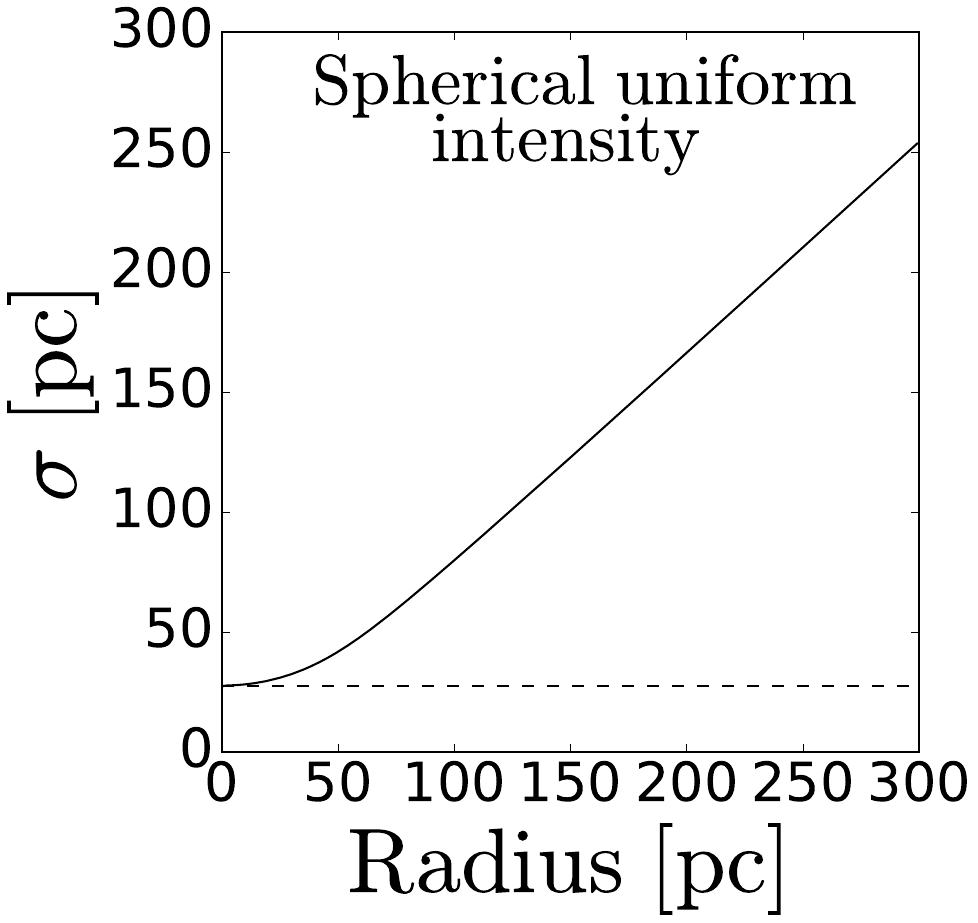}}
\caption{The HII region candidate integrated H$\alpha$ luminosity as a function of their radius $R_{HII}$ obtained when comparing them with Str\"{o}mgren spheres. Only regions with a correlation coefficient R$^2$\,$>$\,0.6 are shown. The dashed curves represent the luminosity and radius relation for theoretical Str\"{o}mgren spheres considering various stellar cluster masses M$_{cl}$, electron temperatures T$_e$ and densities n$_e$, and volume filling factors $\epsilon$, as indicated in the plot.
For each dashed curve, the cluster mass increases from the left to the right. The vertical dashed line is the sensitivity threshold of the H$\alpha$ intensity distribution for the Str\"{o}mgren model extracted from the relation between the $\sigma_g$ of the Gaussian fit for the intensity profile and Str\"{o}mgren radius $R_S$(overplotted in the left lower corner). The horizontal dashed line is the luminosity detection limit.}
\label{reg_size3}
\end{center}
\end{figure}
\begin{figure}
\begin{center}
\includegraphics[width=3.3in]{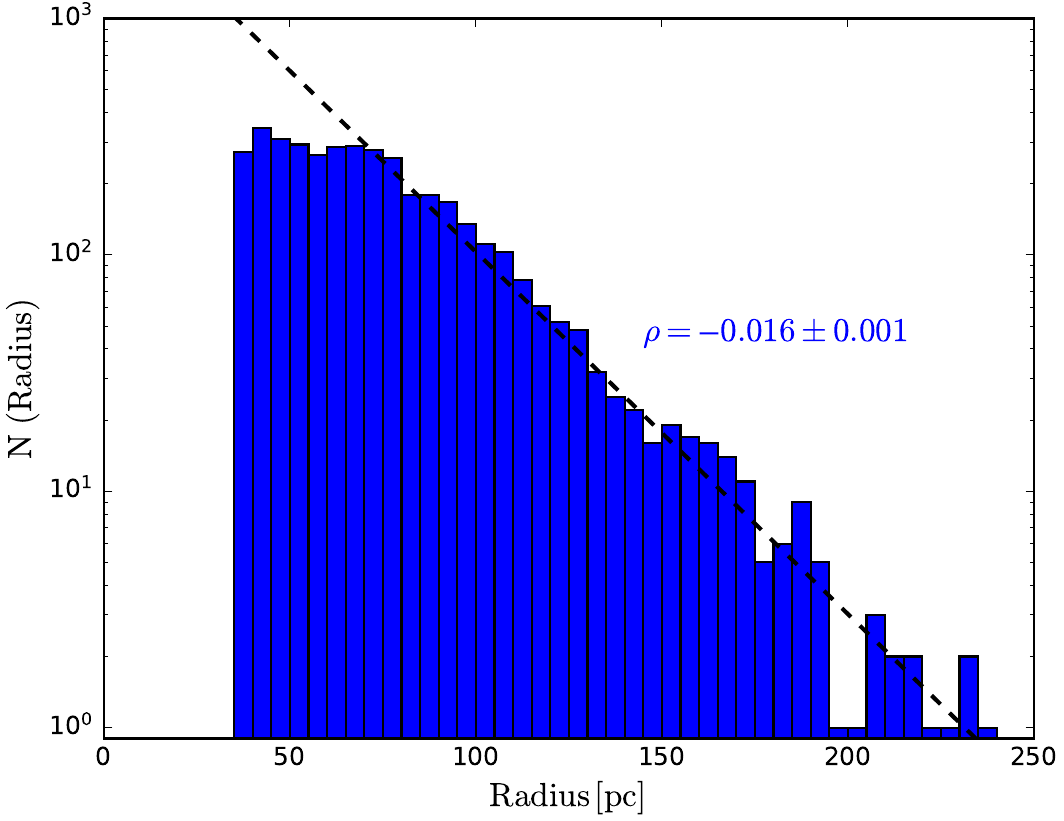}
\caption{Size distribution function of the HII region candidates. The dashed line shows a linear fit with a slope $\rho$ for regions with a radius $R_{HII}$ between 70 and 250\,pc.}
\label{reg_size4}
\end{center}
\end{figure}

Figure\,\,\ref{reg_size3} shows the HII region integrated H$\alpha$ luminosity as a function of their radius $R_{HII}$. Theoretical curves for Str\"{o}mgren spheres with various stellar cluster masses, gas electron temperatures and densities, and volume filling factors are also shown on this plot. It is important to note that no luminosity-size relation, such as the often used L\,$\propto$\,$D^\eta$ (where $D$ is the diameter), follows the well-spread distribution of our measurements. In fact, \cite{s01}, who found $\eta$\,=\,2, stated that this relation is biased by unresolved, superposed, or blended HII regions. \cite{l13} found a value of $\eta$\,=\,3 but also made a careful analysis of the possible misinterpretation of the data, and concluded that the estimation of $\eta$ is strongly affected either by the algorithm used to define the regions or by observational biases. Here, the combination of the good spatial resolution, our emission region identification technique, and radius definition method reveal the complex nature of the luminosity-size distribution of NGC\,628 HII regions. Multiple models for different ISM conditions are required to understand this distribution. From the comparison with the theoretical curves, we can see that the electron temperature has a very small effect on the HII region radius, whereas both the electron density and the volume filling factor can be held responsible for most of the scattering. Also, as the size of a bright HII region increases, the electron density and/or the volume filling factor decreases. From the general distribution of the low-luminosity regions and the models, we can conclude that they are either dense and small, with a higher electron density and volume filling factor, or (most probably) diffuse and large, with a smaller electron density and volume filling factor in agreement with the fact that low-mass HII regions are less dense and ionisation bounded. For bright regions, the size is globally smaller (between 30 and 100\,pc) in comparison with the low-luminosity sample. The bright HII regions have either a high electron density and/or a high volume filling factor.

Figure\,\,\ref{reg_size4} shows the size distribution function of the HII regions. We simply plotted the histogram of log(N($R_{HII}$)d$R_{HII}$) as a function of $R_{HII}$d$R_{HII}$, where N($R_{HII}$)d$R_{HII}$ is the number of regions with a radius in the range $R_{HII}$ to $R_{HII}$+d$R_{HII}$. The ability to recover the size of a region is limited by the spatial resolution of the data and the physical distance between regions. Therefore, the measured size of the smallest regions is only an upper limit. 
Although the physical meaning of this plot is not clear, it is interesting as it presents a relatively simple slope, $\rho$\,=\,0.016$\pm$0.001 (measured between a radius of 70 and 250\,pc). 

\subsection{Luminosity Function of the HII Region Candidates}
\label{lumf}

The total H$\alpha$ flux of an emission region is obtained by summing the flux of all the pixels up to the outer limit radius, and after subtraction of the DIG background. Using a distance of 9.006\,Mpc for the galaxy (NED), the total flux is then converted into the H$\alpha$ luminosity. For some fainter regions, the central peak can be well detected but noise in the region's boundary can produce a total luminosity below the initial detection threshold. Therefore, from the 4285 regions detected, 4158 have a log(L$_{H\alpha}$ in erg\,s$^{-1}$) above the detection limit of 35.65.

\begin{figure}
\begin{center}
\includegraphics[width=3.3in]{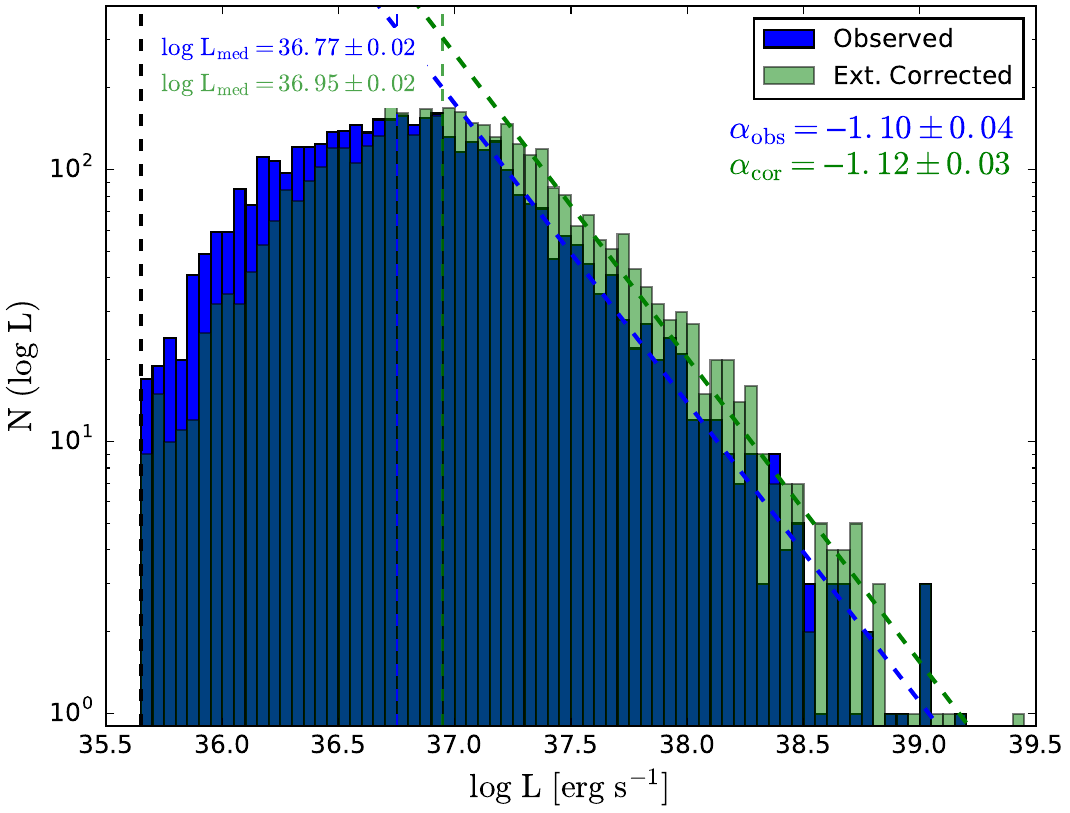} 
\caption{The H$\alpha$ luminosity function of the 4158 HII region candidates before (in blue) and after (in green) the extinction correction. The slope $\alpha$ is evaluated in the luminosity range log(L$_{med}$)+0.2 to 39.3 in both cases. The vertical lines in blue and green indicate the median value for each case. The vertical line in black shows the detection limit.}
\label{lf}
\end{center}
\end{figure}

Considering that the majority of the emission regions selected are HII regions (this is well supported by the BPT diagrams discussed in Section 6), an HII region luminosity function was calculated using
the sample of 4158 emission regions and is shown in Figure\,\,\ref{lf}. It follows the relation N(L)dL\,=\,AL$^{\alpha}$dL, where N(L)dL is the number of regions with a luminosity in the range L to L+dL. The analysis was done by considering regions before and after the extinction correction. Bins of 0.05 in log(L$_{H\alpha}$ in erg\,s$^{-1}$) were selected due to the data quality, allowing us to show a smooth and well-sampled curve. The luminosity function slopes derived with the extinction corrected data, $\alpha_{cor}\,=\,-1.12\pm0.03$, and without the correction, $\alpha_{obs}\,=\,-1.10\pm0.04$, are identical. The median luminosity for the extinction corrected data is 36.95$\pm$0.02, and for the uncorrected data is 36.77$\pm$0.02. No break in the slopes is seen, as previously noticed at high luminosity, e.g. \cite{b06}. 

\begin{figure}
\begin{center}
\includegraphics[width=3.3in]{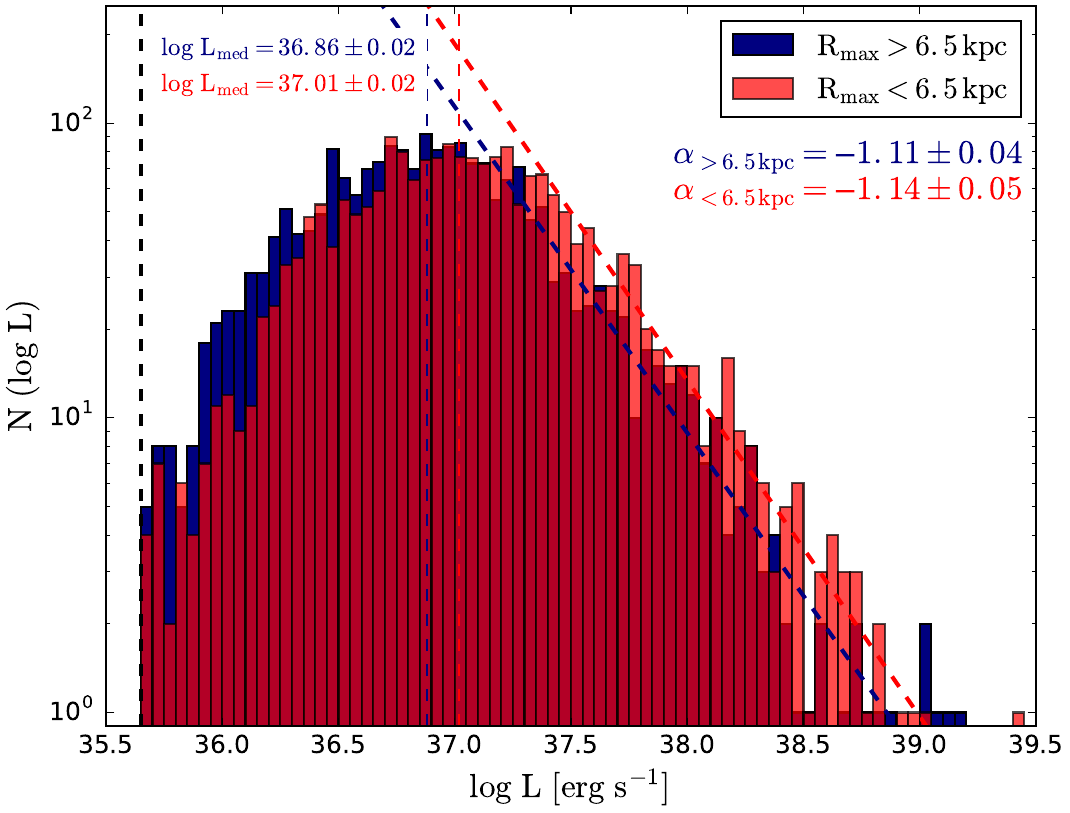} 
\caption{The H$\alpha$ luminosity function of the HII region candidates (extinction corrected) inside and outside a galactocentric radius $R_G\,=\,6.5$\,kpc. The slope $\alpha$ is evaluated in the luminosity range log(L$_{med}$)+0.2 to 39.1 for both samples (inside the 6.5\,kpc radius in red and outside in blue). The vertical lines in red and blue indicate the median value of each sample. The vertical line in black shows the detection limit.}
\label{lf2}
\end{center}
\end{figure}

We looked for variations of the luminosity function with the galactocentric distance of the HII regions, separating them into two samples, one inside and one outside a radius $R_{G}$ of 6.5\,kpc (in the case of the extinction corrected data only). This distance was first selected because it simply gives a nearly equal number of regions in both samples (2075 regions inside and 2083 regions outside 6.5\,kpc). These two samples are also covering a different gas metallicity range, as already known from the metallicity gradient reported by several authors (e.g. \citealt{s11}; \citealt{p14}; \citealt{b15}) and as supported by the line ratios studied in Section\,6. Figure\,\,\ref{lf2} shows the resulting luminosity functions. We do not see a significant change in the slope of the function with respect to the radial position in the disc ($\alpha_{>\,6.5\,kpc}$\,=\,$-1.11\pm0.04$ and $\alpha_{<\,6.5\,kpc}$\,=\,$-1.14\pm0.05$). Regions in the central portion of the disc ($<$\,6.5\,kpc) are in general slightly brighter with a median value of log\,L$_{med}$\,=\,37.01$\pm$0.02 compared to log\,L$_{med}$\,=\,36.86$\pm$0.02 in the external portion of the disc.
Indeed, brighter regions (logL\,$>$\,37.5) and less faint regions (logL\,$<$\,37) are seen in the central 
portion of the disc ($R_G$\,$<$\,6.5\,kpc). This could results from a crowding effect in the central disc, where many small regions overlap and are measured as one bright region. 
To evaluate if the level of crowding is different in both subsamples, the distance between two peaks was evaluated for the closest, the second-closest, and the third-closest peak for all the 4285 peaks in the sample. We found a median value for the distance between peaks of 110.0\,pc at $R_G$\,$<$\,6.5\,kpc and 115.5\,pc at $R_G$\,$>$\,6.5\,kpc for the closest, 159.5\,pc at $R_G$\,$<$\,6.5\,kpc and 168.3\,pc at $R_G$\,$>$\,6.5\,kpc for the second closest, and 198.6\,pc at $R_G$\,$<$\,6.5\,kpc and 211.0\,pc at $R_G$\,$>$\,6.5\,kpc for the third closest peak. The error for all median values is $<$\,1\,pc if evaluated by moving the galactocentric radius threshold by 500\,pc. Although small, this difference supports the idea that a crowding effect could explain the small differences observed in the luminosity functions presented in Figure\,\,\ref{lf2}. 
\begin{figure}
\begin{center}
\includegraphics[width=3.3in]{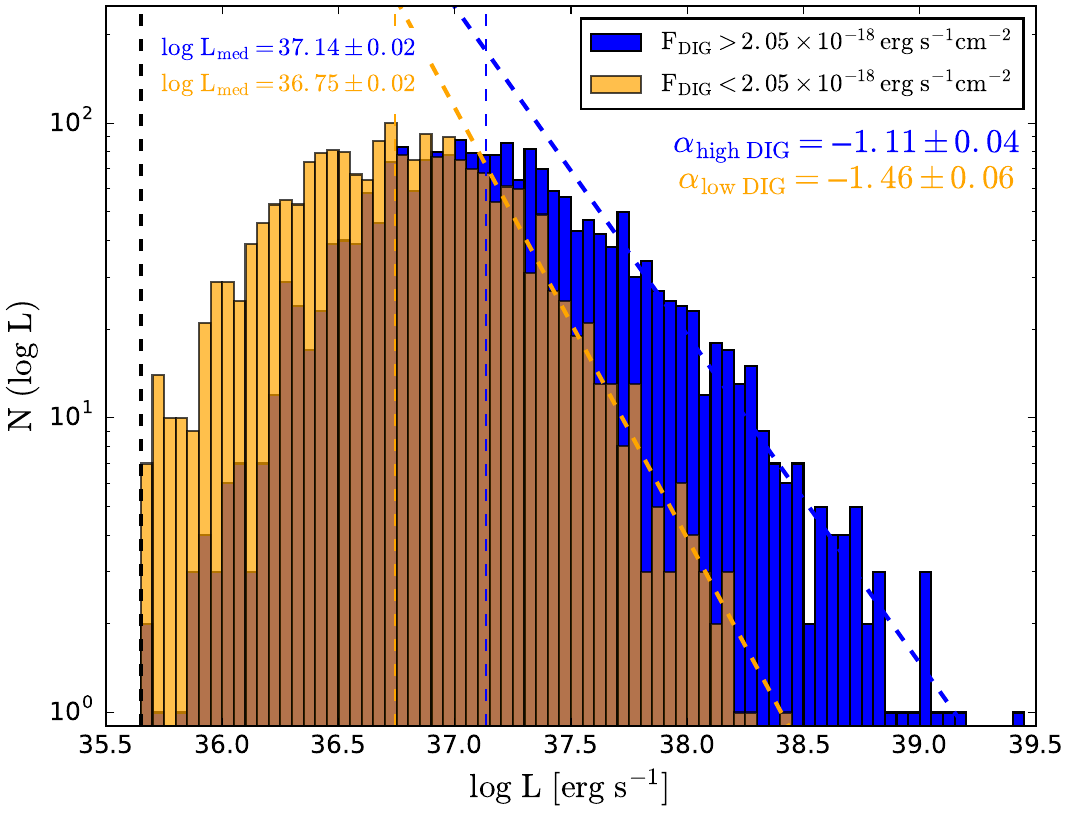} 
\caption{The H$\alpha$ luminosity function of HII region candidates (extinction corrected) as a function of their DIG background. The threshold for the DIG emission if 2.05$\times$10$^{-18}$\,erg\,s$^{-1}$cm$^{-2}$ over one pixel. The slope $\alpha$ is evaluated in the luminosity range log(L$_{med}$)+0.2 to 38.4 for the low DIG background sample (in orange) and in the range log(L$_{med}$)+0.2 to 39.3 for the high DIG background sample (in blue). The vertical lines in gold and blue indicate the median value of each sample. The vertical line in black shows the detection limit.}
\label{lf3}
\end{center}
\end{figure} 

\begin{figure}
\begin{center}
\includegraphics[width=3.3in]{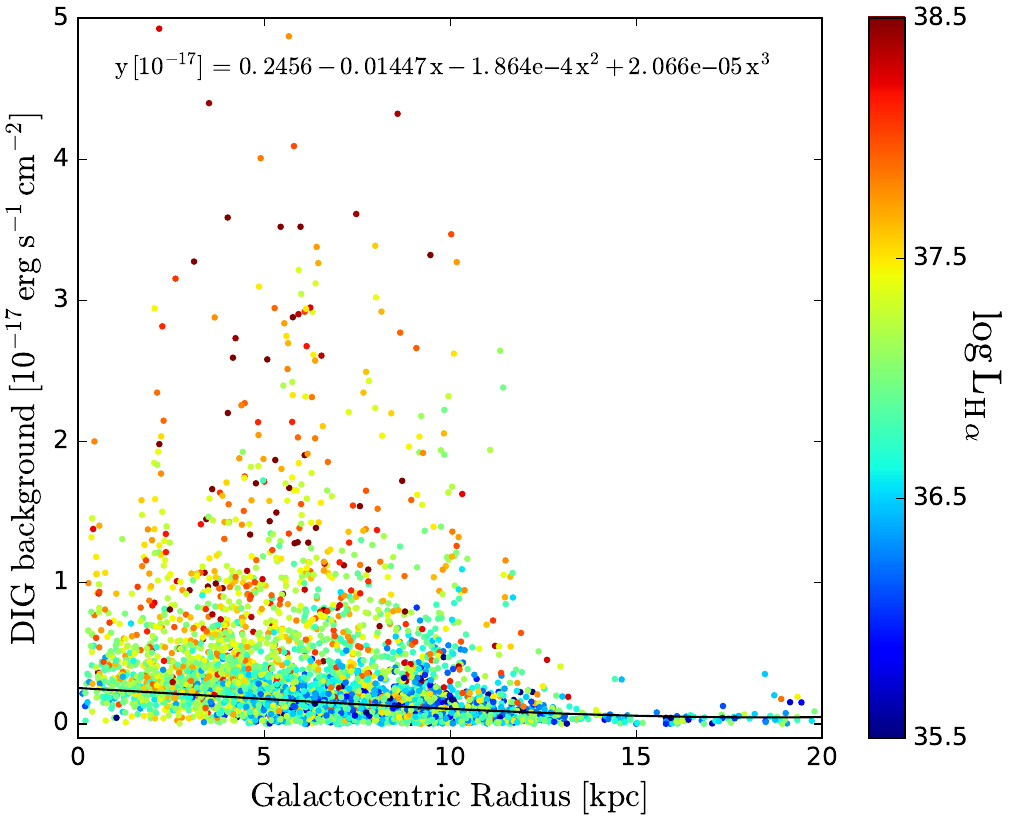} 
\caption{The DIG background from the H$\alpha$ emission flux for each of the 4285 spaxels as a function of their galactocentric radius $R_G$. The colour scale is a function of the total H$\alpha$ luminosity of the region where the spaxel is located, as define in Section\,5.4. The black line shows a polynomial fit to the data while excluding regions above 5 $\times$10$^{-18}$\,erg\,s$^{-1}$cm$^{-2}$ (the relation from the fit is given on the plot).}
\label{ldr}
\end{center}
\end{figure}

Figure\,\,\ref{lf3} shows the luminosity function for two samples of HII region candidates separated according to their DIG background. A threshold for the DIG surface brightness of 2.05 $\times$10$^{-18}$\,erg\,s$^{-1}$cm$^{-2}$ over one pixel was used to separate the two samples. Although some very bright regions with a significant DIG background are not in the arms, the higher DIG background group contains regions mostly located in the core of the spiral arms (as previously found by \citealt{k16}). As the bright regions have brighter DIG background, it suggests that the main ionisation source for the DIG is within the regions themselves, i.e. massive stars. When excluding spiral arm regions, we find that the DIG background level is correlated with the galactocentric radius, as higher DIG emission is measured in the regions closer to the galaxy centre (see Fig.\,\,\ref{ldr}). This suggests that the evolved star background, more important in the galaxy inner region, is also a source of ionisation for the inter-arm regions. The selected DIG threshold also ensures two samples with a similar size (2075 regions above and 2083 below the threshold). In this case, a significant difference is measured for both the median value and the slope of the luminosity functions (Fig.\,\,\ref{lf3}). The sample with a lower DIG background has lower values : log(L$_{med})\,=\,36.75\pm$0.02, compared to 37.14$\pm$0.02 for the high DIG background regions, and a slope $\alpha_{low\,DIG}\,=\,-1.46\pm0.06$, compared to $\alpha_{high\,DIG}\,=\,-1.11\pm0.04$. The slope of the lower DIG background regions is similar to the slope found in M31 ($\alpha_{M\,31}\,=\,-1.52\pm0.07$) by \cite{a11.2}. Therefore there is a clear relation with the DIG background and the average luminosity of the emission regions, here associating the DIG origin with the importance of the massive star population. But in the case of the lower DIG background regions, the change in the slope of the luminosity function and the correlation with the galactocentric radius suggest multiple origins for the DIG background.

\section{LINE RATIOS OF THE EMISSION REGIONS}
\label{profiles}

\subsection{Line Ratios for Individual Spaxels}

Line ratios were calculated using flux maps of the lines extracted with ORCS ($\S$\,\,3.5). Line ratios of [OII]/[NII], [OIII]/[NII], [OIII]/[OII], [OII]/H$\beta$, ([OII]+[OIII])/H$\beta$, [NII]/H$\alpha$, [SII]/H$\alpha$, [SII]/[NII], [OIII]/H$\beta$, and [SII]$\lambda$6716/[SII]$\lambda$6731 were obtained after the subtraction of the stellar population contribution (\S\,3.4) and the extinction correction (\S\,5.1). 
Appendix B contains line ratio maps covering the whole galaxy as well as for four enlarged reference sections of the disc. A detection threshold of 3$\sigma$ for the individual lines was applied to produce these maps. The maps of the four enlarged reference sections show variations of the line ratios within emission regions themselves (the H$\alpha$ emission peaks are all indicated with a cross on the images), and variations of the DIG background. For example, in Figure\,\,\ref{n2ha}, the [NII]/H$\alpha$ ratio can change significantly within a distance of 100\,pc. As seen in Figure~\ref{n2h_reg}, the H$\alpha$ emission peak may be superimposed to low- or high-local values of the ratio and large fluctuations can be found in the DIG surrounding the regions. Particularly in the case of [SII]/H$\alpha$ 
(Fig.\,\,\ref{s2h_reg}), there is a strong correlation between the position of the peaks and local minimum values. This behavior is expected as this ratio is known to increase when the DIG contribution is important (Hafner et al. 1999; Elwert \& Dettmar 2005). But also, with a resolution of 35\,pc, we could as well be seeing the ionisation structure within the HII regions (\citealt{l10}; Alarie, private communication).

\begin{figure*} 
\begin{center}
\includegraphics[width=3.39in]{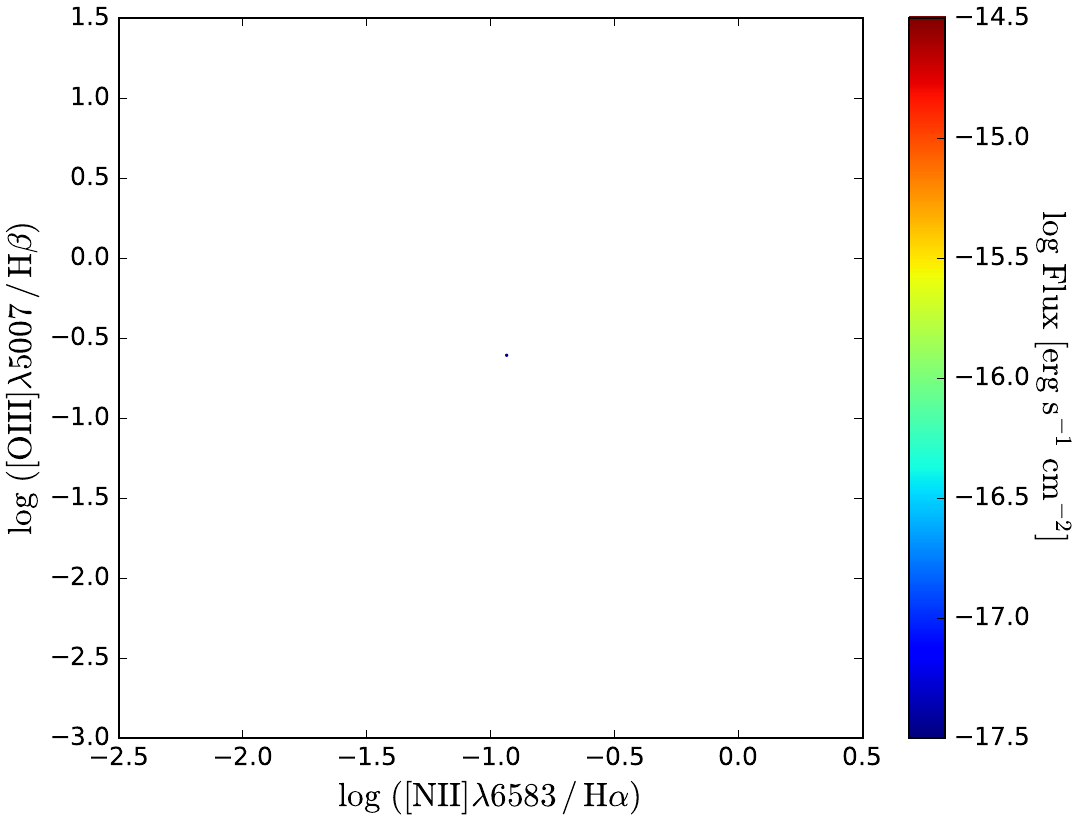} 
\includegraphics[width=3.3in]{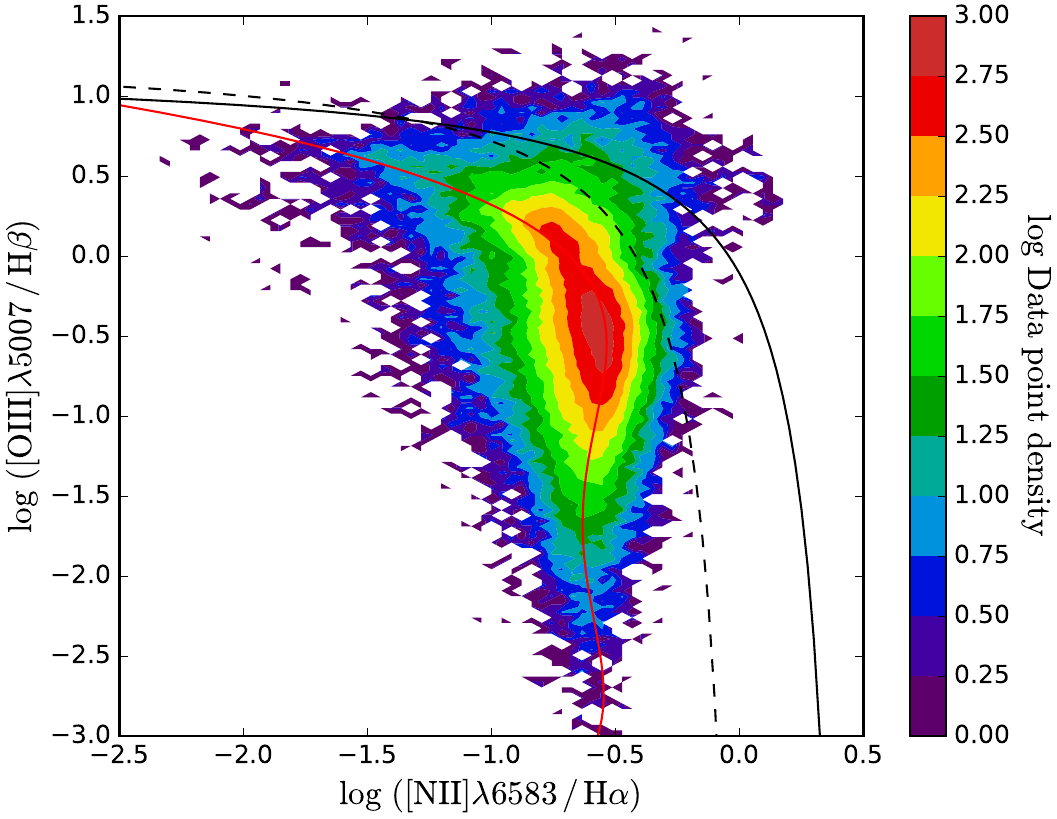} 
\Put(-462,201.1){\includegraphics[width=2.334in, height=2.256in]{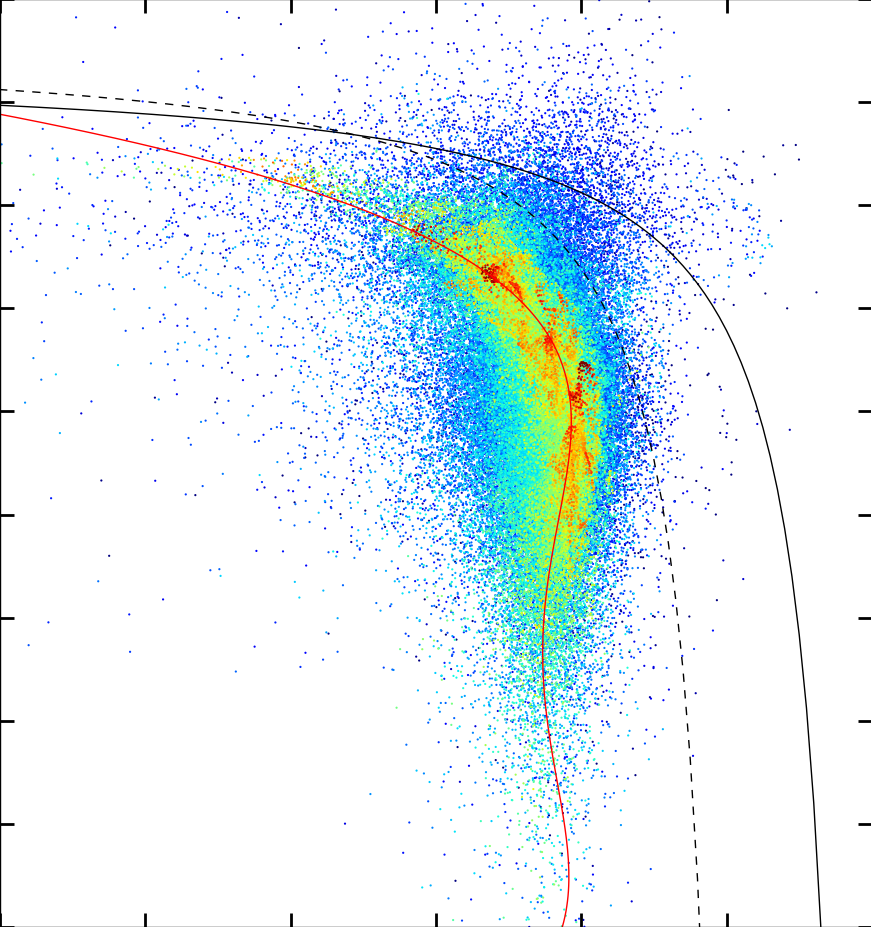}}
\caption{BPT diagram of [OIII]/H$\beta$ vs [NII]/H$\alpha$ for all the spaxels measured, on the left, colour coded according to the H$\alpha$ flux, and on the right, colour coded according to the spaxel density. The dashed curve defines the BPT limit between the HII regions and the transition zone, whereas the black curve defines the BPT limit between the transition zone and the AGN regime. The red curve corresponds to the polynomial fit of the data. In the left panel, spaxel values are stacked over each other from the faintest H$\alpha$ flux region to the brightest one, and therefore, bright spaxels always appear in front.}
\label{bpt_flux_NII}
\end{center}
\end{figure*}

Figure\,\,\ref{bpt_flux_NII} shows the BPT diagram (as initially defined by \citealt{b81}) of [OIII]/H$\beta$ vs [NII]/H$\alpha$ for all the spaxels measured over the whole galaxy disc. Two versions of the same diagram are shown, one colour coded according to the H$\alpha$ flux for each spaxel and the other according to the spaxel density. As expected, most spaxels lie in the HII region ionisation regime area of the BPT diagram and a few spaxels cross the transition zone and AGN limit. The general trend drawn by the bright H$\alpha$ spaxels (red curve in the Fig.\,\,\ref{bpt_flux_NII}) can be represented by the polynomial fit : \begin{equation}
\label{bpt_poly}
 \begin{split}
 \mathrm{log\,([NII]/H\alpha)} = &\,\,-0.6841 - 0.651x - 0.9026x^{2} \\
 & \,\,- 0.4249x^{3} -0.06401x^{4}, \\
 \end{split}
\end{equation} 

\noindent where $x$\,=\,log([OIII]/H$\beta$). Globally, the fit (going from high to low [OIII]/H$\beta$ ratio) follows an increase in the gas metallicity. We can also notice a turnover at log([OIII]/H$\beta$)\,=\,$-$0.5 and log([NII]/H$\alpha$)\,=\,$-$0.5, where the [NII]/H$\alpha$ ratio starts to decrease while the [OIII]/H$\beta$ continues to decrease. This turnover is expected when the metallicity reaches a value above 12+log[O/H]\,$\simeq$\,9 (Fig. 7 from \citealt{k02.2}) while considering that the number of ionizing photons and therefore the ionisation parameter ($q$\,=\,$Q$$_{H}$/$n_H$, the number of hydrogen ionizing photons over the hydrogen atom density) both decrease with increasing metallicity (number of ionizing photons as a function of the metallicity; \citealt{t15}, or ionisation parameter as a function of the metallicity; \citealt{pm14}). A turnover may also be due to a variation in the relative abundances (e.g. N/O) in the galaxy. Therefore, the location of the polynomial fit and the possible turnover are expected to vary for galaxies with different metallicities, relative abundances, and other properties. 

\begin{figure*}
\begin{center}
\includegraphics[width=3.3in]{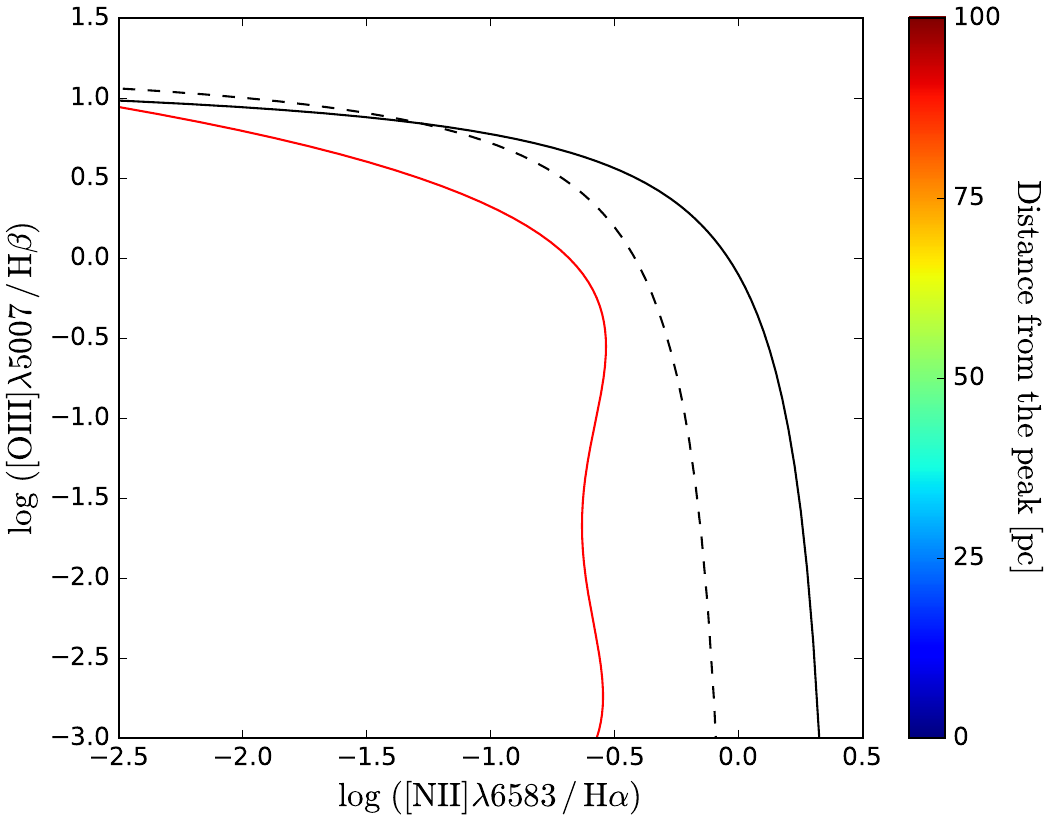} 
\includegraphics[width=3.3in]{NGC628_BPT_distance_peak_tt.pdf} 
\Put(-455.5,201.1){\includegraphics[width=2.334in, height=2.256in]{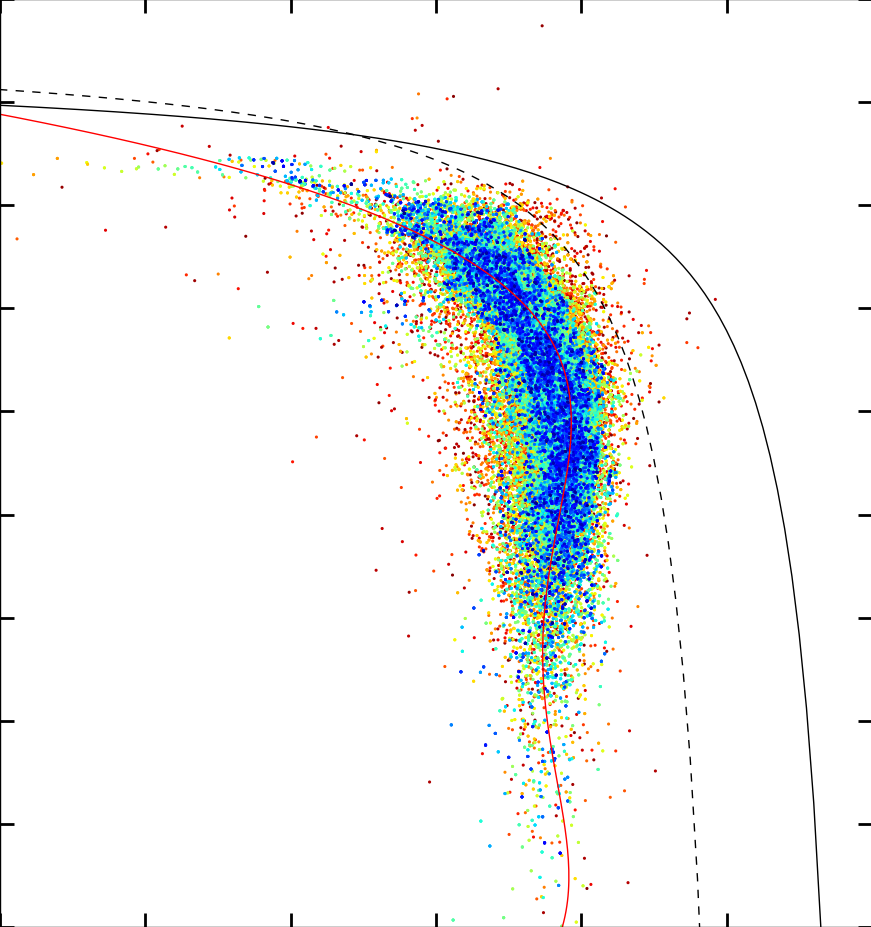}}
\Put(-217,201.1){\includegraphics[width=2.334in, height=2.256in]{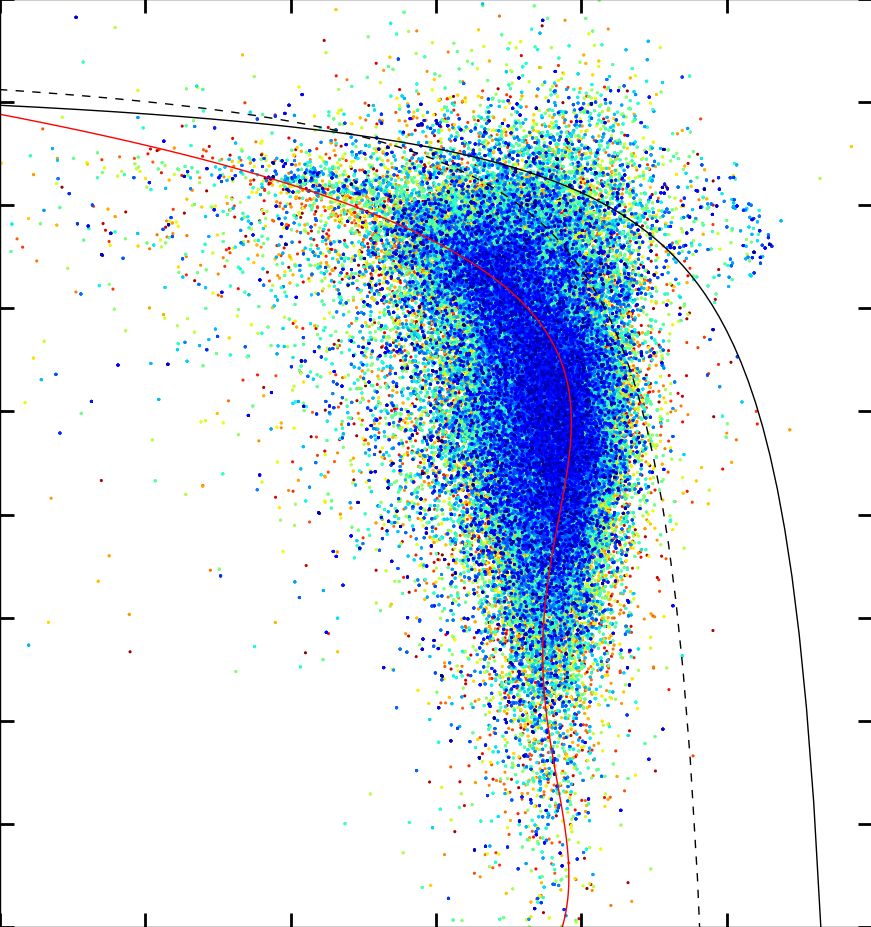}}
\caption{BPT diagram of [OIII]/H$\beta$ vs [NII]/H$\alpha$ for spaxels included in very bright emission regions (i.e. regions with an integrated H$\alpha$ flux above 5$\times$10$^{-15}$erg\,\,s$^{-1}$cm$^{-2}$; on the left) and, for moderately bright spaxels (i.e. spaxels with an H$\alpha$ flux above 5$\times$10$^{-16}$erg\,\,s$^{-1}$cm$^{-2}$; on the right). The plots are colour coded according to the spaxel distance to its emission peak. Spaxel values are stacked over each other from the more distant to the closest one, and therefore, closest regions always appear in front. The curves are as in Figure\,\,\ref{bpt_flux_NII}.}
\label{bpt_dist}
\end{center}
 \end{figure*}
 
\begin{figure*}
\begin{center}
\includegraphics[width=3.3in]{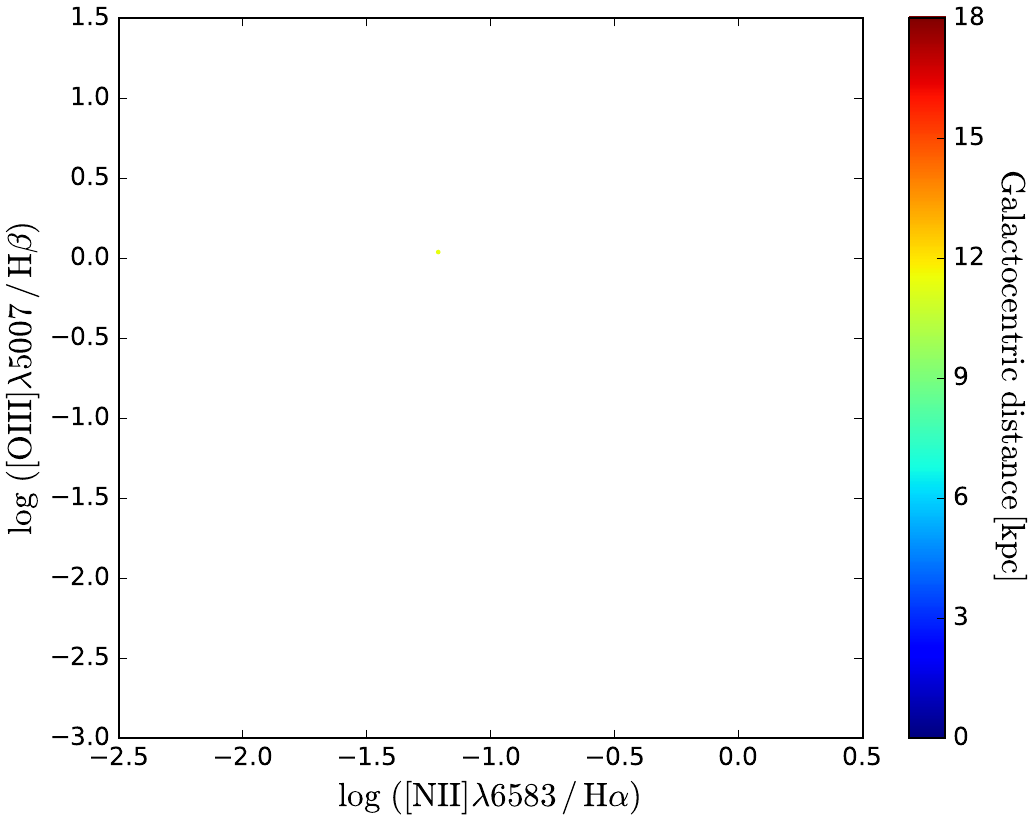} 
\includegraphics[width=3.3in]{NGC628_BPT_distance_gal_tt.pdf} 
\Put(-455.9,204.6){\includegraphics[width=2.39in, height=2.313in]{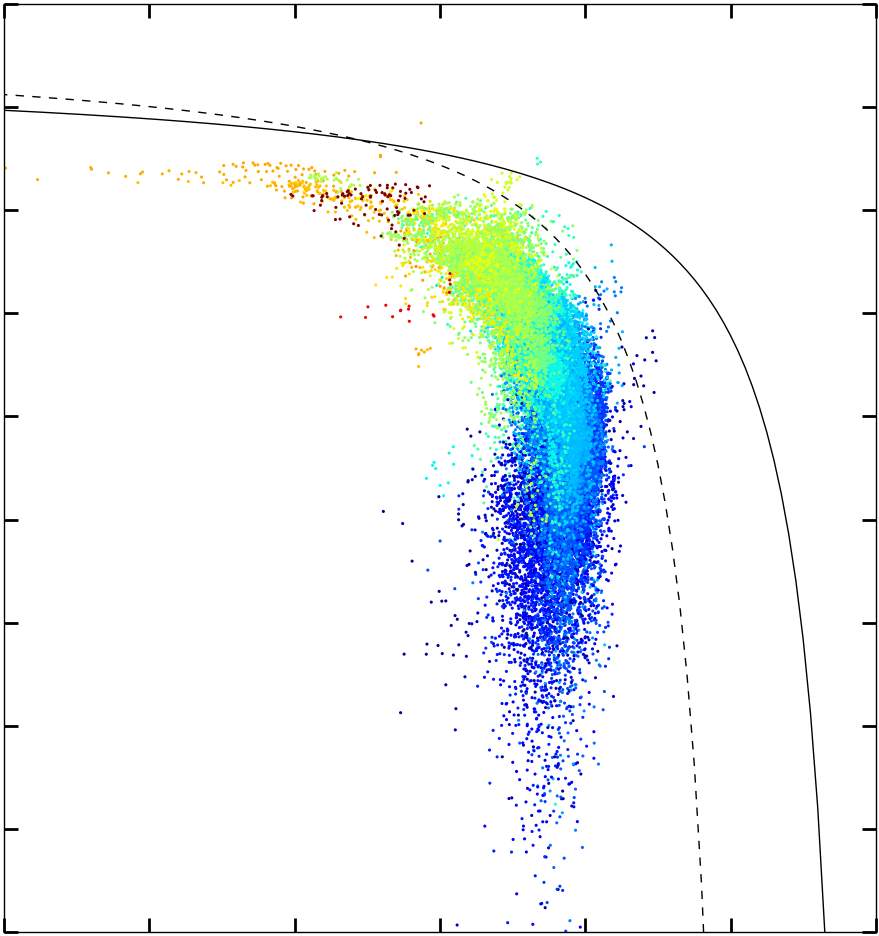}}
\Put(-217.5,204,6){\includegraphics[width=2.39in, height=2.313in]{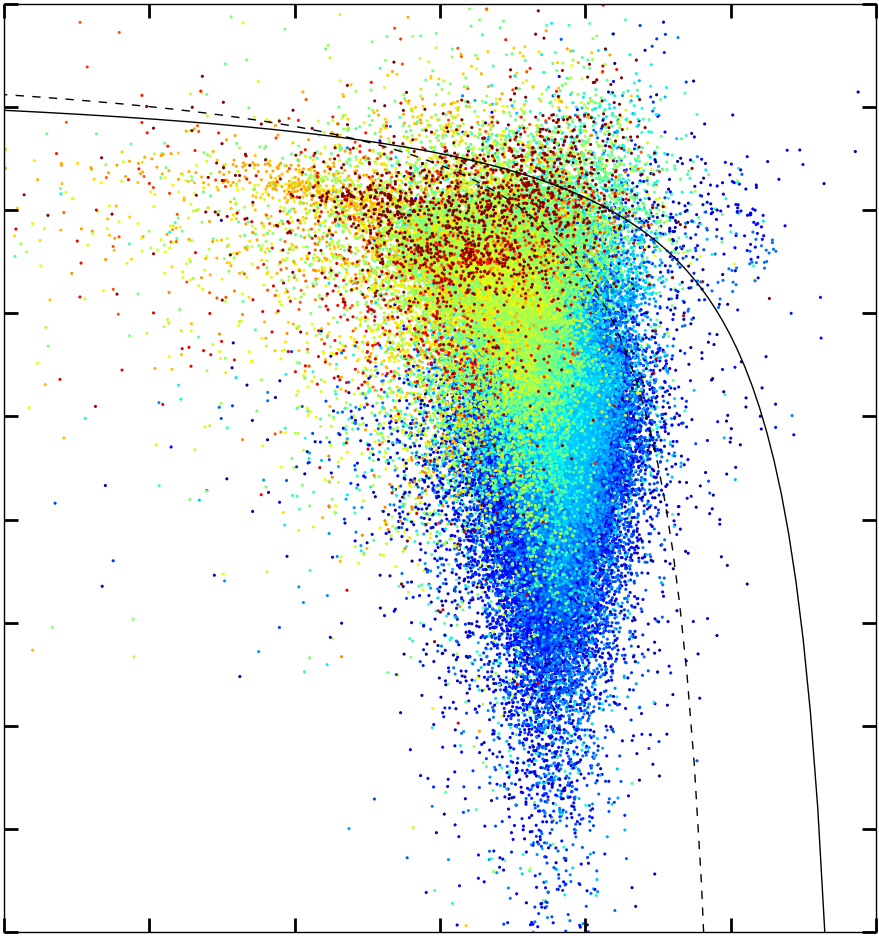}}
\caption{BPT diagram of [OIII]/H$\beta$ vs [NII]/H$\alpha$ for spaxels included in very bright emission regions (i.e. regions with an integrated H$\alpha$ flux above 5$\times$10$^{-15}$erg\,\,s$^{-1}$cm$^{-2}$; on the left), and for faint spaxels (i.e. spaxels with an H$\alpha$ flux above 5$\times$10$^{-17}$erg\,\,s$^{-1}$cm$^{-2}$; on the right). The plots are colour coded according to the spaxel galactocentric radius $R_{G}$. Spaxel values are stacked over each other from the smallest radius to the larger one, and therefore, spaxels at large radius always appear in front. The curves are as in Figure\,\,\ref{bpt_flux_NII}.}
\label{bpt_gal_dist}
\end{center}
\end{figure*}

\begin{figure*}
\begin{center}
\includegraphics[width=3.39in]{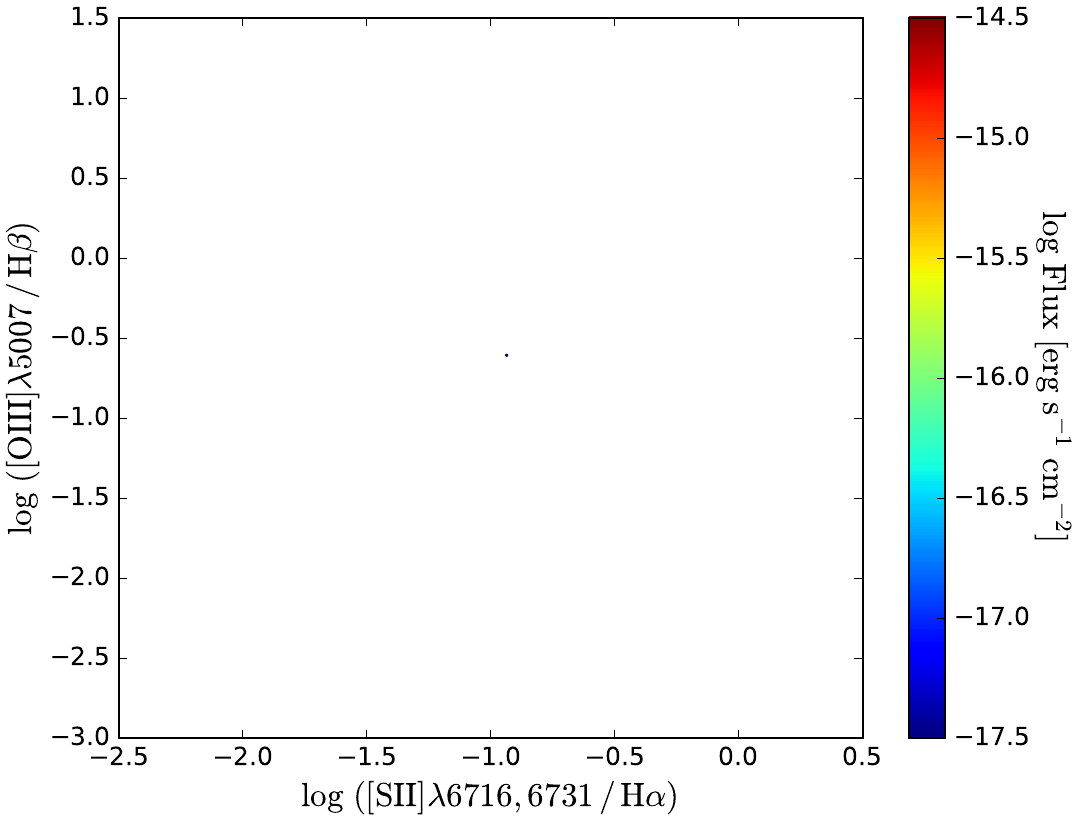} 
\includegraphics[width=3.3in]{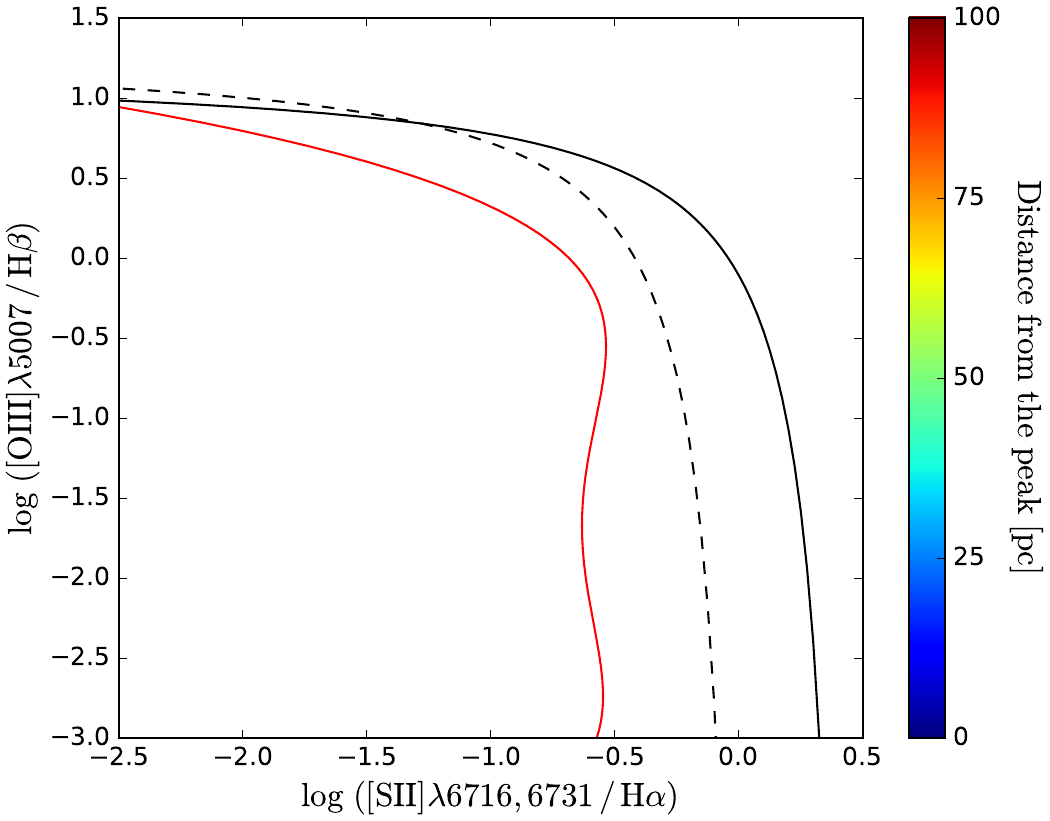} 
\Put(-462.1,201.1){\includegraphics[width=2.334in, height=2.256in]{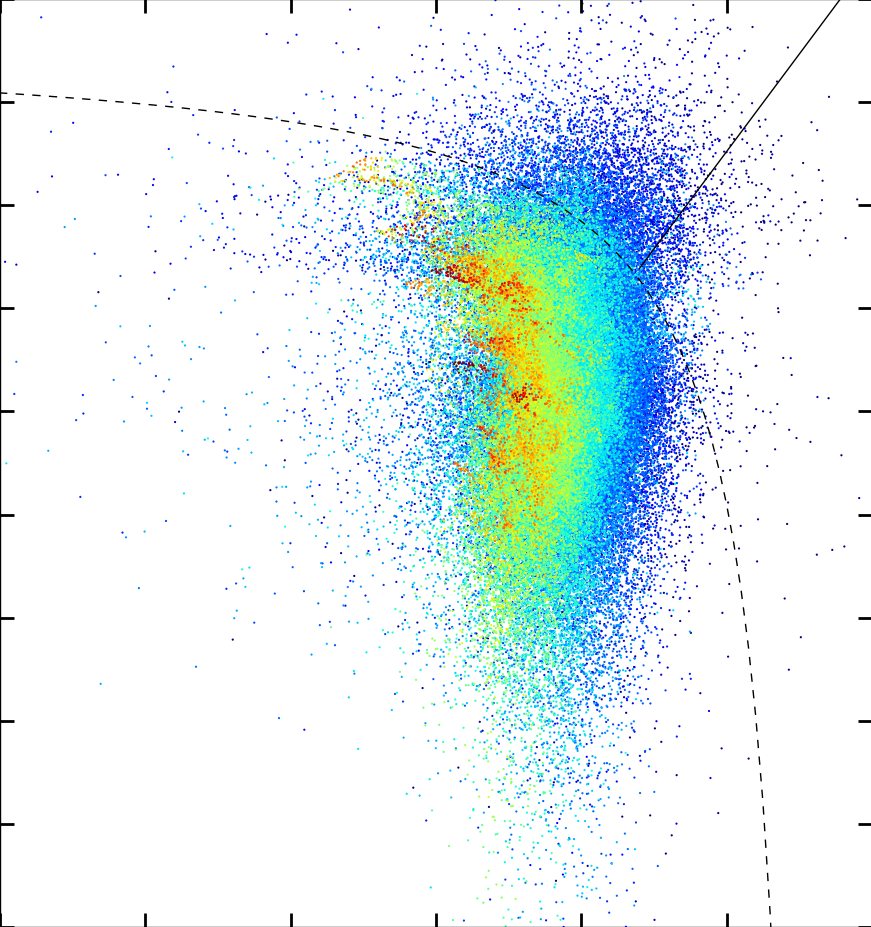}}
\Put(-217,201.1){\includegraphics[width=2.334in, height=2.256in]{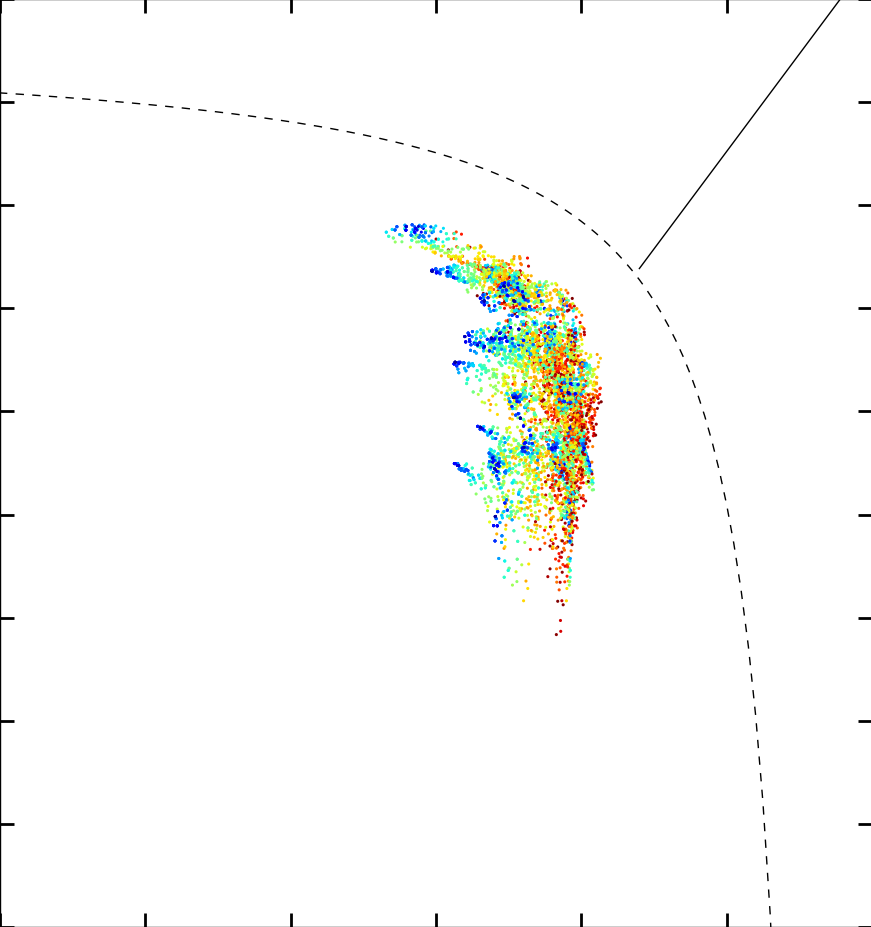}}
\caption{BPT diagram of [OIII]/H$\beta$ vs [SII]/H$\alpha$ for all the spaxels measured. To the left: the spaxels are shown colour coded according to their individual H$\alpha$ intensity. To the right: only the spaxels
from extremely bright HII regions (integrated H$\alpha$ flux above 3$\times$10$^{-14}$erg\,\,s$^{-1}$cm$^{-2}$) are shown colour coded according to their distance to the emission peak. The dashed curve defines the BPT limit between the HII regions and the AGN ionisation regime, whereas the full line defines the BPT limit between the Seyfert and LINER ionisation regime. On the left, spaxel values are stacked over each other from the faintest H$\alpha$ flux to the brightest one, and therefore, bright spaxels always appear in front. On the right, spaxel values are stacked over each other from the more distant to the closest one, and therefore, closest regions always appear in front.}
\label{bpt_flux_SII}
\end{center}
\end{figure*} 

In Figure\,\,\ref{bpt_flux_NII}, the spaxels that are bright in H$\alpha$ are predominantly associated with massive HII regions, where it is most likely that the IMF is well represented (i.e. no stochastic sampling with the theoretical slope and mass limits). The dispersion observed along the fit in the bright H$\alpha$ spaxels could therefore be mainly attributed to different ages (as the ionizing spectrum gets less energetic with time with the death of the most massive stars). On the contrary, the dispersion of faint 
H$\alpha$ spaxels can be due to multiple causes: age, IMF under-sampling, DIG contamination, different sources of ionisation, SNR, etc.

In the right panel of Figure\,\,\ref{bpt_flux_NII}, the density envelop of the BPT diagram shows a 
concentration of spaxels around the ratio values [$-$0.5, $-$0.5], associated with some of the most 
intense peaks, and a more extended cloud of spaxels centered at [$-$0.8, 0.2]. 
Many of the spaxels in this clouds reach the transition and AGN zones, 
suggesting that young stellar clusters are not their main ionizing source.
These spaxels are not associated with any particular structure (like the arms)
in the galaxy, but, as shown in the following figures, are more related to the less 
bright regions.

Figure\,\,\ref{bpt_dist} shows the [OIII]/H$\beta$ vs [NII]/H$\alpha$ BPT diagram by taking into account the spaxel distance to its ionizing source (i.e. the distance between the spaxel and the emission peak within the zone of influence defining an emission region). Diagrams are displayed independently for very bright regions (i.e. with an integrated H$\alpha$ flux above 5$\times$10$^{-15}$erg\,\,s$^{-1}$cm$^{-2}$) and for moderately bright spaxels (i.e. with an integrated H$\alpha$ flux above 5$\times$10$^{-16}$erg\,\,s$^{-1}$cm$^{-2}$. In the case of very bright regions (Fig.\,\,\ref{bpt_dist} left panel), the further away they are from their emission peak, the greater is the dispersion around the best polynomial fit. In the case of the sample including moderately bright regions (Fig.\,\,\ref{bpt_dist} right panel), the relation with the distance to the emission peak is more ambiguous. In this case, for moderately bright regions, the effect of the distance is mixed with other parameters like the age, IMF under-sampling, DIG contribution, and other sources of ionisation, as seen in the previous paragraphs.

Figure\,\,\ref{bpt_gal_dist} shows the [OIII]/H$\beta$ vs [NII]/H$\alpha$ BTP diagram as a function of the spaxel galactocentric radius $R_{G}$, for bright regions and faint regions. The general distribution of the colours in these plots reflects the metallicity gradient of the galaxy. The central spaxels (in blue) are more metal rich than those of the extended disc (in yellow-orange-red), as they have larger values of [NII]/H$\alpha$ and smaller values of [OIII]/H$\beta$ (as expected from the global metallicity indicators, e.g. \citealt{pp04}). Nevertheless, the northern spiral arm, located at a distance larger than 15 kpc (red dots in the plots near log([OIII]/H$\beta$)\,$\simeq$\,0), behaves differently than this somewhat traditional linear-negative metallicity gradient. This could be due to a flattening or an inversion of the slope of the metallicity gradient at large galactocentric radius and/or to a change in the gradient of the relative abundance of N/O. 
 
Figure\,\,\ref{bpt_flux_SII} presents the [OIII]/H$\beta$ vs [SII]/H$\alpha$ BTP diagram.
As shown in the left panel, the brightest spaxels are mostly found within 
the HII region ionizing regime. When considering only the brightest regions 
(the right panel), the spaxels that are closer to the emission peak display a smaller [SII]/H$\alpha$ ratio in general. Larger values for this ratio are a signature of DIG and/or of the ionisation structure, as mentioned already. 

 \begin{figure*}
\begin{center}
\includegraphics[width=3.25in]{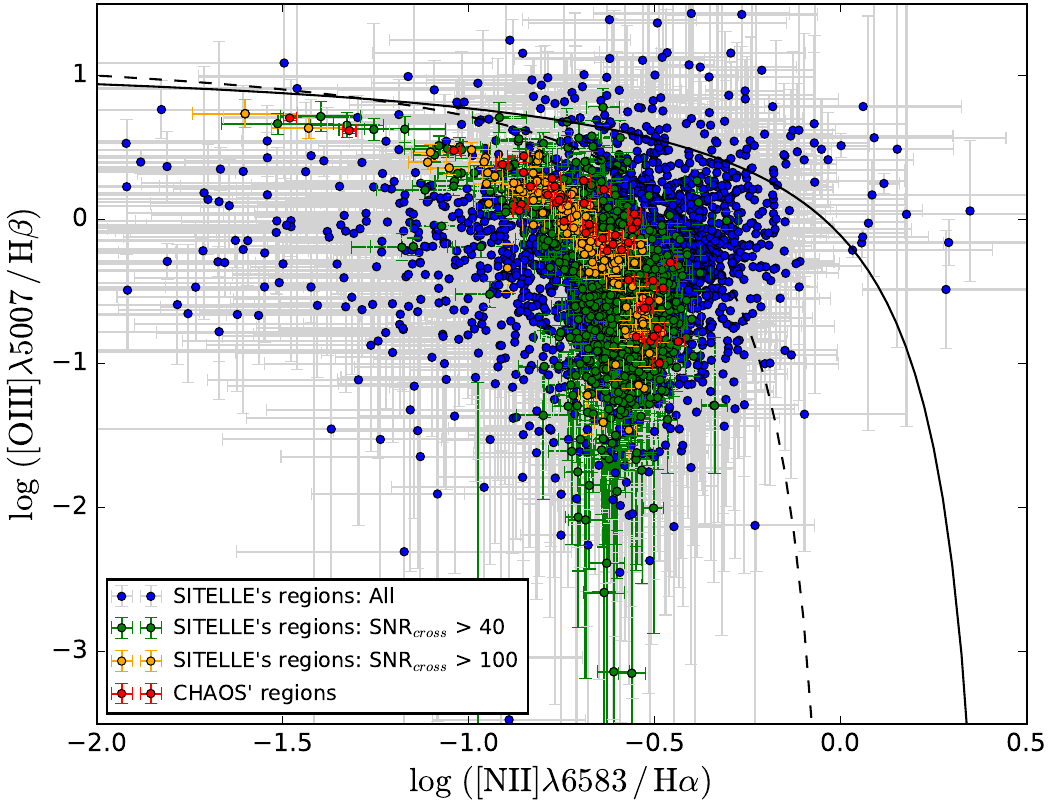} 
\includegraphics[width=3.25in]{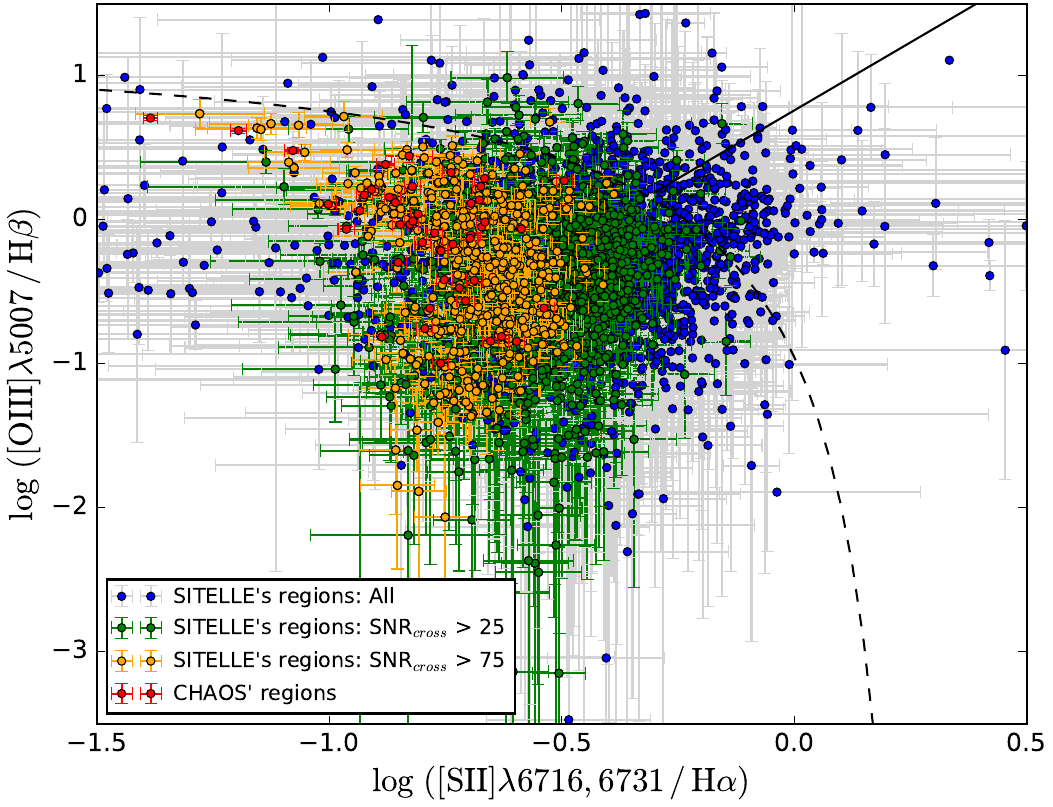} 
\includegraphics[width=3.25in]{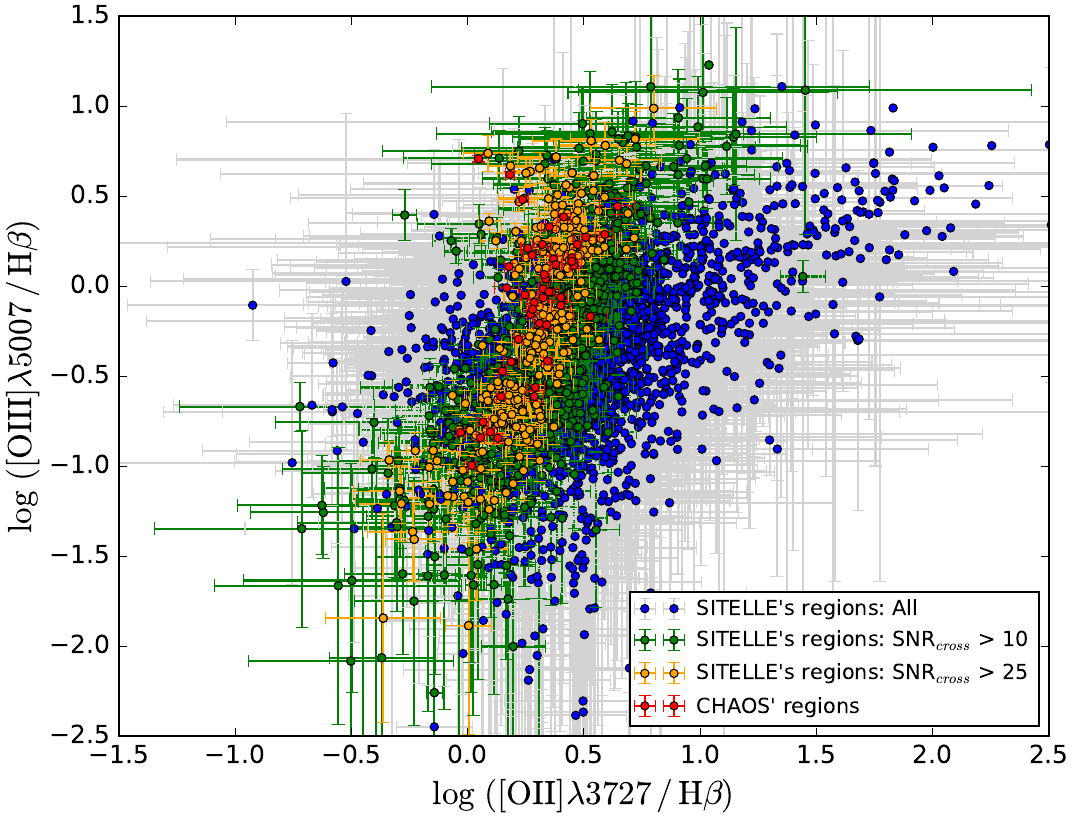} 
\includegraphics[width=3.25in]{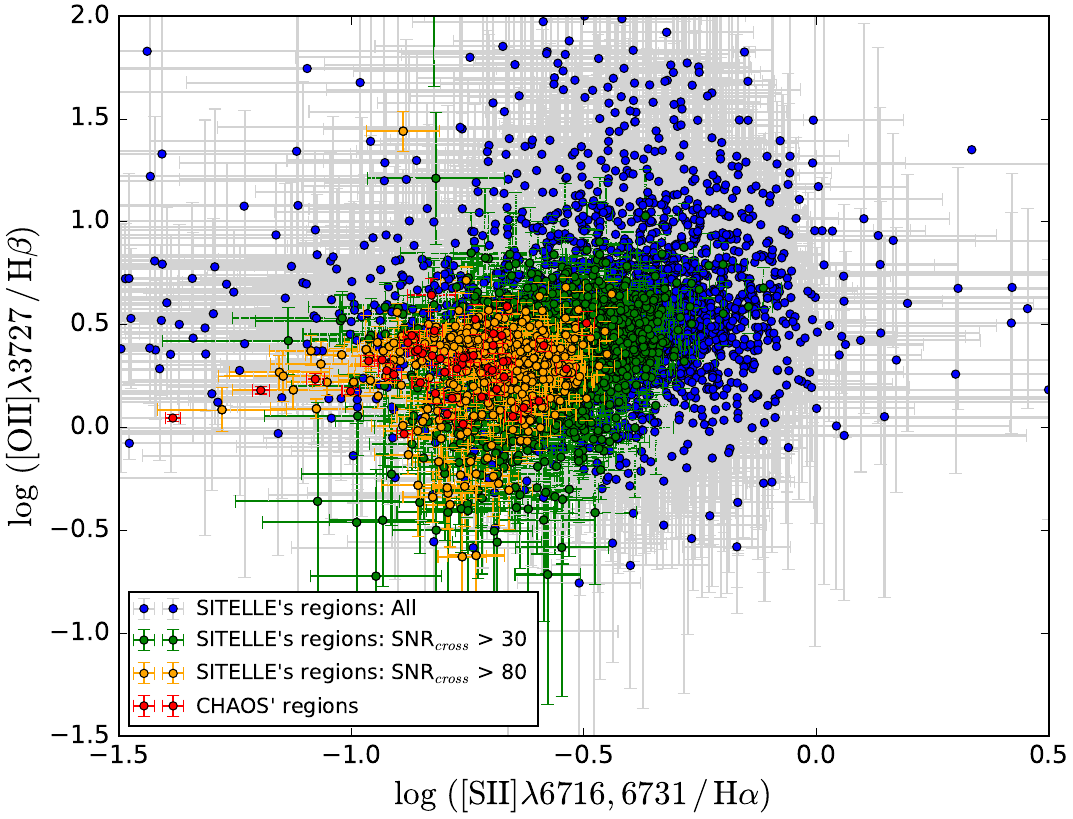} 
\includegraphics[width=3.25in]{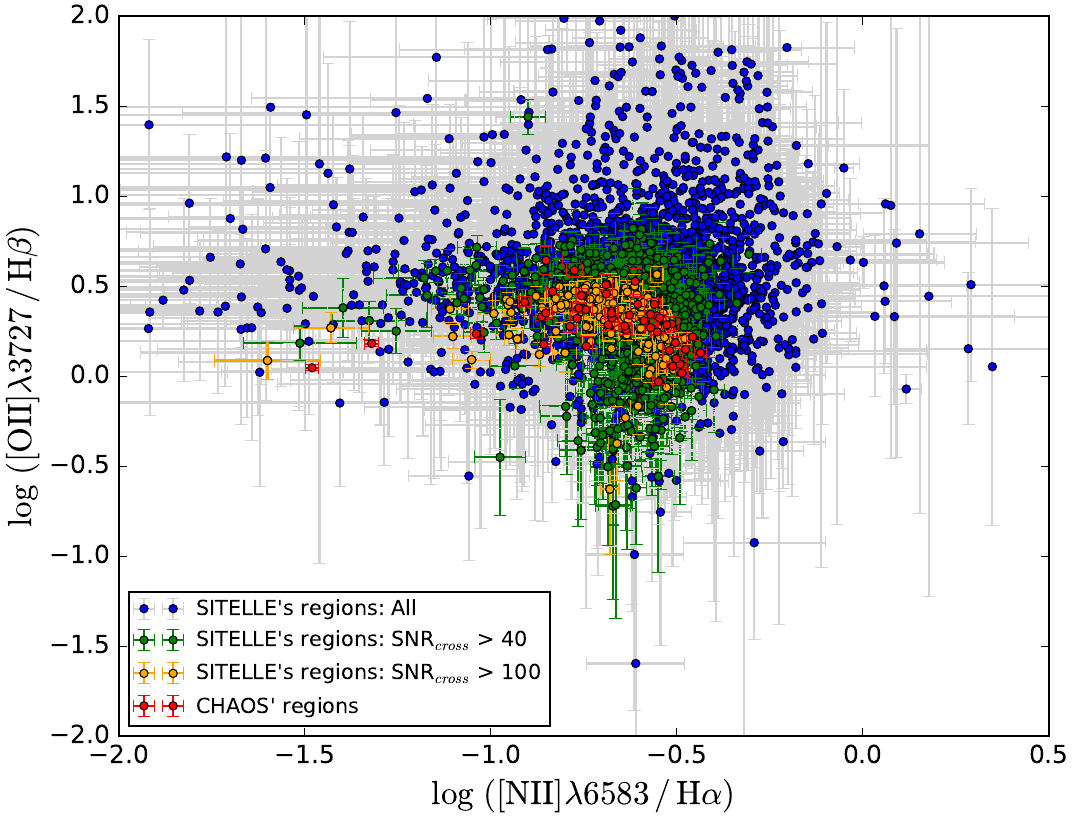} 
\caption{BPT diagrams of [OIII]/H$\beta$ vs [NII]/H$\alpha$, [OIII]/H$\beta$ vs [SII]/H$\alpha$, [OIII]/H$\beta$ vs [OII]/H$\beta$, [OII]/H$\beta$ vs [SII]/H$\alpha$, and [OII]/H$\beta$ vs [NII]/H$\alpha$ based on the integrated spectra of each region identified with SITELLE. Blue dots with gray error bars are for line ratios where all the lines have an SNR\,$>$\,3. Green and orange symbols identify line ratios with different SNR$_{cross}$, as indicated on the diagrams. The regions observed for the CHAOS project by Berg et al. (2015) are plotted in red. Black curves in the top panels are as in Figures\,\,\ref{bpt_flux_NII} and \ref{bpt_flux_SII}.}
\label{bpt_regions}
\end{center}
\end{figure*}

\subsection{Line Ratios for Each Emission Region}
\label{lrer}

For each region identified, as described in Section\,\,\ref{Proc}, global line ratios were evaluated by fitting the integrated flux that is encompassed within an aperture radius corresponding to the half-width 
$\sigma$ of the region pseudo-Voigt fitted intensity profile. All spaxels within this aperture have been corrected for the stellar population (\S\,3.4) and the extinction (\S\,5.1), but not for the DIG background. 
We adopted the half-width $\sigma$ radius instead of the regionÕs outer limit in order to minimise effects related to the DIG background and to the crowding of the regions.
All lines in the integrated spectrum of each region were fitted using ORCS as described in Section\,\,\ref{msrt}. Figure\,\,\ref{bpt_regions} shows the BPT diagrams of [OIII]/H$\beta$ vs [NII]/H$\alpha$, [OIII]/H$\beta$ vs [SII]/H$\alpha$, [OIII]/H$\beta$ vs [OII]/H$\beta$, [OII]/H$\beta$ vs [SII]/H$\alpha$, and [OII]/H$\beta$ vs [NII]/H$\alpha$ separating the regions according to the SNR. 
We defined the parameter SNR$_{cross}$ to consider a signal threshold which takes into account the 
different SNR of the lines involved in a diagram: first all the lines used in a diagram must have an SNR\,$>$\,3 and then, at least one of these lines must have an SNR greater than SNR$_{cross}$. In the diagrams, different colours are used for the emission regions according to different values of SNR$_{cross}$ (as specified in Fig.~\ref{bpt_regions}). A larger SNR$_{cross}$ basically points to a brighter region. The bright HII regions of the CHAOS project (\citealt{b15}) are also shown in our diagrams.

Again, most regions identified fall in the ionizing zone for the HII regions. The comparison with the CHAOS sample, where only the highest H$\alpha$ surface brightness knots have been selected (\citealt{b15}), demonstrates SITELLE's ability to reproduce line ratios previously observed for these regions, i.e. the CHAOS data are well superimposed to the SITELLEÕs regions with the highest SNR$_{cross}$ threshold selected. As these bright regions probably represent the most massive and youngest HII regions, our larger sample includes objects with a wider range of physical properties (masses, ages, stellar content, ISM properties, etc.) allowing us to see an unbiased distribution in the BPT diagrams. 

The Appendix B also presents plots of line ratios as a function of the galactocentric radius. While the plots for [NII]/H$\alpha$ and [SII]/H$\alpha$ show decreasing ratios at large radius, the other ratios have a tendency to increase or stay flat at large radius. These behaviours indicate changes in the
physical gas parameters, like the abundances, which will be discussed in a following paper. The comparison with the CHAOS bright HII regions sample (superposed in red on the SITELLE data in these plots) shows a larger dispersion for the SITELLE data since they contain a broader diversity of emission regions detected. For the bright regions, SITELLE is in good agreement with CHOAS for most line ratios, except for those involving the [SII] lines (Fig.~B3 and B5). In these plots, the SITELLE's ratios [SII]/H$\alpha$ and [SII]/[NII] are higher, but also the SNR$_{cross}$ is somewhat smaller than in Figure~\ref{bpt_regions}. The discrepancy with the CHAOS data may be explained by the fact that the [SII] lines are semi-strong lines, are more dependent on the electron density, and are more sensitive to the DIG (Hafner et al. 1999; Elwert \& Dettmar 2005) and its background subtraction. Berg et al. (2015) mention a sky subtraction for the CHAOS data using a 2D background fit within the slit used to observe an HII region; this background may include a significant contribution from the DIG. In our case, we measured the line ratios using a limited aperture size (i.e. the half-width of the pseudo-Voigt fitted on the region intensity profile) to minimise the DIG without doing any subtraction. But it could also be due to a selection bias, as bright regions (with high ionisation parameter) are more likely to present low [SII]/H$\alpha$ ratios. Figure~B19 presents the ratio [SII]$\lambda$6716/[SII]$\lambda$6731 as a function of the galactocentric radius where we are considering a higher SNR limit (from 5 to 10). For the brightest regions with SNR\,$>$\,10, the agreement with the CHAOS sample is then excellent suggesting indeed that the electron density in bright HII regions is often lower than the limit of sensitivity of the ratio (i.e.[SII]/[SII]\,$\simeq$\,1.4 when $n_e$\,$>$\,$10^2$\,electrons/cm$^3$). This does not exclude as well the fact that a larger SNR may be required when measuring semi-strong line ratios due to the propagation of the uncertainties of both lines.

\section{Summary}

In this first paper of a large program focusing on star formation and chemical evolution in disc galaxies, we have shown the high potential of SITELLE for the study of nearby extragalactic HII regions from observations on the well-known spiral NGC\,628. Our results are as follows:

\,\begin{itemize}
\vspace{-0.25cm}
\item[{1)}]{From our datacubes covering the whole disc of NGC\,628, we identified 4285 HII region candidates. Ten strong and semi-strong emission lines have been sampled in about 4 billion spaxels using automated routines specifically created for SITELLE. We developed a new identification procedure for the regions, i.e. to find out all the emission peaks and define their zone of influence. Using both the spatial and spectral information, this new procedure optimizes the number of parameters extracted for each region (RA/DEC position, dust extinction, velocity, H$\alpha$ profile, DIG background, luminosity, size, morphological type, integrated line fluxes, etc.). } 

\item[{2)}]{By comparing the intensity profile of the regions, four morphological categories, from spherical to diffuse, were defined. These morphologies point out to different types of stellar aggregates (dense or sparse), different gas distributions (homogeneous to filamentary), and multiple sources of ionisation. }

\item[{3)}]{A well-sampled luminosity function of the HII region candidates was obtained and we studied its variation as a function of the location of the regions along the galactic radius and the DIG background flux. Using the complete sample, a constant slope of $\alpha_{cor}\,=\,-1.12\pm0.03$ (corrected for the dust extinction, using the Balmer lines, and for the galaxy disc stellar populations on the line of sight to each region) was found; no evidence of a break in the distribution was noticed. The slope remains the same while selecting subsamples of regions located at different galactic radii (which can be seen as different metallicity ranges). A significant change of the slope is seen when considering only regions with a low DIG background. We noticed that high DIG regions are close or belong to the spiral arms of the galaxy (in agreement with the findings of \citealt{k16}). A correlation between the DIG background flux and the total H$\alpha$ luminosity of the bright regions was found, suggesting that ionizing photons primarily related to massive stars and escaping directly from these regions are the main source for the DIG. We also showed that the global DIG background (in fainter regions outside the spiral arms) increases slowly toward the galaxy centre. This behavior rather points to an additional source for the DIG which is related to evolved stars, as these are more numerous closer to the galaxy centre.}

\item[{4)}]{The high spatial resolution of SITELLE allowed us to compare the integrated H$\alpha$ luminosity of the regions with their intensity profile width (i.e. $\sigma$ of a pseudo-Voigt fit). Diffused/extended high luminosity regions are not seen while low luminosity regions present a wide range of compactness. This is in agreement with the proposition that denser clouds are the progenitors of the most massive HII regions (\citealt{mmmm15}) and the fact that massive regions are density bounded rather than ionisation bounded. In less massive HII regions, the number of ionizing photons is subject to IMF fluctuations and the Str\"{o}mgren radius of regions in a lower density medium is ionisation bounded. Therefore low luminosity HII regions present a wider range in size.}

\item[{5)}]{The size of the regions was evaluated using the H$\alpha$ profile and a simple Str\"{o}mgren sphere model. The size distribution function (i.e. number of regions versus their radius) shows a well-defined slope $\gamma$\,=\,1.81$\pm$0.02 for radii between 70 to 250\,pc, as found in other studies.}

\item[{6)}]{We looked at the distribution of the line ratios for the individual spaxels and for the integrated regions in different BPT diagrams. Bright spaxels show a distinct turnover in the [OIII]/H$\beta$ vs [NII]/H$\alpha$ diagram. A larger dispersion is seen for the fainter H$\alpha$ spaxels which can be due to their age, IMF under-sampling, DIG contamination, different sources of ionisation, SNR, etc. High-definition maps of the line ratios reveal the complex variations of the ionisation conditions within the emission regions themselves. A polynomial fit of each line ratios as a function of the galactocentric radius is now available for comparative studies.}

\item[{7)}]{And finally, detailed information on the emission line regions of NGC\,628 have been extracted from the SITELLE data. We have consolidated the information for all the 4285 regions identified in a catalogue.}
\end{itemize}

 Table\,\,3 shows an example of the catalogue content for the HII region candidates in NGC\,628. The objective of this new catalogue is to provide a complete database to study the vast parameter space covered by star-forming regions. Further work on NGC 628 will address HII region characteristics and nebular physics (abundances, mass, age, ionisation parameter, density) in a more detailed manner. In a following paper, we will use photoionisation models to investigate the ionisation conditions in the HII regions of NGC\,628.

\section*{acknowledgments}

This research is based on observations obtained at the Canada-France-Hawaii Telescope (CFHT) which is operated from the summit of Maunakea by the National Research Council of Canada, the Institut National des Sciences de l'Univers of the Centre National de la Recherche Scientifique of France, and the University of Hawaii. The observations at the Canada-France-Hawaii Telescope were performed with care and respect from the summit of Maunakea which is a significant cultural and historic site. The observations were obtained with SITELLE, a joint project between Universit\'e Laval, ABB-Bomem, Universit\'e de Montr\'eal and the CFHT with funding support from the Canada Foundation for Innovation (CFI), the National Sciences and Engineering Research Council of Canada (NSERC), Fonds de Recherche du Qu\'ebec - Nature et Technologies (FRQNT), and CFHT.

LRN, CR, and LD are grateful to the Fonds de recherche du Qu\'ebec - Nature et Technologies (FRQNT) for individual and team financial support. LD thanks the Canada Research Chair program. CR and LD are grateful to the Natural Sciences and Engineering Research Council of Canada (NSERC). This research has made use of NASA's Astrophysics Data System and of the VizieR catalogue access tool, CDS, Strasbourg, France.

\begin{landscape}
\begin{table}
\centering
\caption{Catalogue Parameters}
\label{cp}
\begin{tabular}{lc|c|c|c|c|c|c|c|c|c|c|c|c|c|c|}
\hline
\hline
ID & RA & DEC & $R_{G}$ & Log L$_{H\alpha}$ & log DIG$_{SB}$ & Cat & $I_{0~profile}$ & A$_{profile}$ & $\sigma$$_{profile}$ & $\alpha$$_{profile}$ & R$^2$$_{profile}$ & Size & E(B$-$V) & E(B$-$V)$_{err}$ \\
\hline
\# & -- & -- & kpc & -- & erg\,s$^{-1}$\,cm$^{-2}$ & \# & erg\,s$^{-1}$ & erg\,s$^{-1}$ & pc & \% & -- & pc & mag & $\pm$ \\
 &  &  & &  & spaxel$^{-1}$ &  & cm$^{-2}$ & cm$^{-2}$ & & & & & &   \\
\hline
2430 & 24.1818 & 15.7876 & 1.39 & 37.20 & -17.26 & 1 & -15.89 & -14.22 & 39.34 & 0.72 & 0.96 & 46.0 & 0.408 & 0.172 \\
2431 & 24.1375 & 15.7867 & 5.48 & 37.21 & -16.85 & 3 & -15.87 & -13.92 & 64.66 & 0.50 & 0.95 & 64.3 & 0.227 & 0.048 \\
2432 & 24.1368 & 15.7868 & 4.60 & 36.49 & -17.65 & 2 & -16.92 & -14.84 & 84.41 & 0.44 & 0.51 & 84.2 & 1.265 & 0.090 \\
2433 & 24.2040 & 15.7882 & 3.89 & 37.10 & -16.91 & 3 & -16.33 & -14.34 & 77.23 & 0.60 & 0.62 & 97.0 & 0.347 & 0.147 \\
2434 & 24.1891 & 15.7879 & 0.83 & 37.87 & -17.06 & 3 & -15.38 & -13.61 & 43.87 & 0.50 & 0.96 & 43.7 & 0.356 & 0.053 \\
2435 & 24.1773 & 15.7876 & 5.64 & 37.70 & -16.89 & 2 & -15.49 & -13.54 & 65.68 & 0.50 & 0.95 & 83.0 & 0.286 & 0.029 \\
2436 & 24.2416 & 15.7892 &10.28& 37.22 & -17.36 & 1 & -15.95 & -14.21 & 40.61 & 0.50 & 0.97 & 48.0 & 0.308 & 0.057 \\
2437 & 24.2121 & 15.7885 & 5.82 & 38.63 & -16.72 & 1 & -14.46 & -12.79 & 34.29 & 0.50 & 0.99 & 35.0 & 0.349 & 0.026 \\
2438 & 24.2021 & 15.7883 & 4.31 & 37.35 & -17.81 & 3 & -16.02 & -14.25 & 59.17 & 1.00 & 0.87 & 75.0 & 0.479 & 0.275 \\
2439 & 24.1997 & 15.7881 & 4.06 & 37.06 & -17.34 & 3 & -16.40 & -14.21 &114.01& 0.48 & 0.30 &140.0& 0.000 & 0.091 \\
2440 & 24.1990 & 15.7881 & 2.52 & 37.04 & -17.45 & 3 & -16.09 & -14.39 & 44.32 & 0.77 & 0.84 & 54.00 & 0.261 & 0.188 \\
2441 & 24.1959 & 15.7881 & 1.68 & 37.28 & -17.28 & 3 & -16.23 & -14.08 &102.78& 0.50 & 0.44 &127.00& 0.067 & 0.168\\
2442 & 24.1840 & 15.7878 & 0.67 & --         & --          & 4 & -16.61 & -14.70 & 60.00 & 0.50 & 0.00 & 76.00 & 0.607 & 0.994\\
2443 & 24.1736 & 15.7876 & 0.79 & 37.45 & -16.89 & 3 & -15.64 & -13.82 & 48.25 & 0.50 & 0.92 & 60.00 & 0.448 & 0.080 \\
2444 & 24.1702 & 15.7876 & 0.91 & 37.06 & -17.26 & 4 & -16.51 & -14.48 & 79.00 & 0.50 & 0.48 & 99.00 & 0.000 & 0.212 \\
2445 & 24.1630 & 15.7874 & 1.68 & 36.67 & -17.51 & 4 & -16.47 & -14.48 & 72.77 & 0.50 & 0.86 & 72.45 & 1.154 & 0.708 \\
... & ... & ... & ... & ... & ... & ... & ... & ... & ... & ... & ... & ... & ... & ... \\
\hline
\hline
\end{tabular}
\end{table}

\begin{table}
\centering
\label{cp}
\begin{tabular}{|c|c|c|c|c|c|c|}
\hline
\hline
log\,([NII]/H$\alpha$) & log\,([NII]/H$\alpha$)$_{err}$ & log\,([NII]/H$\alpha$)\,SNR$_{cross}$ & log\,([SII]/H$\alpha$) & log\,([SII]/H$\alpha$)$_{err}$ & log\,([SII]/H$\alpha$)\,SNR$_{cross}$ & ...$^*$ \\
 \hline
  -- & $\pm$ & --  & -- & $\pm$ & -- & ...\\
 \hline
 -0.450 & 0.069 & 20.919 &   -0.657 &  0.114 &20.919& ... \\
 -0.629 & 0.023 & 56.179 &   -0.615 &  0.042 & 56.179 & ... \\
 -0.898 & 0.162 & 8.021 &  -0.413&  0.329 & 8.021& ... \\
 -0.487 & 0.048 & 29.180 &  -0.452 &  0.086 & 29.180& ... \\
 -0.597 & 0.024 & 54.618 &  -0.768 &  0.042 & 54.618 & ... \\
 -0.618 & 0.014 & 88.427 &  -0.627 &  0.026 & 88.427 & ... \\
 -0.769 & 0.023 & 52.319 &  -0.504 &  0.047 & 52.319& ... \\
 -0.586 & 0.017 & 77.199 &  -0.696 &  0.030 & 77.199 & ... \\
 -0.592 & 0.069 & 19.538 &  -0.443 &  0.130 & 19.538& ... \\
 -0.493 & 0.056 & 25.023 &   -0.396 &  0.104 & 25.023& ... \\
  -0.698 & 0.067 & 19.171 &  -0.444 &  0.133 & 19.171& ...\\
  -0.610 & 0.061 & 21.629 &  -0.505 &  0.117 & 21.629& ... \\
  -0.330 & 0.302 &    5.915 &   0.034 & 0.631  & 6.344 & ...\\
  -0.461 & 0.035 &  40.596 & -0.801 &  0.056 & 40.596 & ... \\
  -0.109 & 0.182 &  10.803 & -0.213 &  0.274 & 10.803& ...\\
  -0.243 & 0.178 &    9.849 & -0.201 &  0.400  &9.849 & ...\\
... & ... & ... & ... & ... & ...  \\
 \hline
\hline
\multicolumn{7}{l}{*All line ratios listed in Section\,\,\ref {lrer} are included in the catalogue.}  \\
\multicolumn{7}{l}{Note: More parameters can be provided upon request.}  \\
\end{tabular}
\end{table}
\end{landscape}

\appendix
\section{Flux and Velocity Maps}
\label{flma}
\begin{figure*}
\begin{center}
\includegraphics[width=7in]{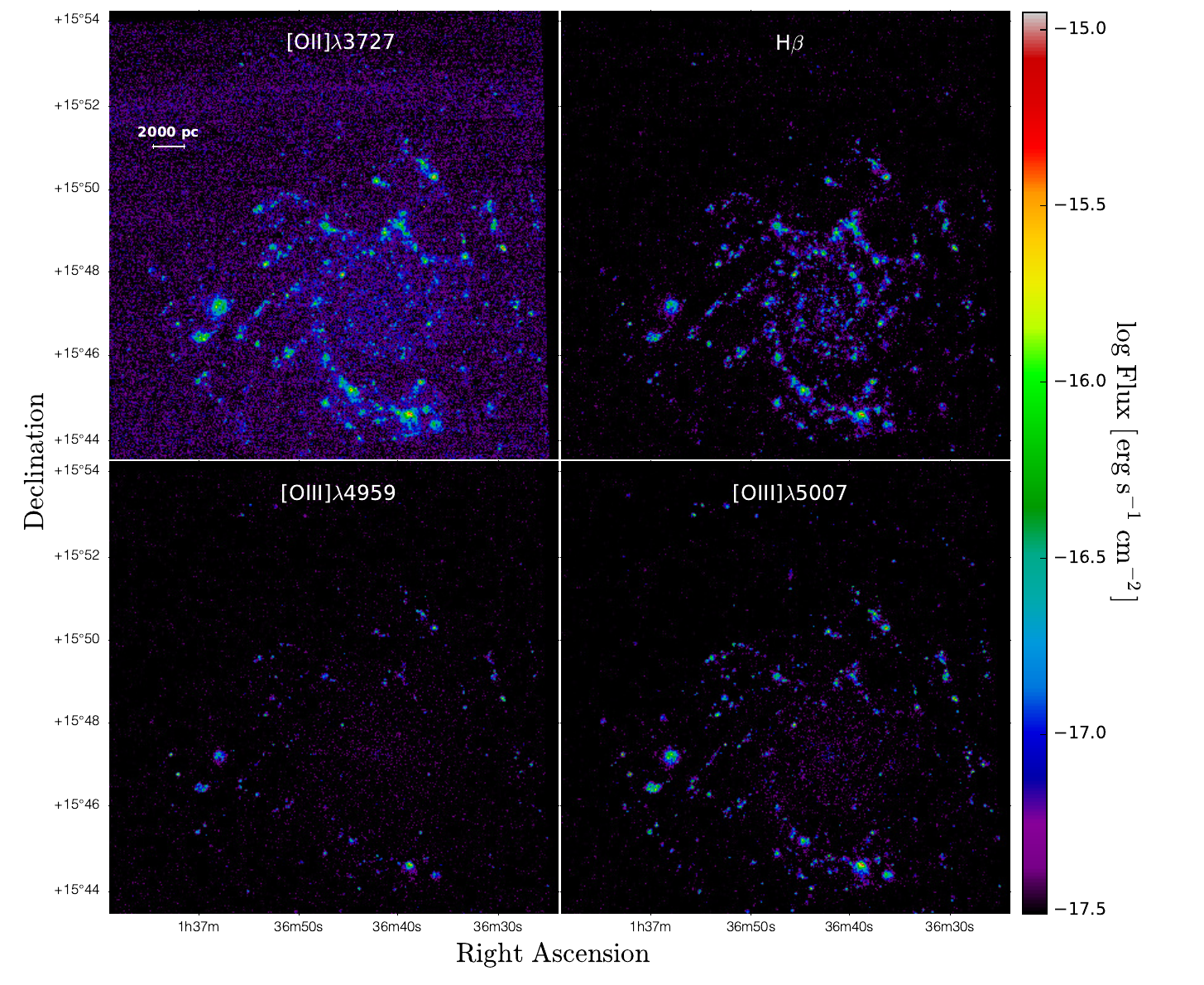} 
\caption{Flux map of the emission lines measured in the SN1 and SN2 datacubes. The correction for the internal extinction is not applied.}
\label{SN2SN1_flux_reg}
\end{center}
\end{figure*}

\begin{figure*}
\begin{center}
\includegraphics[width=7in]{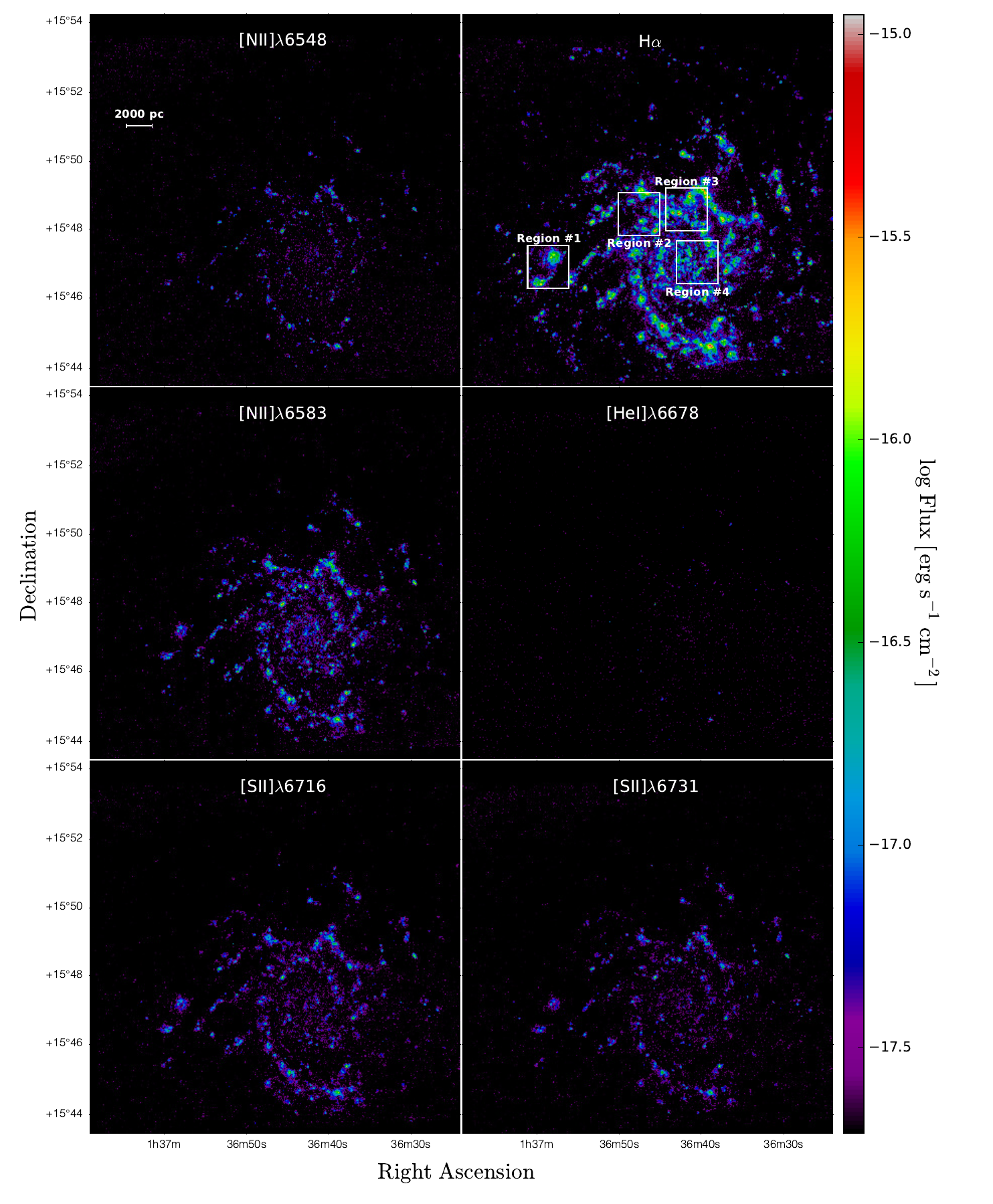} 
\caption{Flux map of the emission lines measured in the SN3 datacube. The correction for the internal extinction is not applied.}
\label{SN3_flux_reg}
\end{center}
\end{figure*}

The flux map for all the emission lines seen in the three filters are shown in Figures \ref{SN2SN1_flux_reg} and \ref{SN3_flux_reg}. Emission lines from the same filter have been fitted simultaneously using the routine from ORCS $\S$\,\,\ref{lfproc}) applied to all the spaxels individually. The subtraction of the galaxy stellar population spectrum on the line of sight for each spaxels was done ($\S$\,\,\ref{substel}). The correction for the internal extinction has not been applied for these figures. For these figures, all spaxels have been kept regardless of the SNR measured for the lines.

\begin{figure*}
\begin{center}
\includegraphics[width=6in]{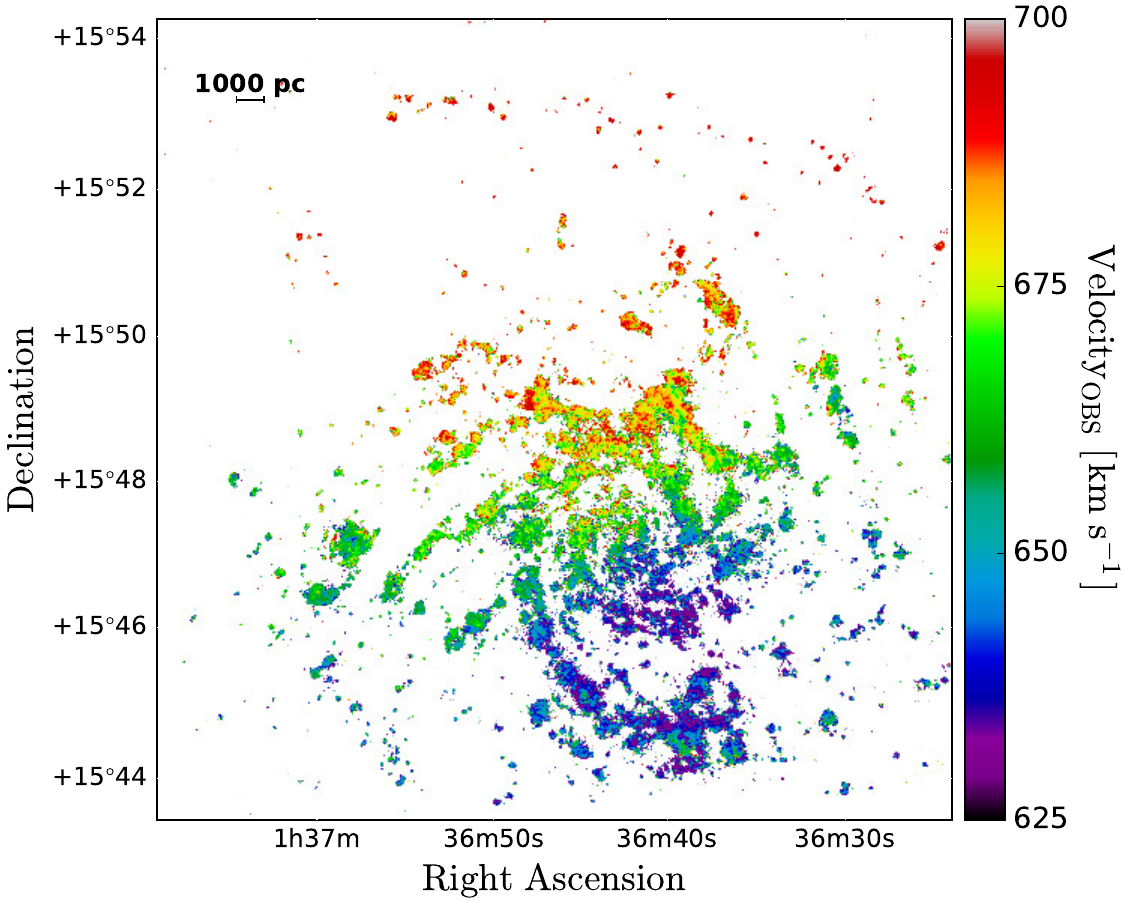} 
\caption{The final velocity map of NGC\,628 based on the SN3 filter (which includes the strong H$\alpha$ emission line). Spaxels with SNR$_{H\alpha}$\,$>$\,3 have been selected.}
\label{v}
\end{center}
\end{figure*}

The velocity map in Figure\,\,\ref{v} was produced after a refinement of the wavelength calibration of the SN3 datacube using a sky line ($\S$\,\,\ref{scr}). All the emission lines from this high-spectral resolution datacube (R $\simeq$ 1800; Tab.~\ref{op}) have been used simultaneously to extract the velocity in each spaxel individually using the line fitting routine from ORCS ($\S$\,\,\ref{lfproc}). The consideration of all the lines simultaneously, as well as the fact that the lines have a sinc profile, improve significantly the precision of the velocity measurements (Martin et al. 2016, 2018). For example, using all the lines in the SN3 filter for spaxels with SNR$_{{\rm H}\alpha}$ between 3 and 20 gives a velocity precision between 30 and 5 km s$^{-1}$. All spaxels with SNR$_{{\rm H}\alpha} > 3$ have been kept in Figure\,\,\ref{v}.

\section{Line Ratios}
\label{lrr}
\begin{figure*}
\begin{center}
\hspace{0.35cm}
\vspace{-0.20cm}
\includegraphics[height=2.54in]{NGC628_Av-eps-converted-to.pdf} 
\includegraphics[height=2.54in]{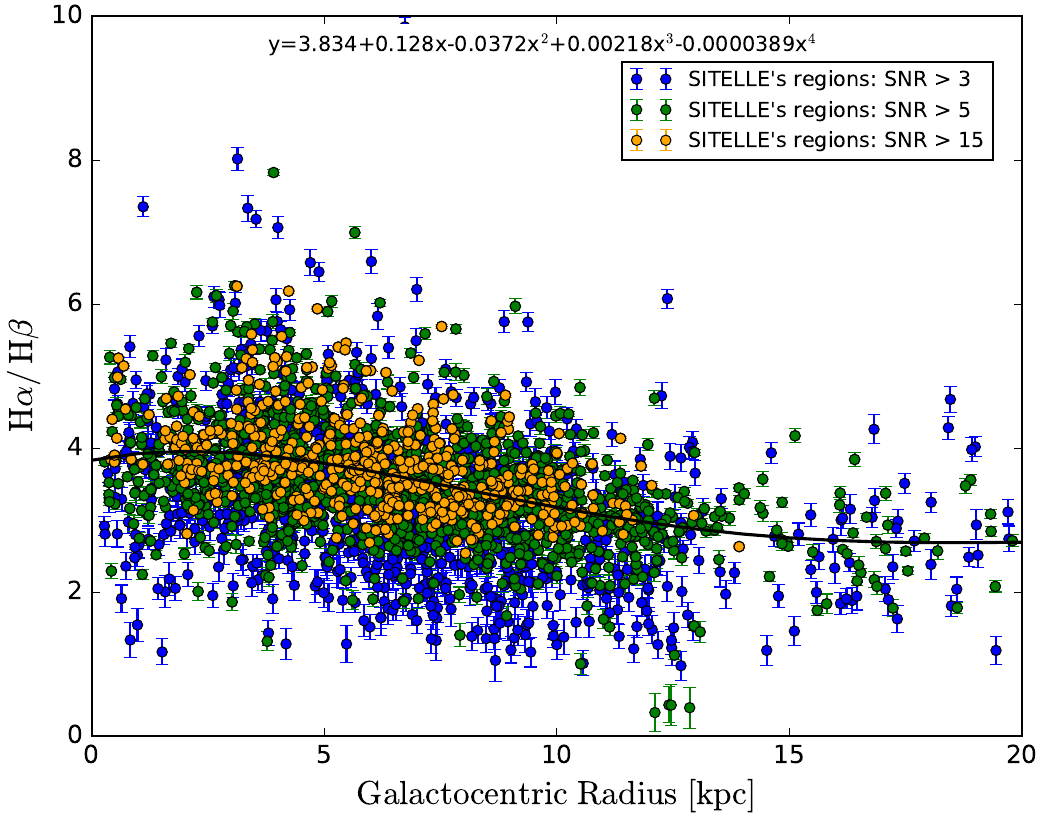} 
\vspace{-0.20cm}
\caption{The extinction A$_v$ and the H$\alpha$/H$\beta$ line ratio. On the right, line ratio for the integrated spectrum of each region as a function of their galactocentric radius. Three thresholds of the SNR have been considered. The black curve is a fit to the highest SNR subsample of regions.}
\label{hahb}
\end{center}
\end{figure*}

\begin{figure*}
\begin{center}
\vspace{-0.85cm}
\includegraphics[width=7.5in]{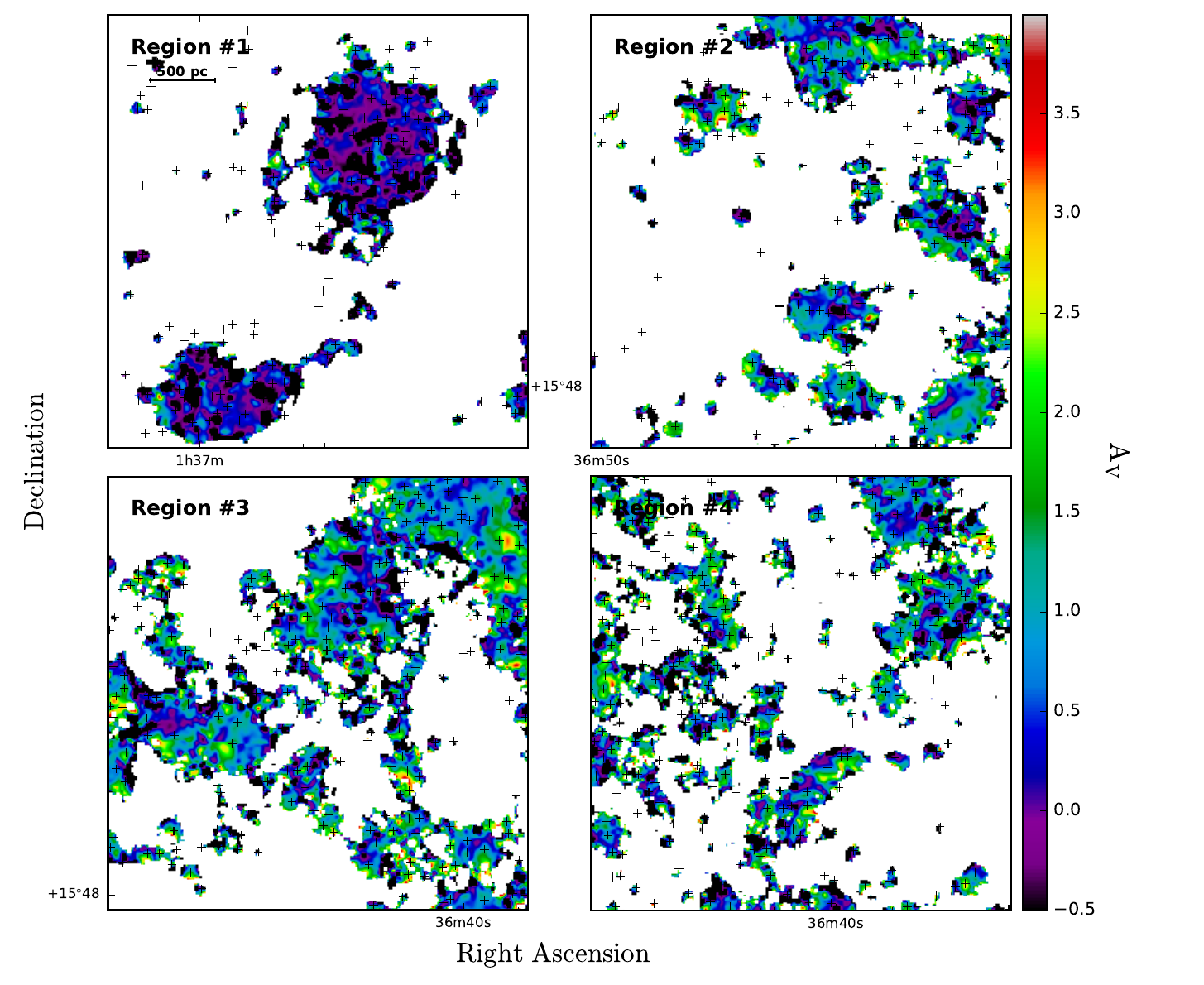} 
\vspace{-0.85cm}
\caption{Map of the extinction A$_v$ for areas defined in Figure\,\,\ref{SN3_flux_reg}. Black crosses indicate the location of the emission peaks (identified in $\S$\,\,\ref{Proc}).}
\label{av_reg}
\end{center}
\end{figure*}

Appendix \ref{lrr} contains maps of the different line ratios, for individual spaxels covering the whole galaxy and for four enlarged reference sections in the disc (their exact location may be seen in Fig.\,\,\ref{SN3_flux_reg}). A detection threshold of 3$\sigma$ on the individual lines was applied to produce the maps. The stellar population contribution (\S\,3.4) was subtracted and theextinction correction was done (\S\,5.1) before the lines have been measured. The first set of figures (Fig.\,\,\ref{hahb} and \ref{av_reg}) presents the H$\alpha$/H$\beta$ emission line ratio along with the extinction map for the regions.

This appendix also presents plots of line ratios as a function of the galactocentric radius for the integrated line flux ratios in each emission region. The regions are encompassed within an aperture radius corresponding to the half-width $\sigma$ of the region pseudo-Voigt fitted intensity profile. All spaxels within this aperture have been corrected for the stellar population (\S\,3.4) and the extinction (\S\,5.1), but not for the DIG background. These plots are colour coded according to different SNR thresholds. The best polynomial fit obtained in the case of the highest SNR$_{cross}$ subsample is drawn over each plot (and its corresponding equation is written at the top of each plot). The HII regions from the CHAOS project (\citealt{b15}) are superimposed to the SITELLE's regions in these plots.

\clearpage

\clearpage

\begin{figure*}
\begin{center}
\hspace{0.35cm}
\vspace{-0.15cm}
\includegraphics[height=2.543in]{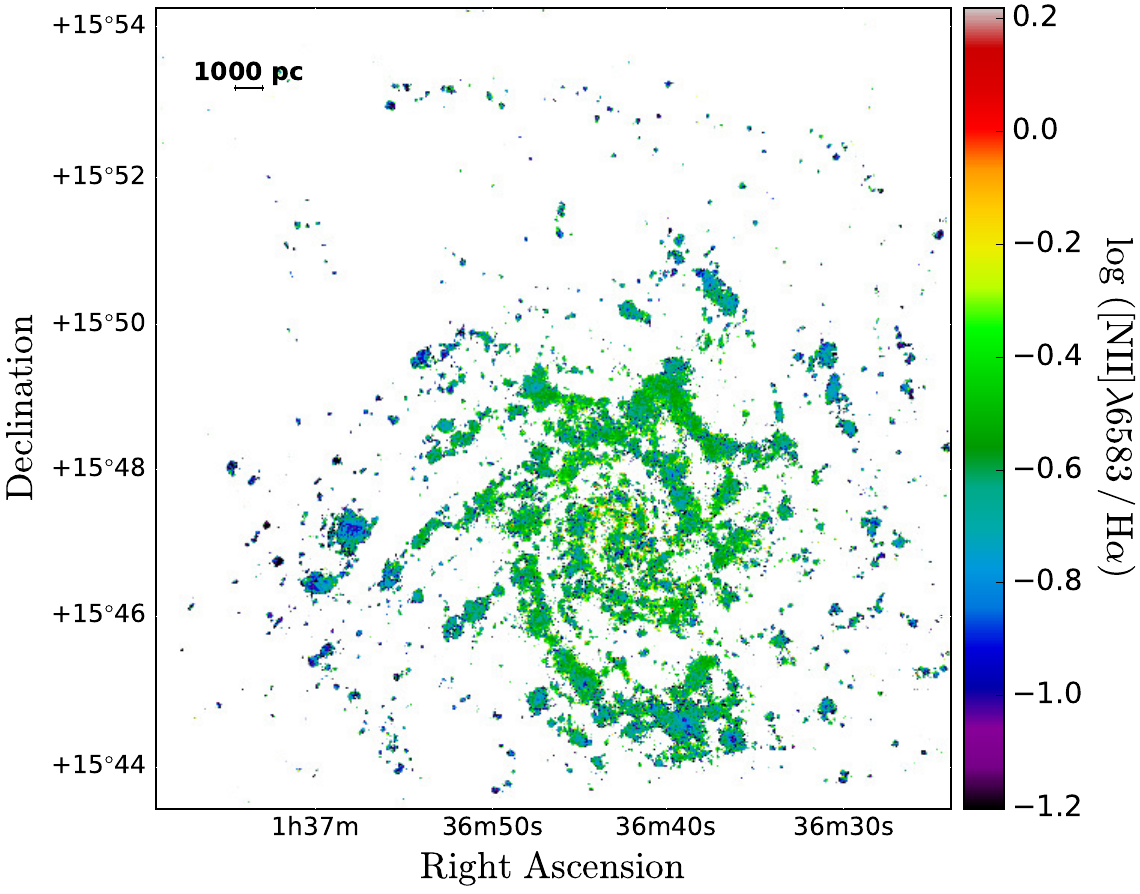} 
\includegraphics[height=2.543in]{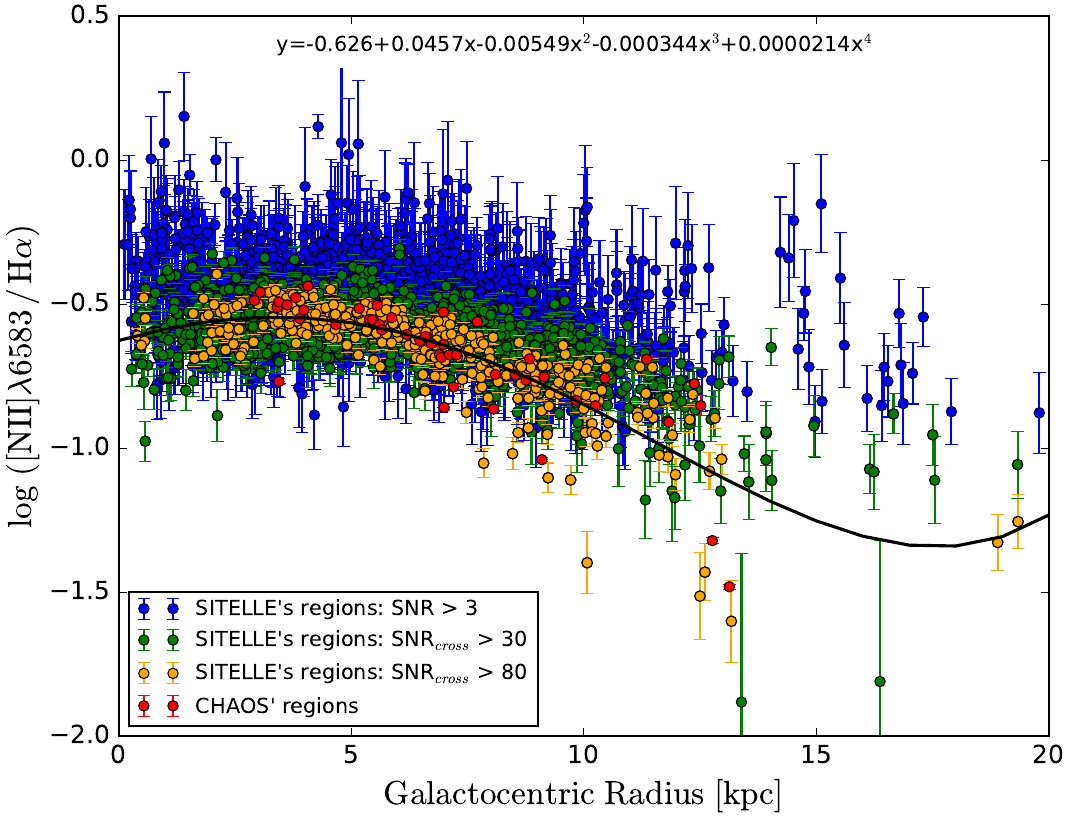} 
\vspace{-0.20cm}
\caption{[NII]$\lambda$6583/H$\alpha$ line ratio. On the right, line ratio for the integrated spectrum of each region as a function of their galactocentric radius. Three thresholds of the SNR have been considered and the CHAOS regions have been superimposed to the SITELLE's data. The black curve is a fit to the highest SNR$_{cross}$ subsample of regions.}
\label{n2ha}
\end{center}
\end{figure*}

\begin{figure*}
\begin{center}
\vspace{-0.85cm}
\includegraphics[width=7.5in]{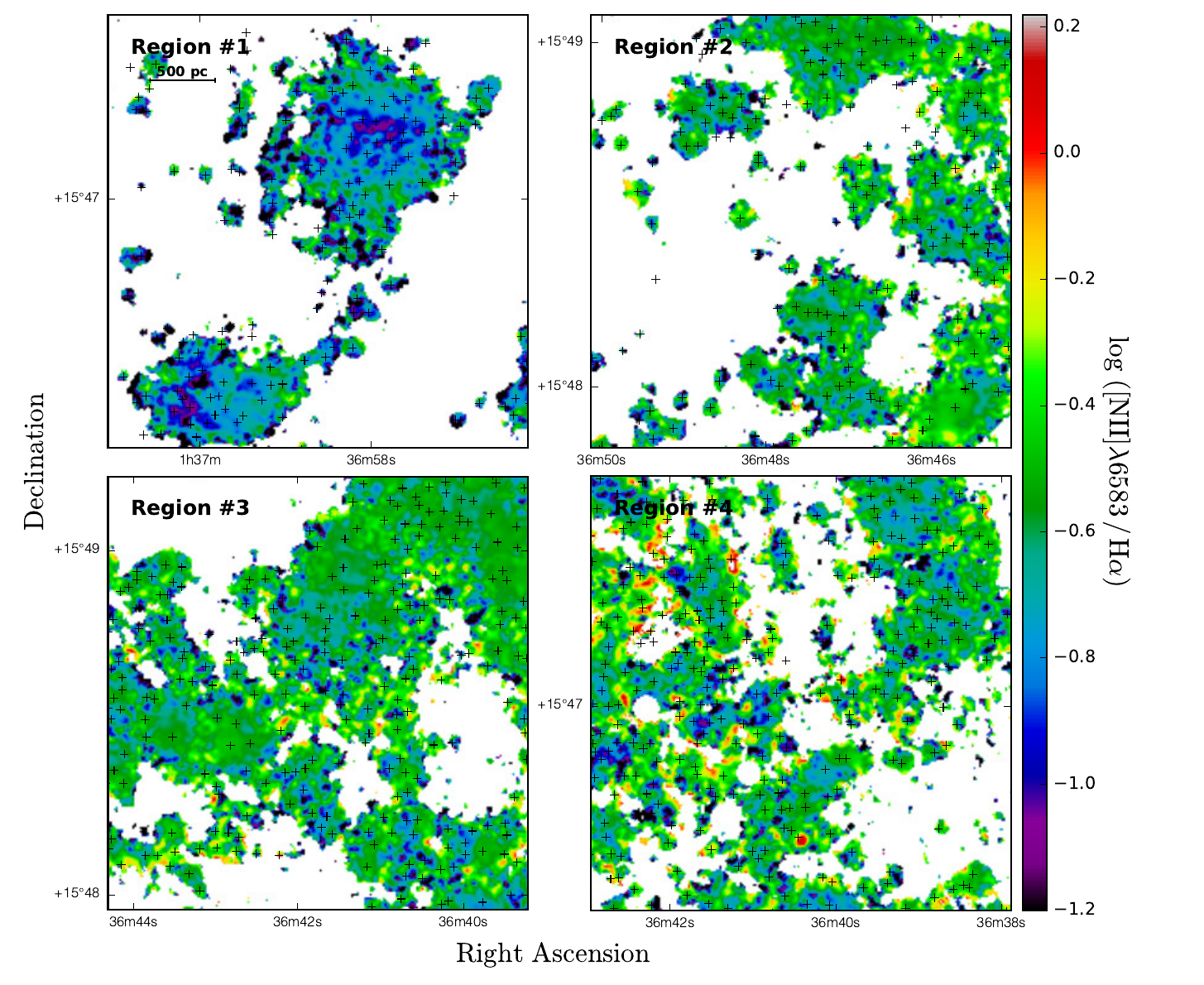} 
\vspace{-0.85cm}
\caption{Map of the [NII]$\lambda$6583/H$\alpha$ line ratio for areas defined in Figure\,\,\ref{SN3_flux_reg}. Black crosses indicate the location of the emission peaks (identified in $\S$\,\,\ref{Proc}).}
\label{n2h_reg}
\end{center}
\end{figure*}

\clearpage

\begin{figure*}
\begin{center}
\hspace{0.35cm}
\vspace{-0.15cm}
\includegraphics[height=2.543in]{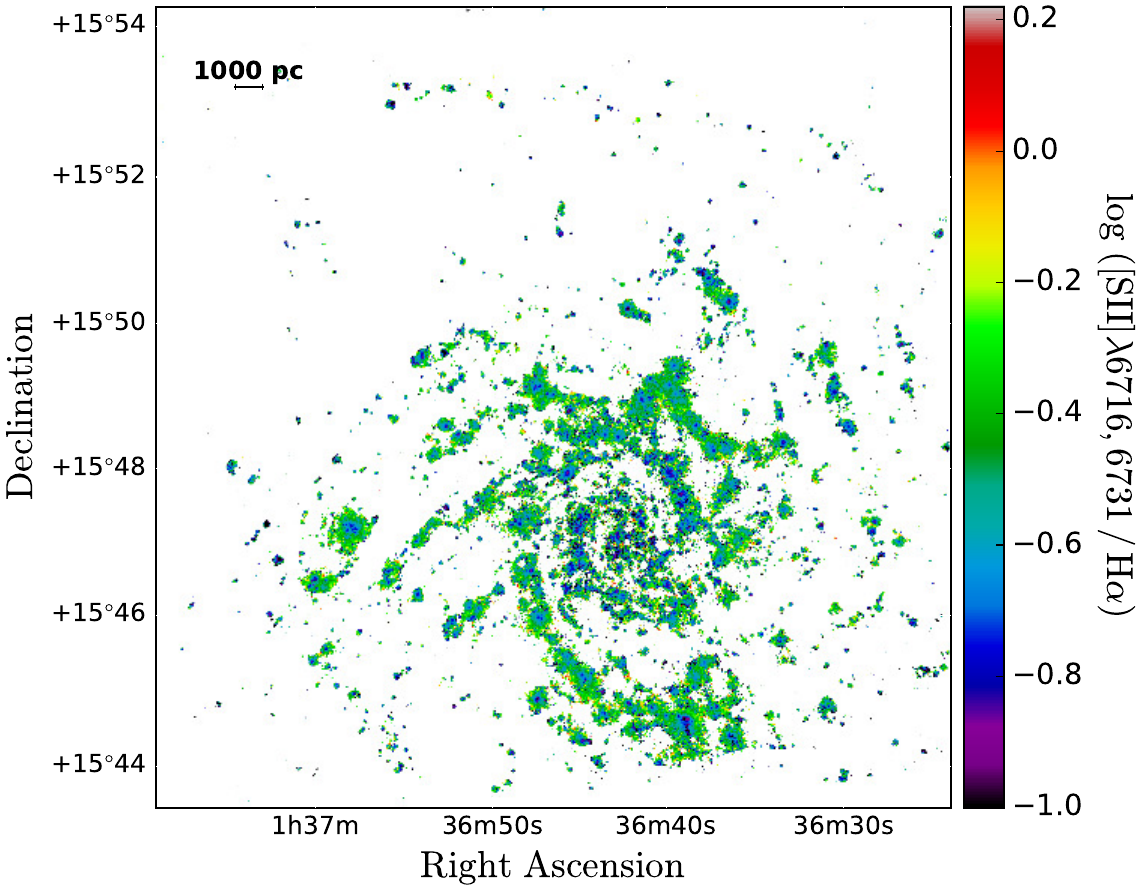} 
\includegraphics[height=2.543in]{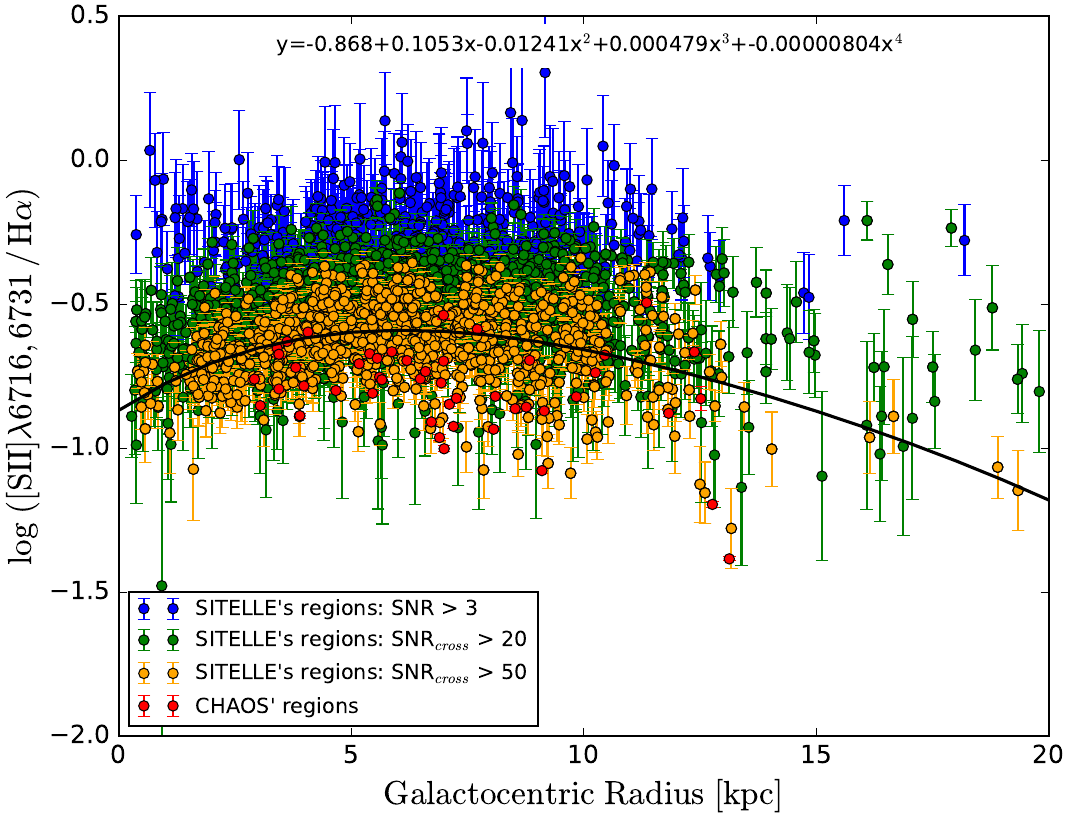} 
\vspace{-0.20cm}
\caption{[SII]$\lambda$6716,6731/H$\alpha$ line ratio. On the right, line ratio for the integrated spectrum of each region as a function of their galactocentric radius. Three thresholds of the SNR have been considered and the CHAOS regions have been superimposed to the SITELLE's data. The black curve is a fit to the highest SNR$_{cross}$ subsample of regions.}
\label{s2ha}
\end{center}
\end{figure*}

\begin{figure*}
\begin{center}
\vspace{-0.85cm}
\includegraphics[width=7.5in]{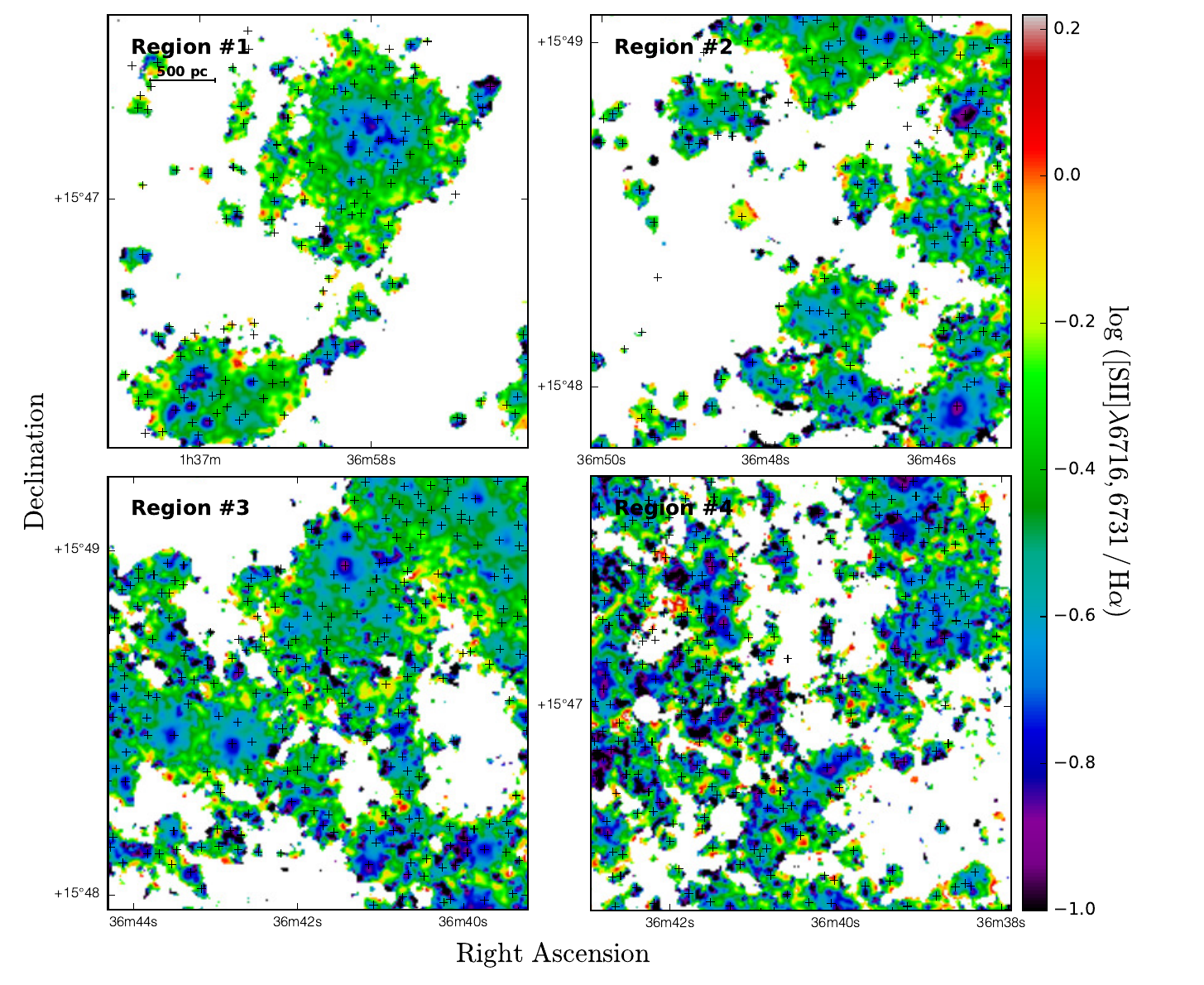} 
\vspace{-0.85cm}
\caption{Map of the [SII]$\lambda$6716,6731/H$\alpha$ line ratio for areas defined in Figure\,\,\ref{SN3_flux_reg}. Black crosses indicate the location of the emission peaks (identified in $\S$\,\,\ref{Proc}).}
\label{s2h_reg}
\end{center}
\end{figure*}

\clearpage

\begin{figure*}
\begin{center}
\hspace{0.35cm}
\vspace{-0.15cm}
\includegraphics[height=2.543in]{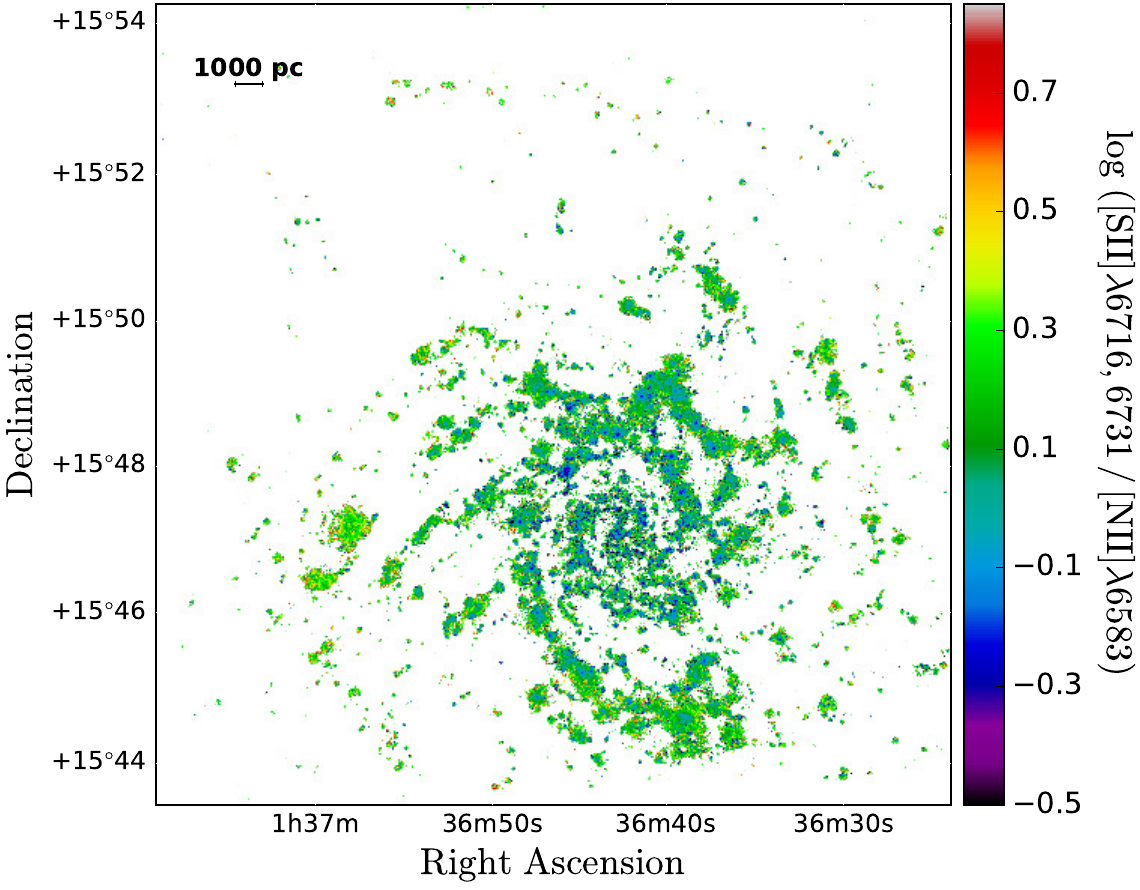} 
\includegraphics[height=2.543in]{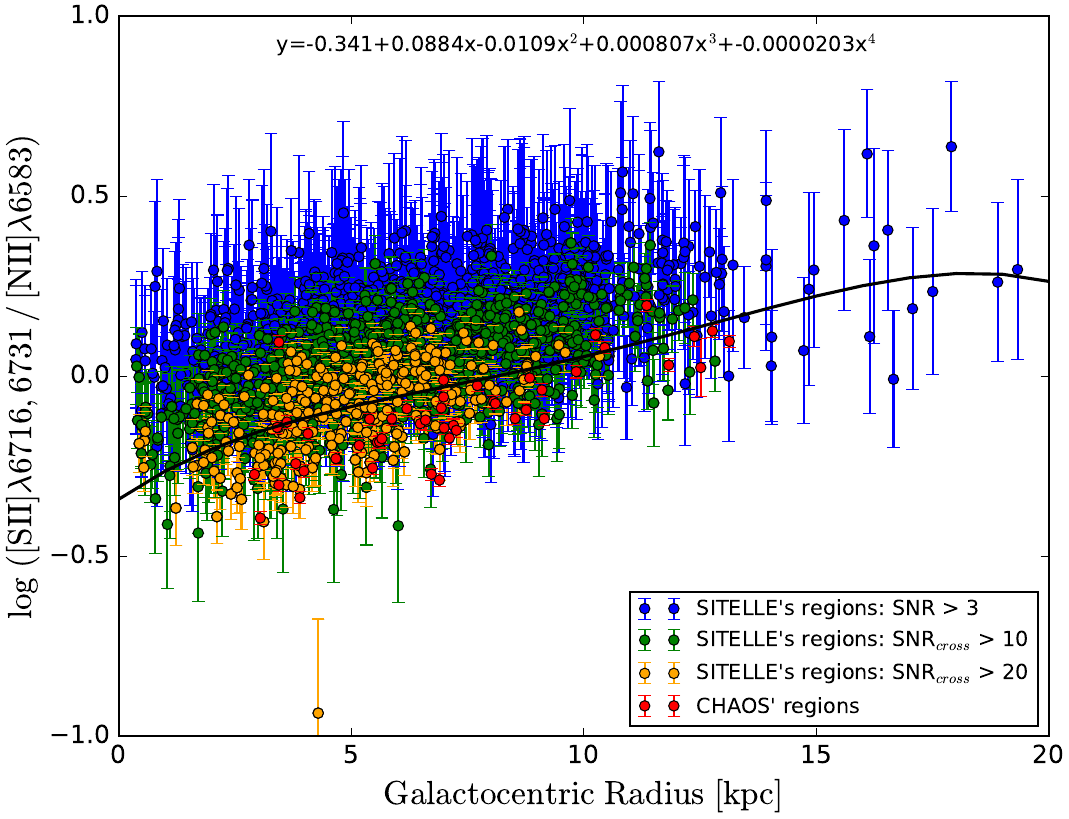} 
\vspace{-0.20cm}
\caption{[SII]$\lambda$6716,6731/[NII]$\lambda$6583 line ratio. On the right, line ratio for the integrated spectrum of each region as a function of their galactocentric radius. Three thresholds of the SNR have been considered and the CHAOS regions have been superimposed to the SITELLE's data. The black curve is a fit to the highest SNR$_{cross}$ subsample of regions.}
\label{s2n2}
\end{center}
\end{figure*}

\begin{figure*}
\begin{center}
\vspace{-0.85cm}
\includegraphics[width=7.5in]{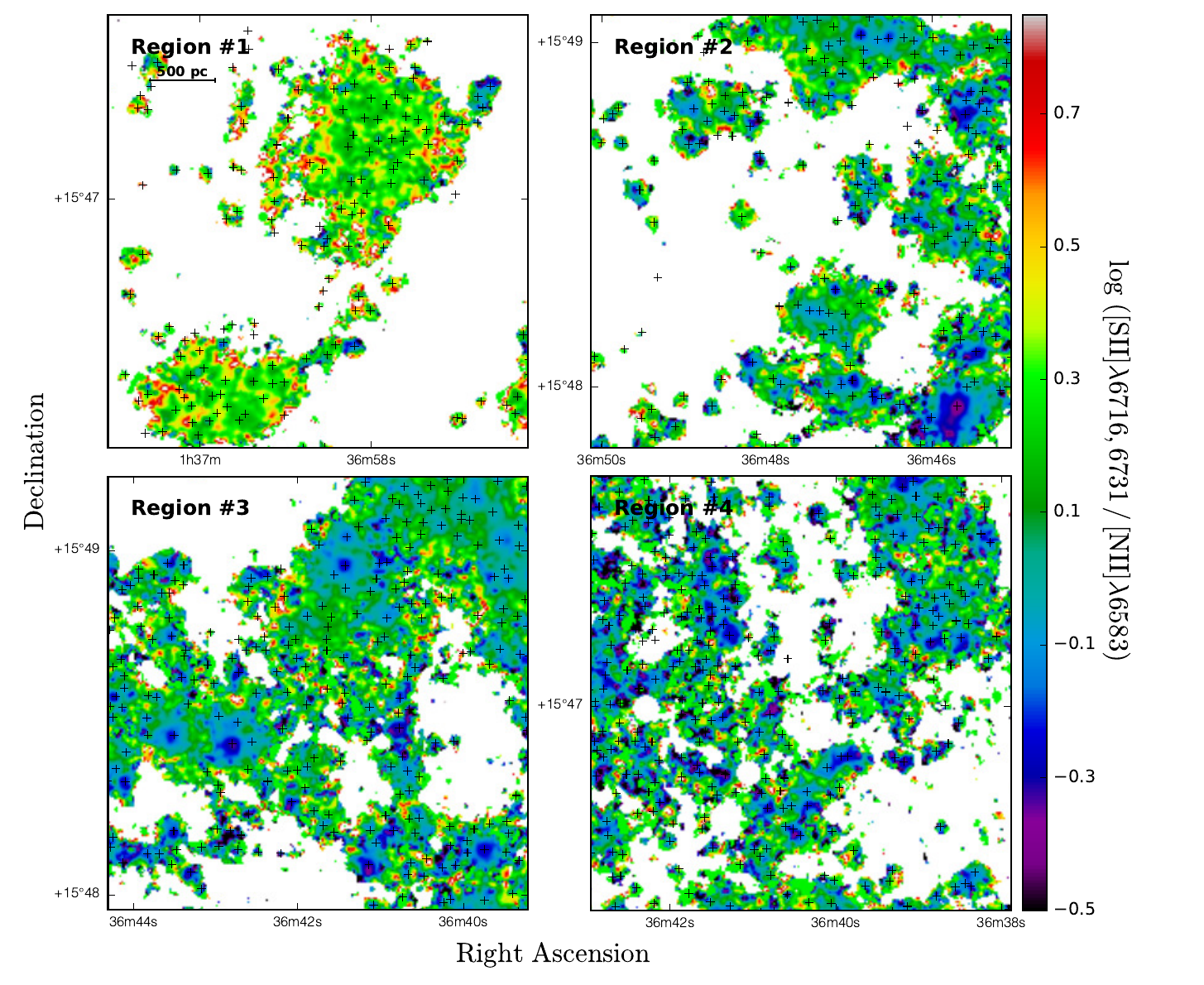} 
\vspace{-0.85cm}
\caption{Map of the [SII]$\lambda$6716,6731/[NII]$\lambda$6583 line ratio for areas defined in Figure\,\,\ref{SN3_flux_reg}. Black crosses indicate the location of the emission peaks (identified in $\S$\,\,\ref{Proc}).}
\label{s2n2_reg}
\end{center}
\end{figure*}

\clearpage

\begin{figure*}
\begin{center}
\hspace{0.35cm}
\vspace{-0.15cm}
\includegraphics[height=2.543in]{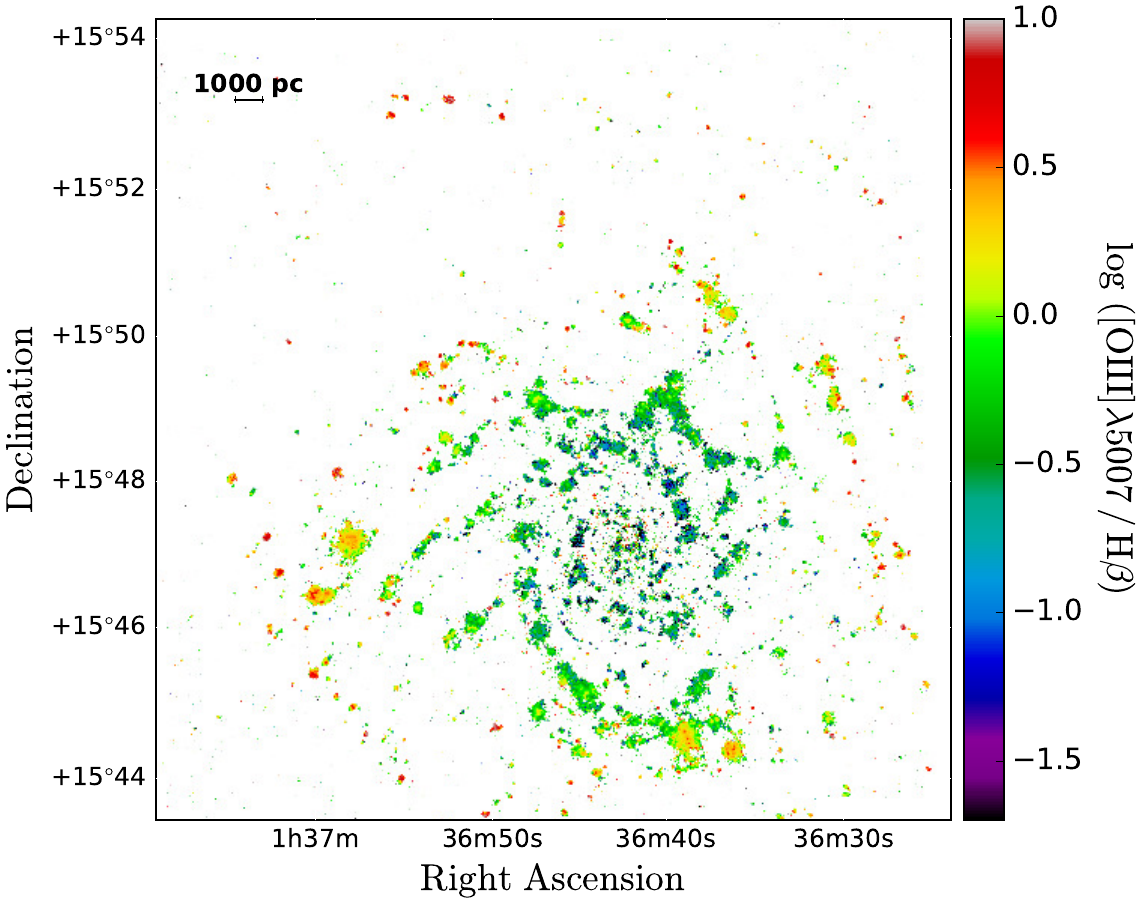} 
\includegraphics[height=2.543in]{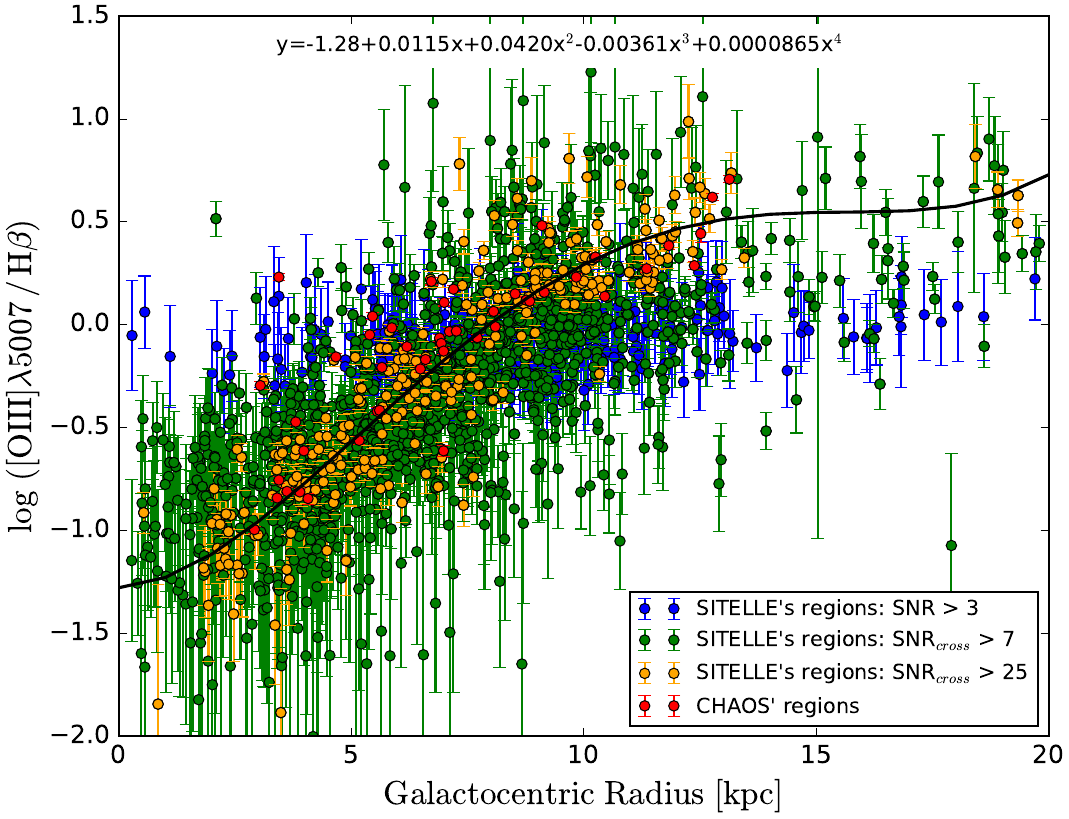} 
\vspace{-0.20cm}
\caption{[OIII]$\lambda$5007/H$\beta$ line ratio. On the right, line ratio for the integrated spectrum of each region as a function of their galactocentric radius. Three thresholds of the SNR have been considered and the CHAOS regions have been superimposed to the SITELLE's data. The black curve is a fit to the highest SNR$_{cross}$ subsample of regions.}
\label{o3h}
\end{center}
\end{figure*}

\begin{figure*}
\begin{center}
\vspace{-0.85cm}
\includegraphics[width=7.5in]{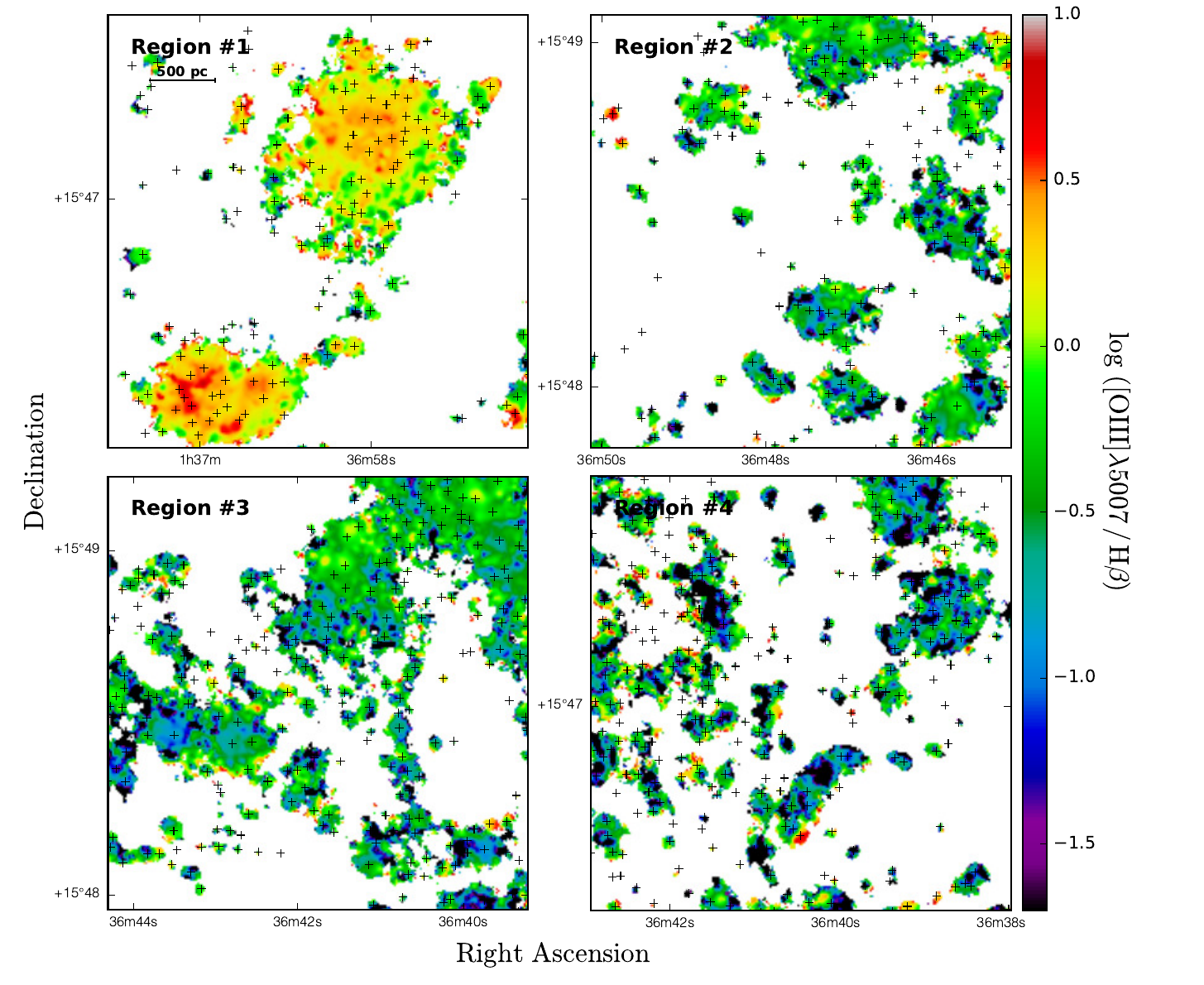} 
\vspace{-0.85cm}
\caption{Map of the [OIII]$\lambda$5007/H$\beta$ line ratio for areas defined in Figure\,\,\ref{SN3_flux_reg}. Black crosses indicate the location of the emission peaks (identified in $\S$\,\,\ref{Proc}).}
\label{o3h_reg}
\end{center}
\end{figure*}

\clearpage

\begin{figure*}
\begin{center}
\hspace{0.35cm}
\vspace{-0.15cm}
\includegraphics[height=2.543in]{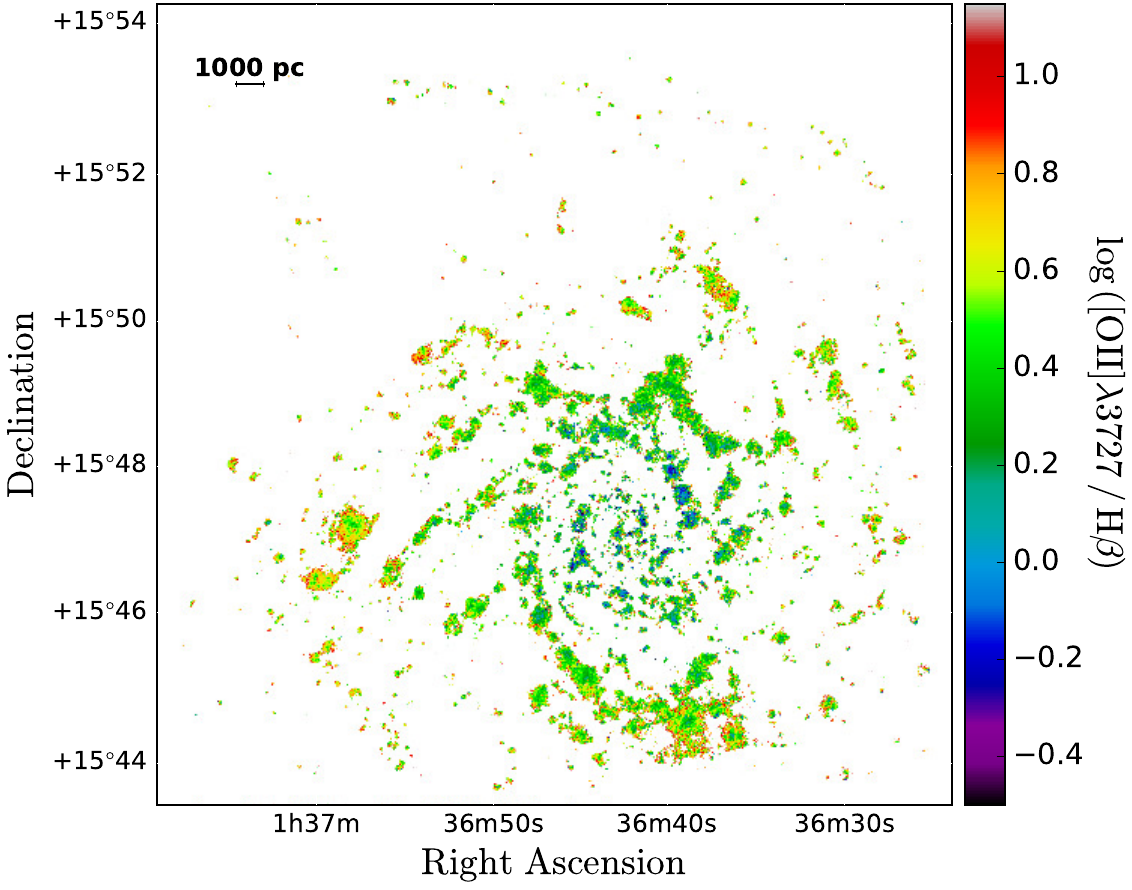} 
\includegraphics[height=2.543in]{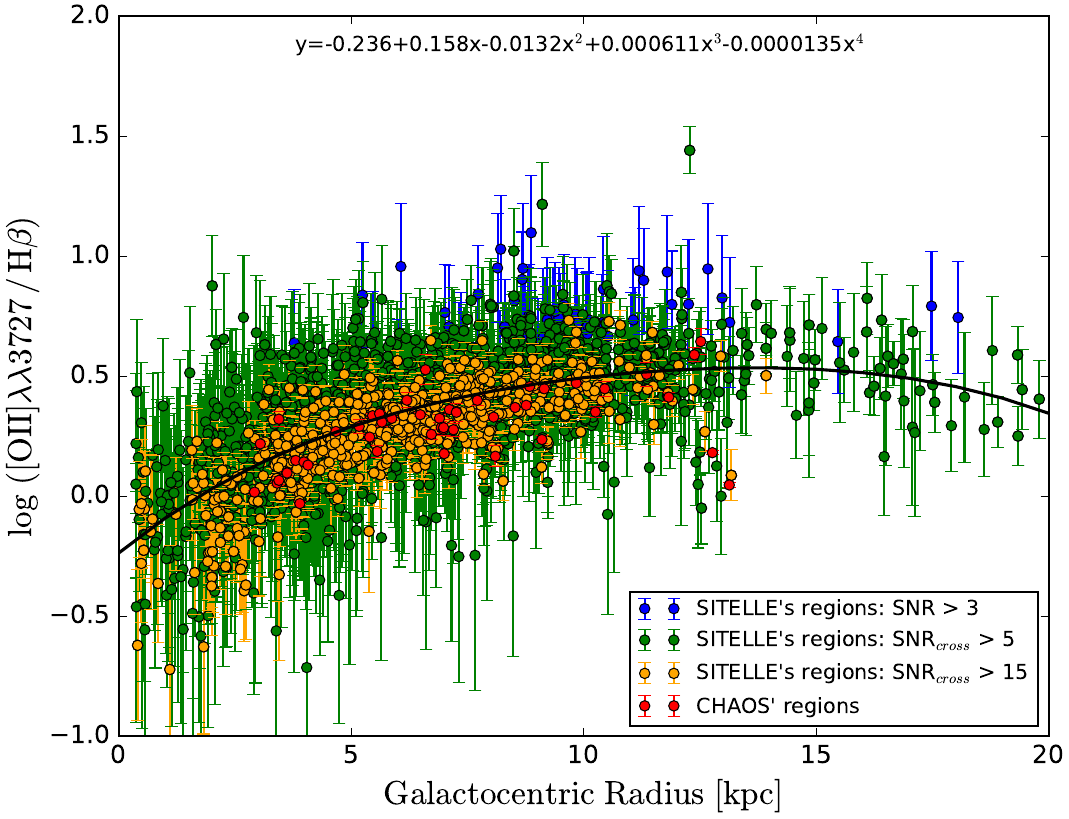} 
\vspace{-0.20cm}
\caption{[OII]$\lambda$3727/H$\beta$ line ratio. On the right, line ratio for the integrated spectrum of each region as a function of their galactocentric radius. Three thresholds of the SNR have been considered and the CHAOS regions have been superimposed to the SITELLE's data. The black curve is a fit to the highest SNR$_{cross}$ subsample of regions.}
\label{o2h}
\end{center}
\end{figure*}

\begin{figure*}
\begin{center}
\vspace{-0.85cm}
\includegraphics[width=7.5in]{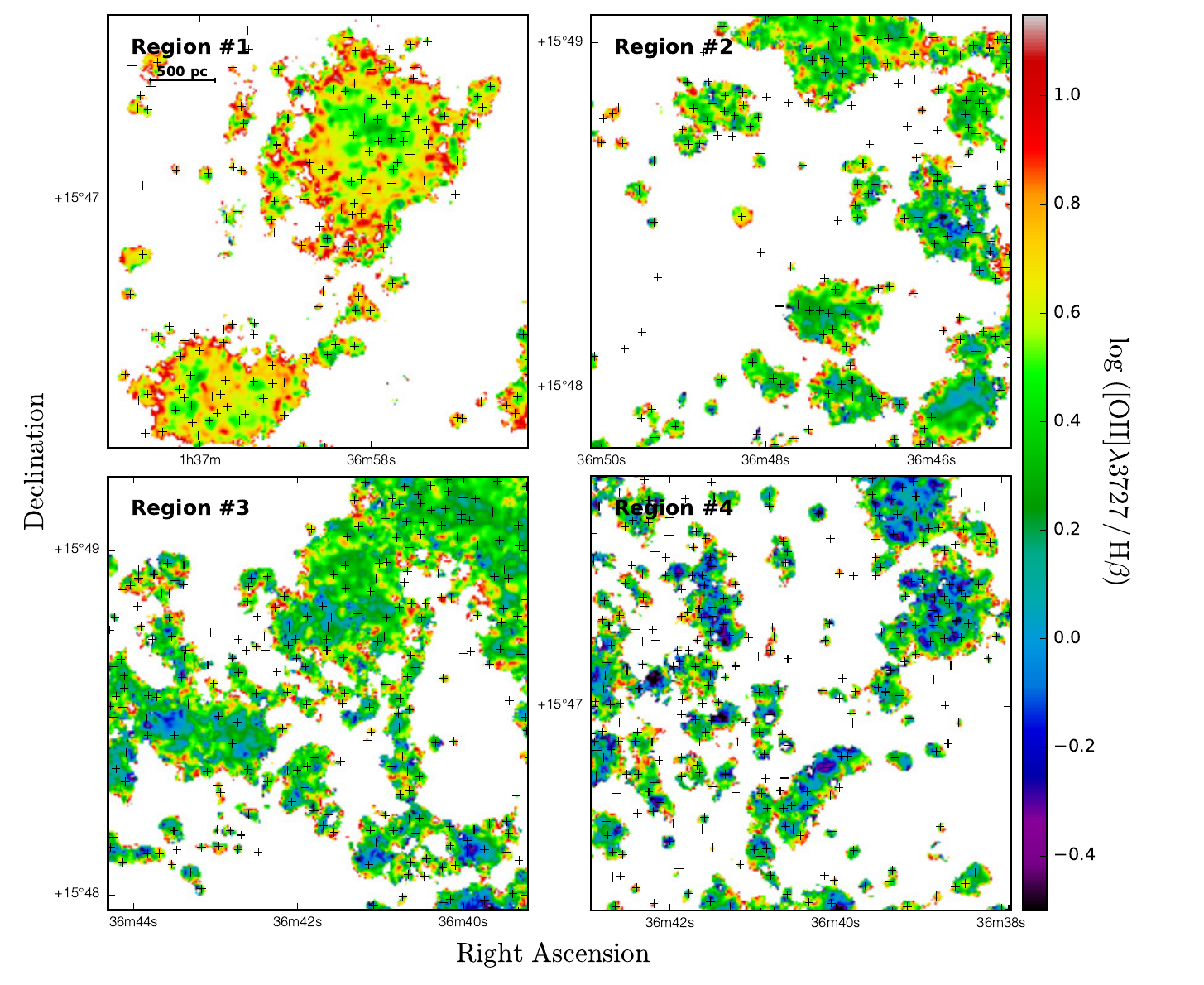} 
\vspace{-0.85cm}
\caption{Map of the [OII]$\lambda$3727/H$\beta$ line ratio for areas defined in Figure\,\,\ref{SN3_flux_reg}. Black crosses indicate the location of the emission peaks (identified in $\S$\,\,\ref{Proc}).}
\label{o2h_reg}
\end{center}
\end{figure*}

\clearpage

\vspace{-0.35cm}
\begin{figure*}
\begin{center}
\hspace{0.35cm}
\vspace{-0.15cm}
\includegraphics[height=2.543in]{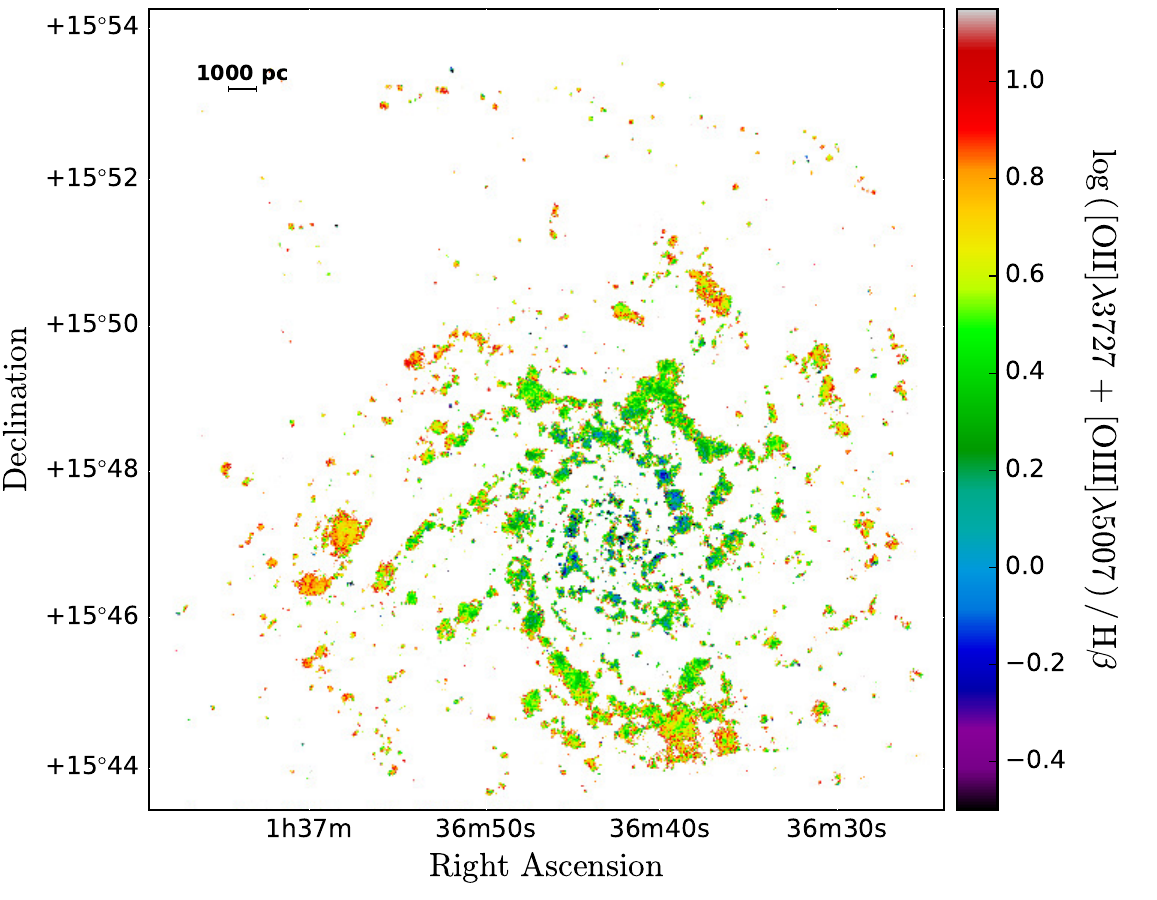} 
\includegraphics[height=2.543in]{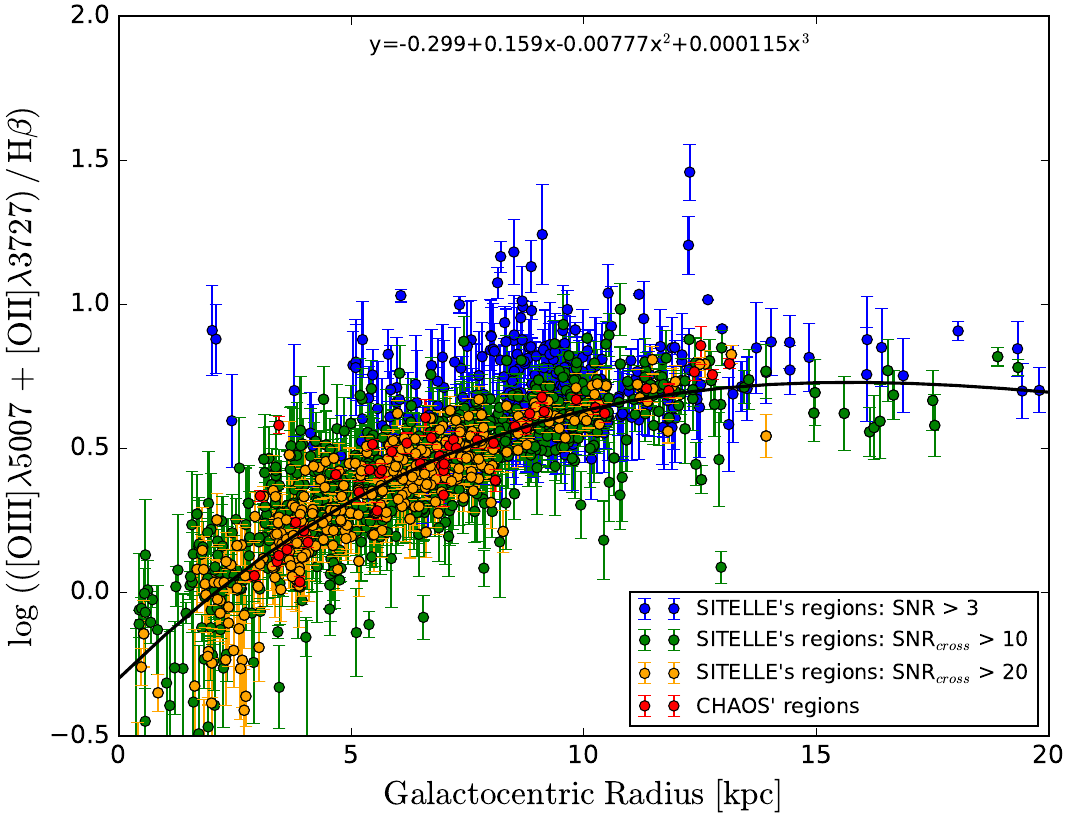} 
\vspace{-0.20cm}
\caption{([OII]$\lambda$3727+[OIII]$\lambda$5007)/H$\beta$ line ratio. On the right, line ratio for the integrated spectrum of each region as a function of their galactocentric radius. Three thresholds of the SNR have been considered and the CHAOS regions have been superimposed to the SITELLE's data. The black curve is a fit to the highest SNR$_{cross}$ subsample of regions.}
\label{o2h}
\end{center}
\end{figure*}

\begin{figure*}
\begin{center}
\vspace{-0.85cm}
\includegraphics[width=7.5in]{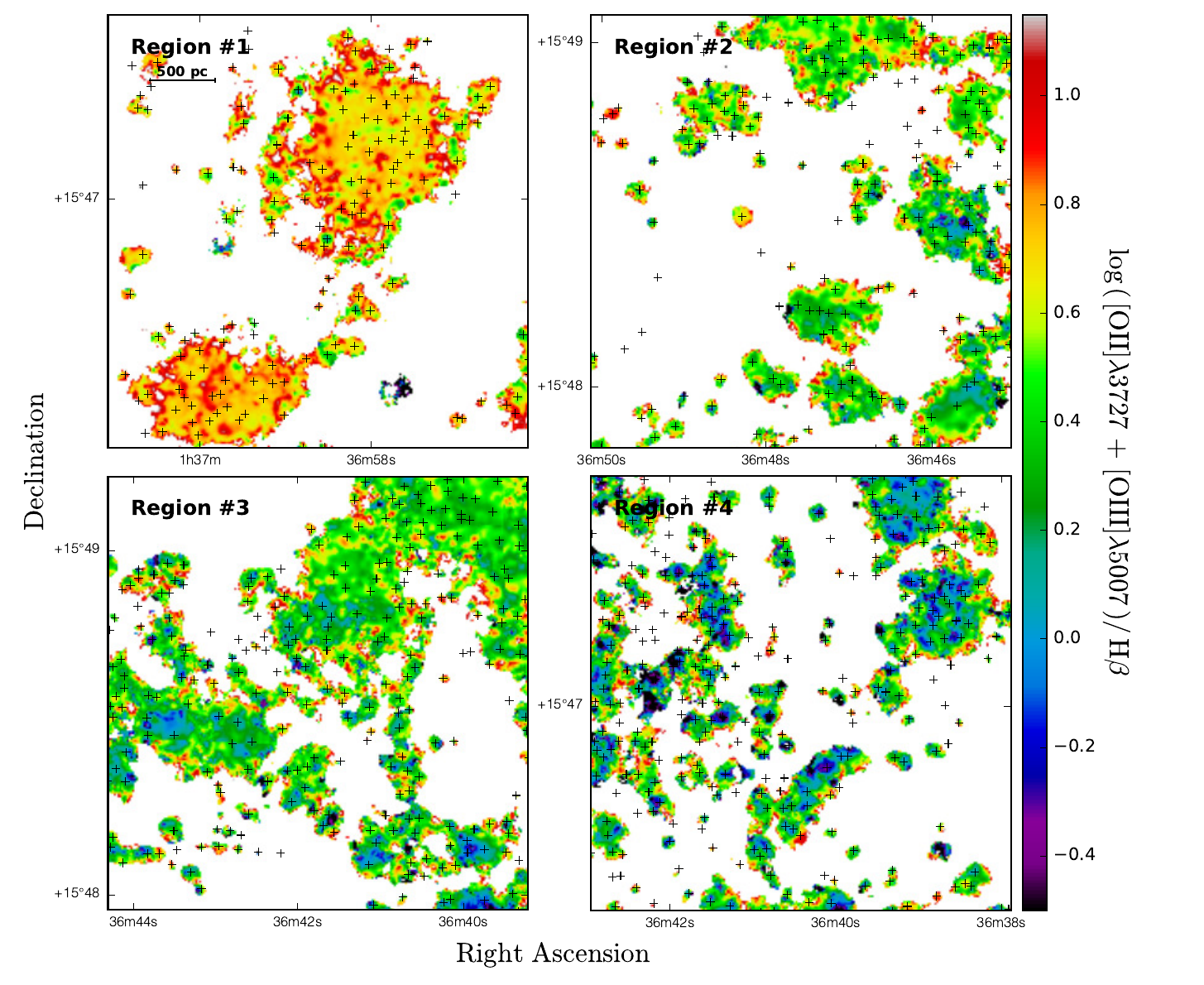} 
\vspace{-0.85cm}
\caption{Map of the ([OII]$\lambda$3727+[OIII]$\lambda$5007)/H$\beta$ line ratio for areas defined in Figure\,\,\ref{SN3_flux_reg}. Black crosses indicate the location of the emission peaks (identified in $\S$\,\,\ref{Proc}).}
\label{o2h_reg}
\end{center}
\end{figure*}

\clearpage

\begin{figure*}
\begin{center}
\hspace{0.35cm}
\vspace{-0.15cm}
\includegraphics[height=2.543in]{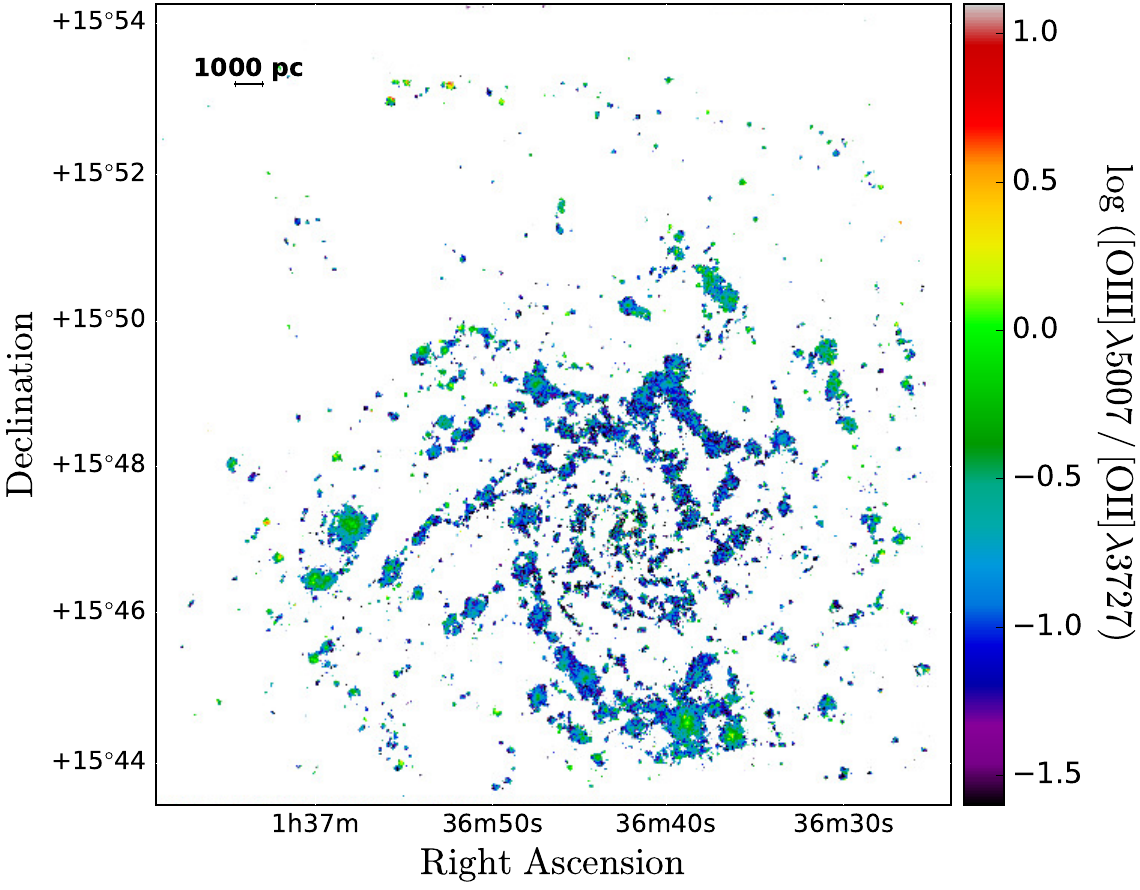} 
\includegraphics[height=2.543in]{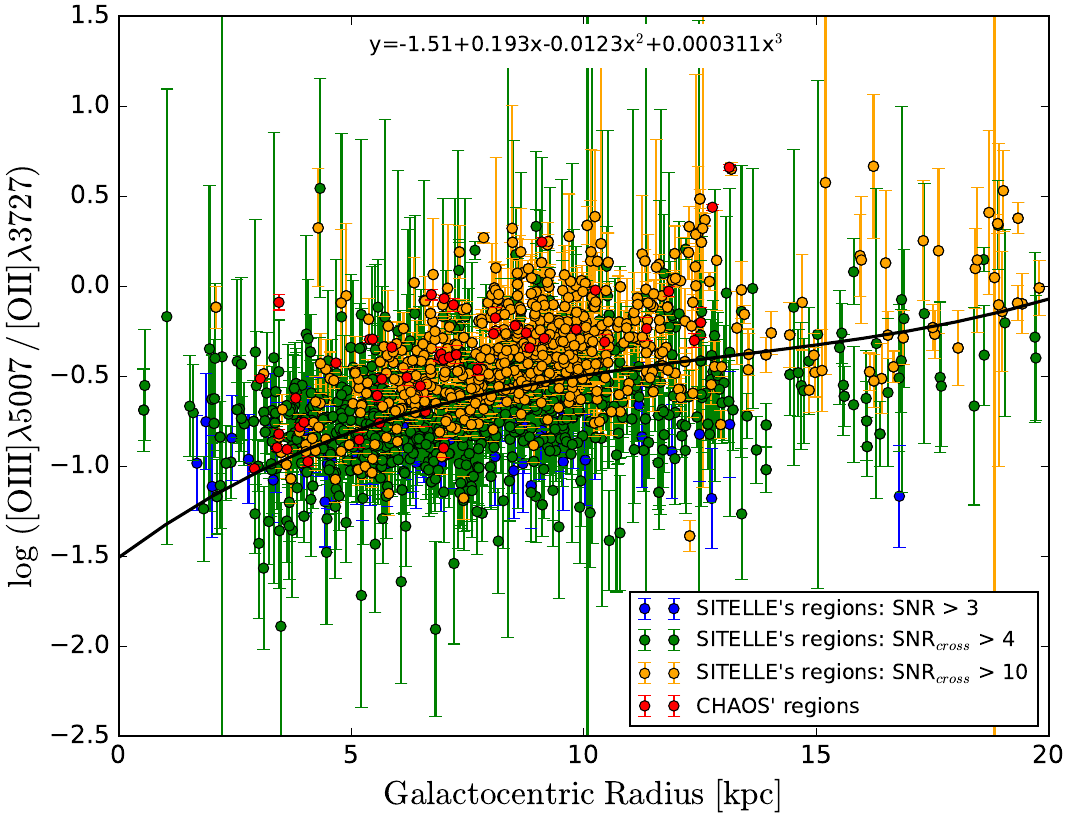} 
\vspace{-0.20cm}
\caption{[OIII]$\lambda$5007/[OII]$\lambda$3727 line ratio. On the right, line ratio for the integrated spectrum of each region as a function of their galactocentric radius. Three thresholds of the SNR have been considered and the CHAOS regions have been superimposed to the SITELLE's data. The black curve is a fit to the highest SNR$_{cross}$ subsample of regions.}
\label{o2h}
\end{center}
\end{figure*}

\begin{figure*}
\begin{center}
\vspace{-0.85cm}
\includegraphics[width=7.5in]{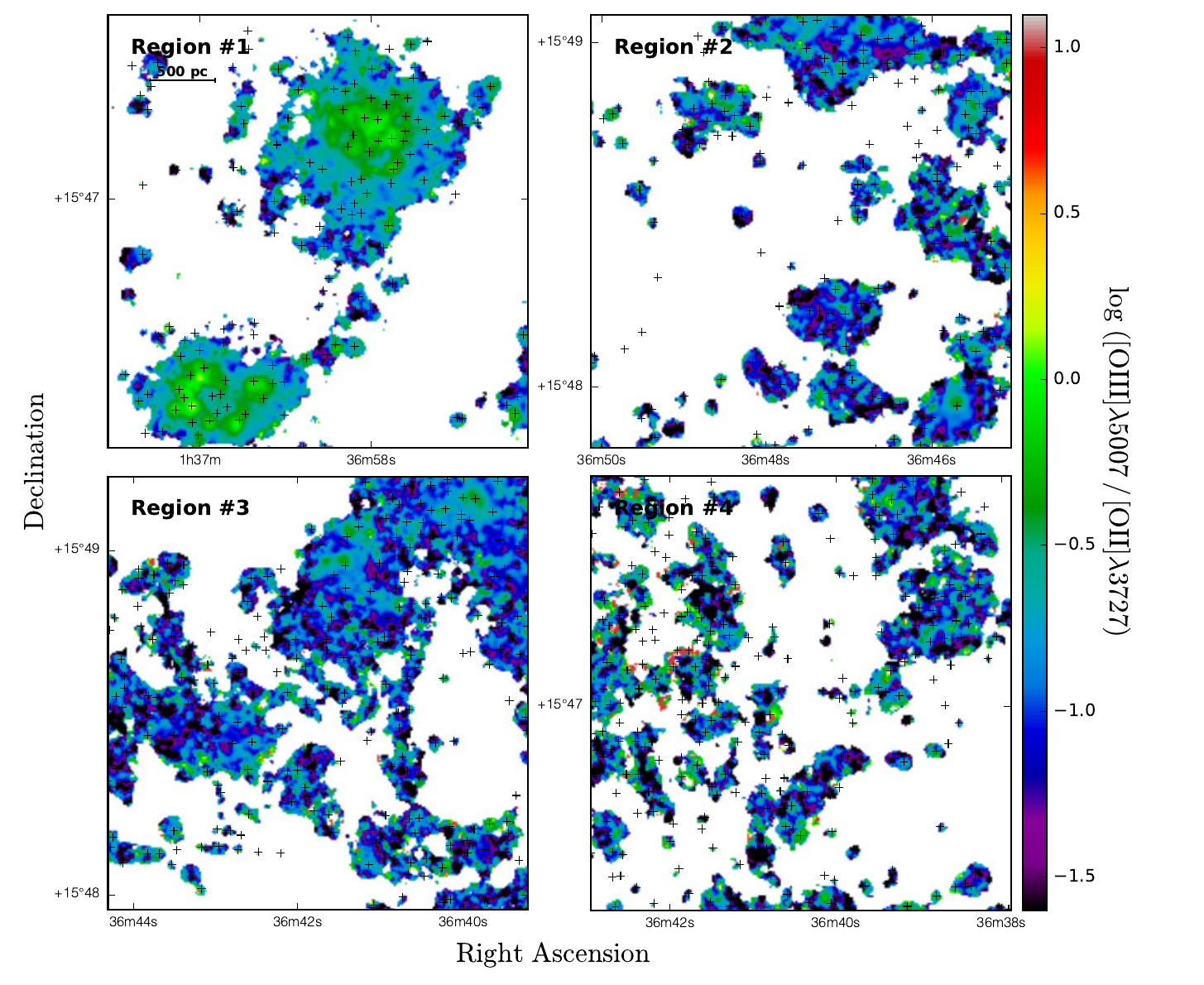} 
\vspace{-0.85cm}
\caption{Map of the ([OIII]$\lambda$5007/[OII]$\lambda$3727 line ratio for areas defined in Figure\,\,\ref{SN3_flux_reg}. Black crosses indicate the location of the emission peaks (identified in $\S$\,\,\ref{Proc}).}
\label{o2h_reg}
\end{center}
\end{figure*}

\clearpage

\begin{figure*}
\begin{center}
\hspace{0.35cm}
\vspace{-0.15cm}
\includegraphics[height=2.543in]{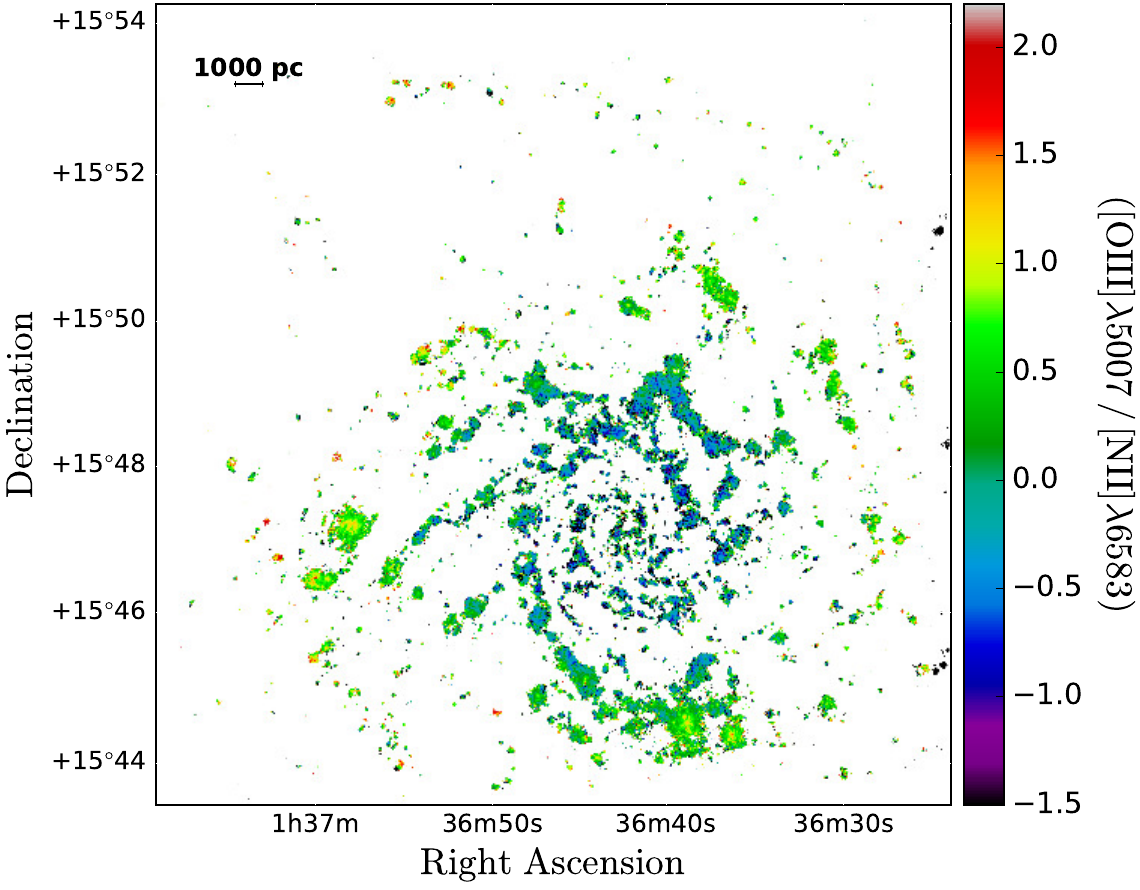} 
\includegraphics[height=2.543in]{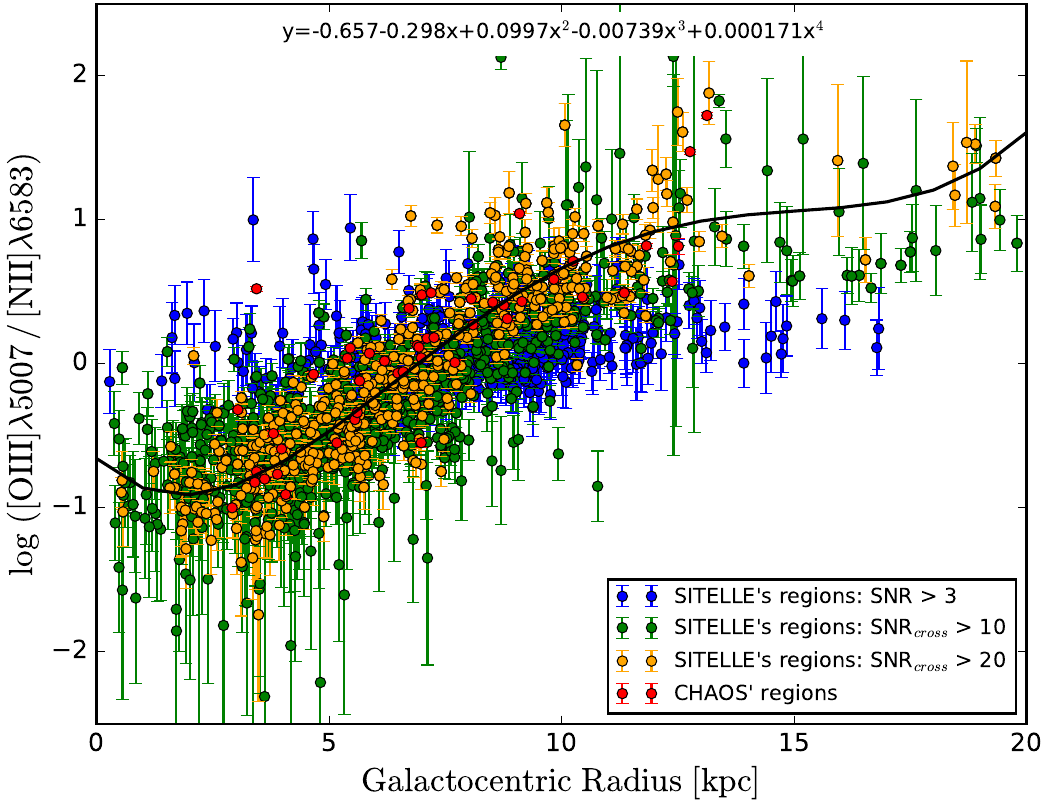} 
\vspace{-0.20cm}
\caption{[OIII]$\lambda$5007/[NII]$\lambda$6583 line ratio. On the right, line ratio for the integrated spectrum of each region as a function of their galactocentric radius. Three thresholds of the SNR have been considered and the CHAOS regions have been superimposed to the SITELLE's data. The black curve is a fit to the highest SNR$_{cross}$ subsample of regions.}
\label{o3n2}
\end{center}
\end{figure*}

\begin{figure*}
\begin{center}
\vspace{-0.85cm}
\includegraphics[width=7.5in]{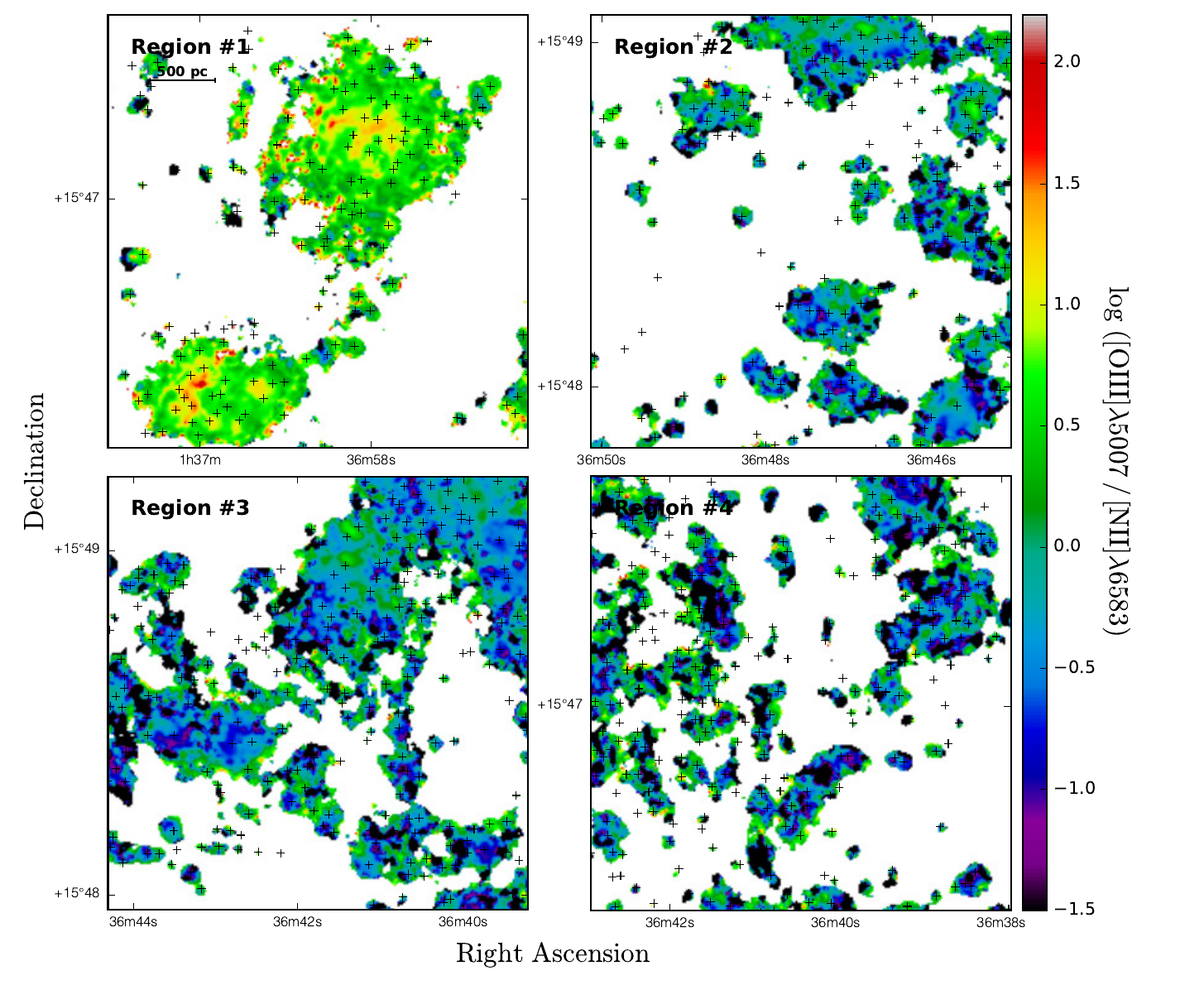} 
\vspace{-0.85cm}
\caption{Map of the [OIII]$\lambda$5007/[NII]$\lambda$6583 line ratio for areas defined in Figure\,\,\ref{SN3_flux_reg}. Black crosses indicate the location of the emission peaks (identified in $\S$\,\,\ref{Proc}).}
\label{o3n2_reg}
\end{center}
\end{figure*}

\clearpage

\begin{figure*}
\begin{center}
\hspace{0.35cm}
\vspace{-0.15cm}
\includegraphics[height=2.543in]{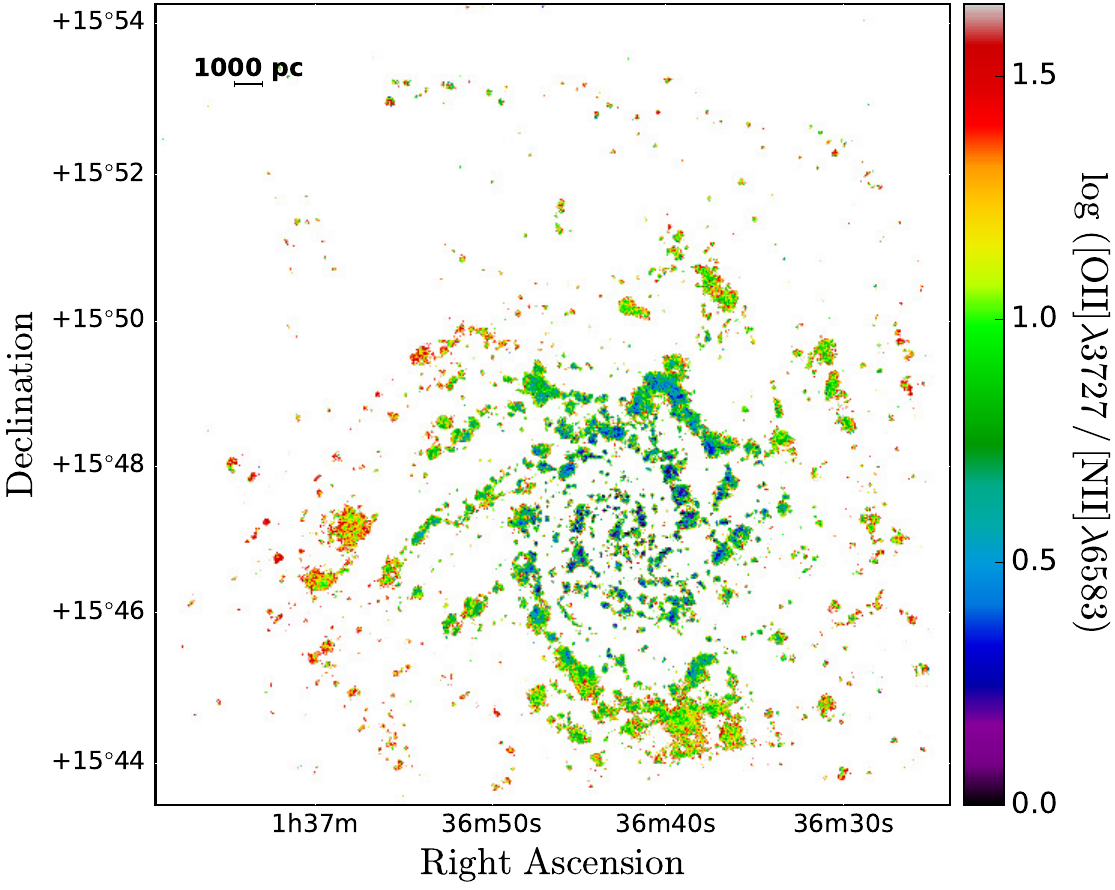} 
\includegraphics[height=2.543in]{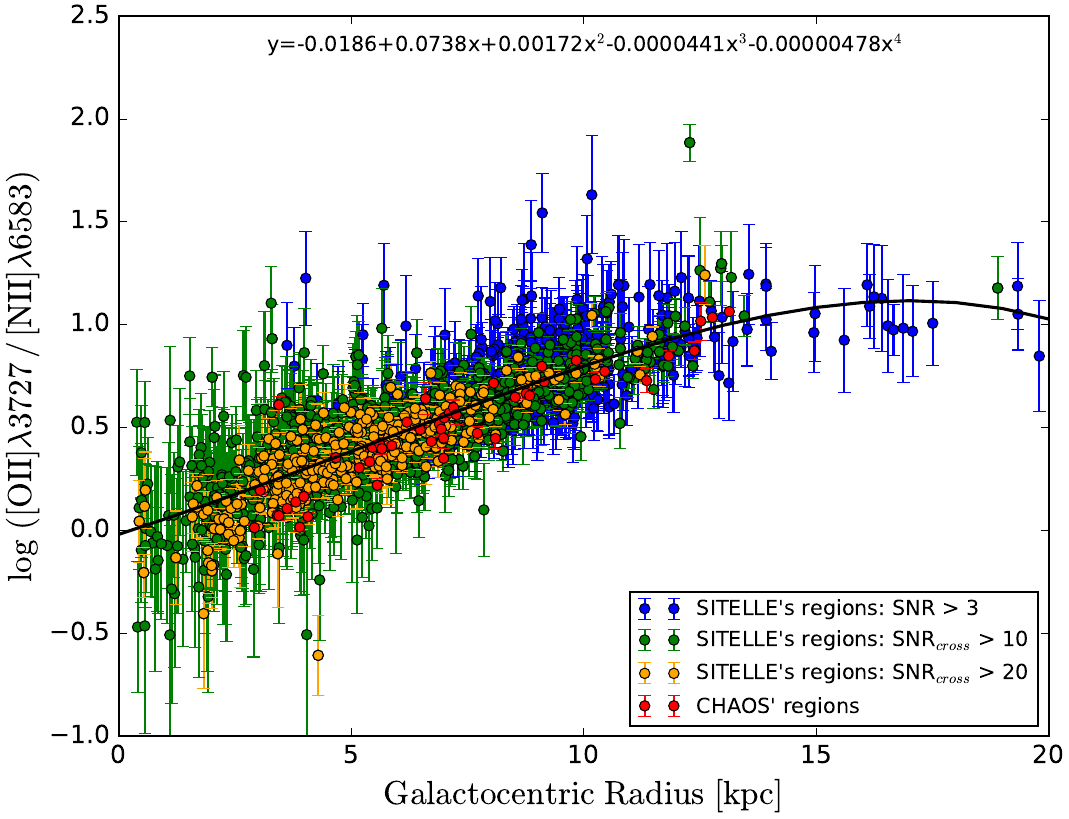} 
\vspace{-0.20cm}
\caption{[OII]$\lambda$3727/[NII]$\lambda$6583 line ratio. On the right, line ratio for the integrated spectrum of each region as a function of their galactocentric radius. Three thresholds of the SNR have been considered and the CHAOS regions have been superimposed to the SITELLE's data. The black curve is a fit to the highest SNR$_{cross}$ subsample of regions.}
\label{o2n2}
\end{center}
\end{figure*}

\begin{figure*}
\begin{center}
\vspace{-0.85cm}
\includegraphics[width=7.5in]{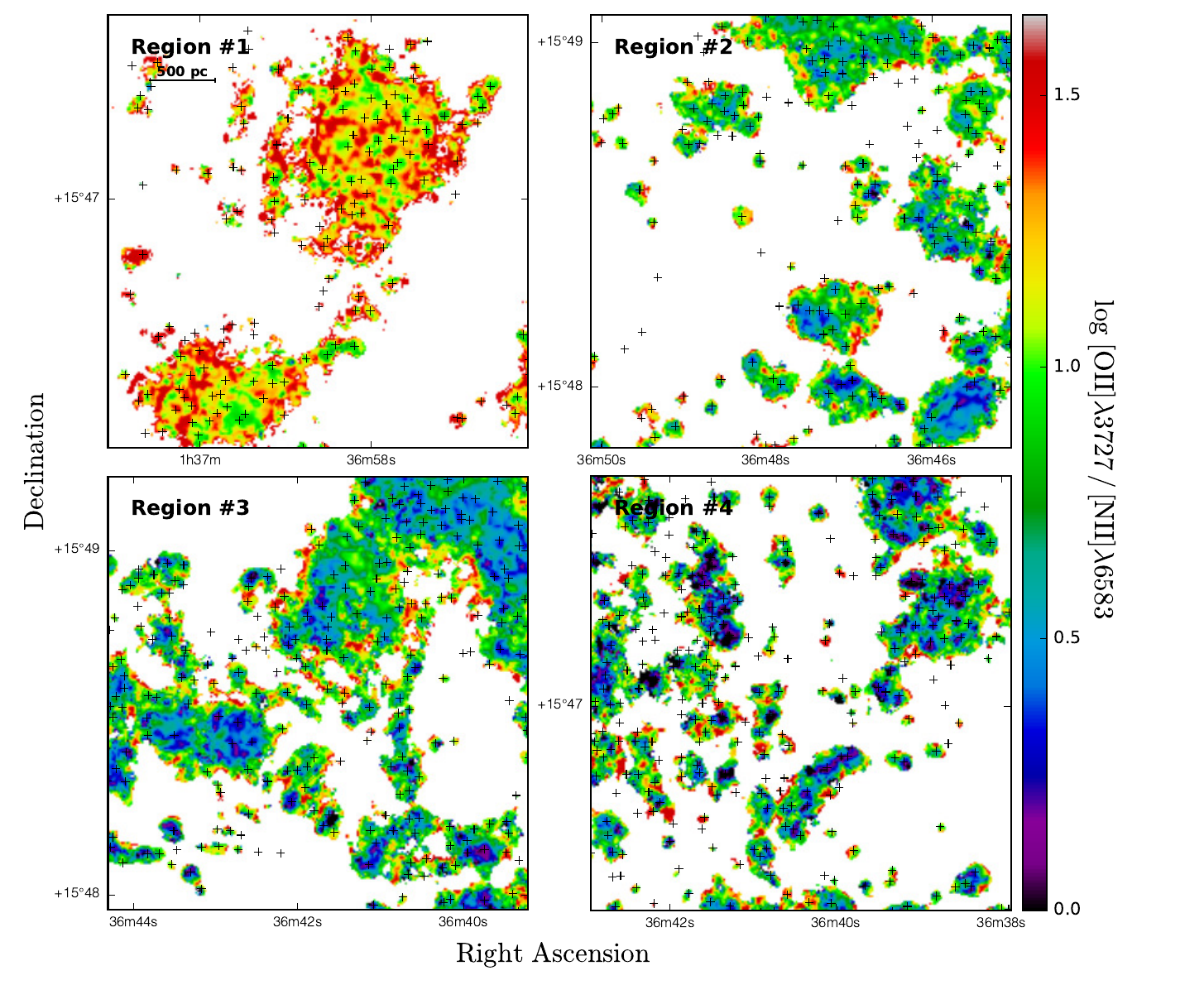} 
\vspace{-0.85cm}
\caption{Map of the [OII]$\lambda$3727/[NII]$\lambda$6583 line ratio for areas defined in Figure\,\,\ref{SN3_flux_reg}. Black crosses indicate the location of the emission peaks (identified in $\S$\,\,\ref{Proc}).}
\label{o2n2_reg}
\end{center}
\end{figure*}

\clearpage

\begin{figure*}
\begin{center}
\hspace{0.35cm}
\vspace{-0.15cm}
\includegraphics[height=2.543in]{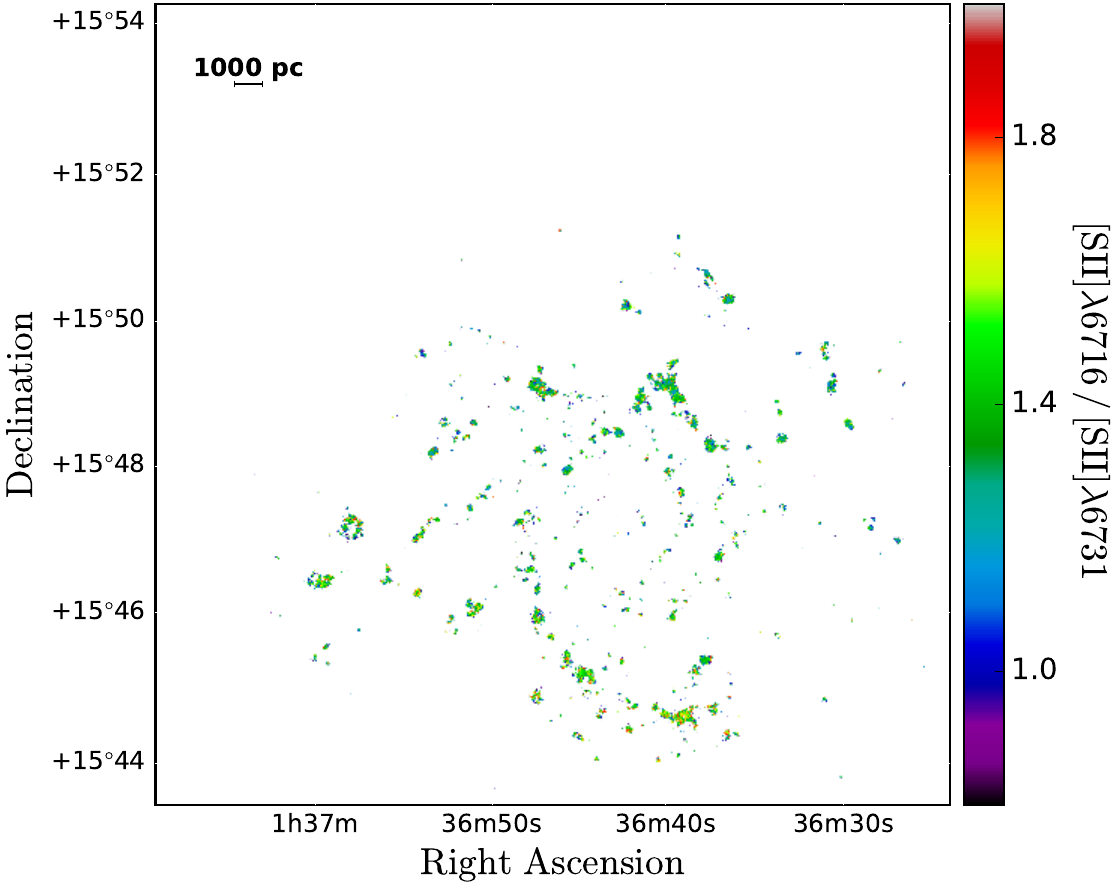} 
\includegraphics[height=2.543in]{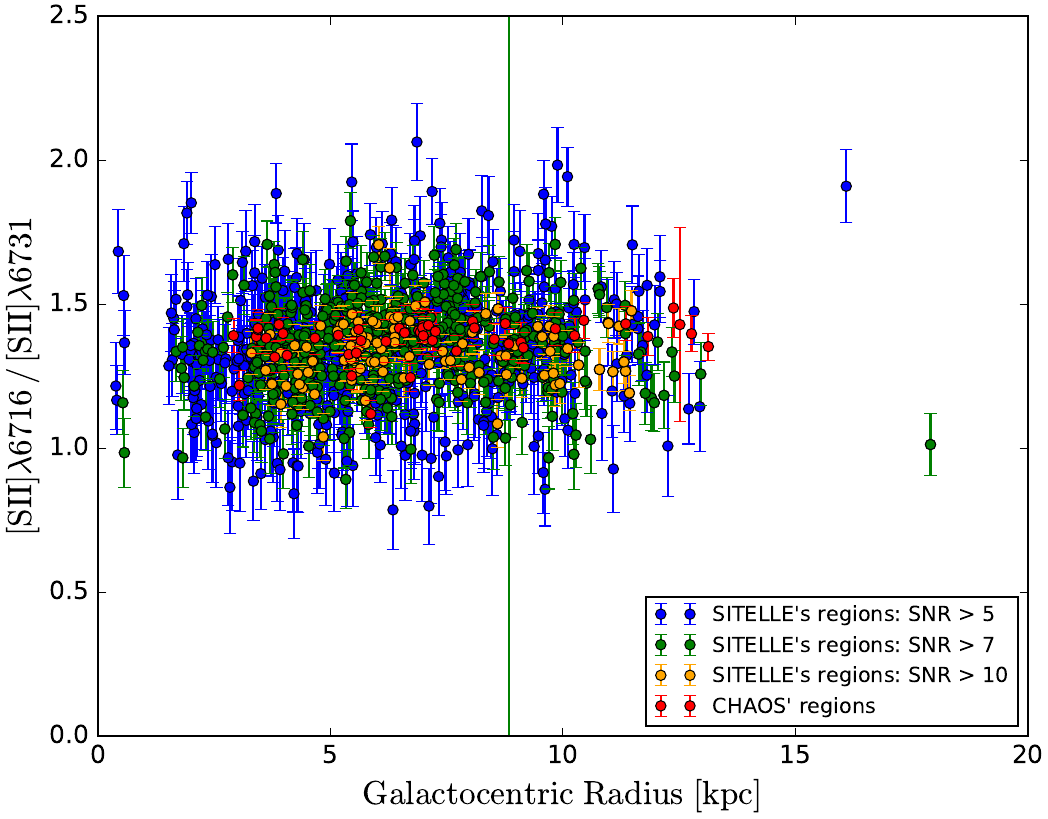} 
\vspace{-0.20cm}
\caption{[SII]$\lambda$6716/[SII]$\lambda$6731 line ratio. On the right, line ratio for the integrated spectrum of each region as a function of their galactocentric radius. Three thresholds of the SNR have been considered and the CHAOS regions have been superimposed to the SITELLE's data.}
\label{o2n2}
\end{center}
\end{figure*}

\begin{figure*}
\begin{center}
\vspace{-0.85cm}
\includegraphics[width=7.5in]{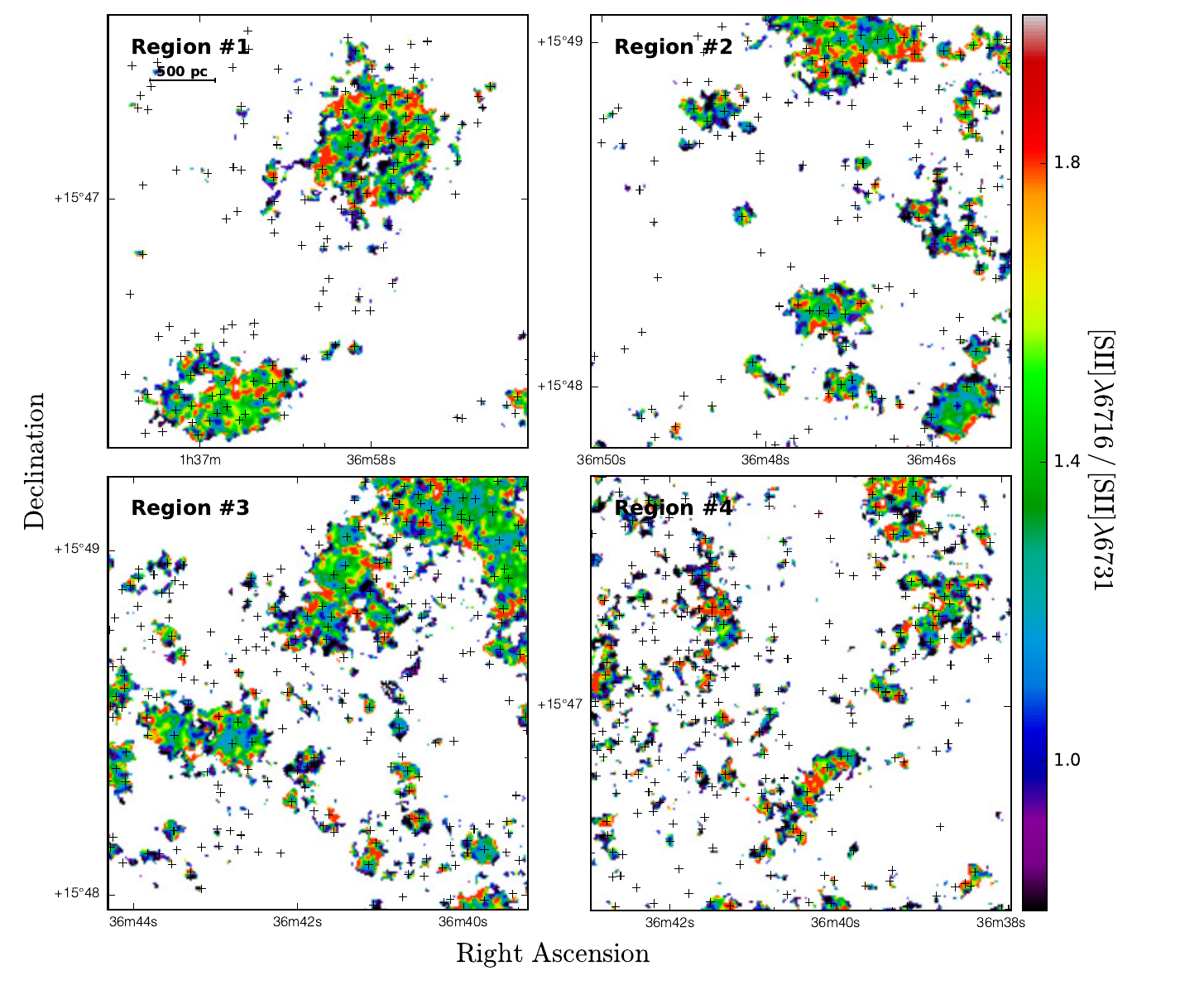} 
\vspace{-0.85cm}
\caption{Map of the [SII]$\lambda$6716/[SII]$\lambda$6731 line ratio for areas defined in Figure\,\,\ref{SN3_flux_reg}. Black crosses indicate the location of the emission peaks (identified in $\S$\,\,\ref{Proc}).}
\label{o2n2_reg}
\end{center}
\end{figure*}

\label{lastpage}

\end{document}